\documentclass[a4paper,11pt]{article}
\pdfoutput=1 

\usepackage{jcappub} 

\usepackage[T1]{fontenc} 
\usepackage{lineno}
\usepackage{xcolor}
\usepackage{orcidlink}
\usepackage{ulem} 
\numberwithin{equation}{section}

\title{\boldmath The Gravitational Wave Bias Parameter from Angular Power Spectra: Bridging Between Galaxies and Binary Black Holes}

\author[a,b]{Amir Dehghani \orcidlink{0009-0005-0395-9553},}
\author[c,d]{J. Leo Kim \orcidlink{0000-0001-8699-834X},} 
\author[a,b]{Dorsa Sadat Hosseini \orcidlink{0009-0008-2800-2725},} 
\author[e,a,*]{Alex Krolewski,}
\author[f]{Suvodip Mukherjee \orcidlink{0000-0002-3373-5236},}
\author[a,b]{and Ghazal Geshnizjani \orcidlink{0000-0002-2169-0579}}

\affiliation[a]{Perimeter Institute for Theoretical Physics, Waterloo, ON N2L 2Y5, Canada}
\affiliation[b]{Department of Applied Mathematics, University of Waterloo, Waterloo, ON N2L 3G1, Canada}
\affiliation[c]{Department of Physics, Engineering Physics \& Astronomy, Queen's University, Kingston ON K7L 3N6, Canada}
\affiliation[d]{Arthur B. McDonald Canadian Astroparticle Physics Research Institute, Kingston ON K7L 3N6, Canada}
\affiliation[e]{Waterloo Centre for Astrophysics, University of Waterloo, Waterloo, ON N2L 3G1, Canada}
\affiliation[f]{Department of Astronomy \& Astrophysics, Tata Institute of Fundamental Research, Colaba, Mumbai 400005, India}
\affiliation[*]{CITA National Fellow}

\emailAdd{a8dehgha@uwaterloo.ca}
\emailAdd{leo.kim@queensu.ca}
\emailAdd{d2sadath@uwaterloo.ca}
\emailAdd{akrolews@uwaterloo.ca}
\emailAdd{suvodip@tifr.res.in}
\emailAdd{ggeshnizjani@pitp.ca}

\abstract{
This study presents the modeling of the gravitational wave (GW) bias parameter by bridging a connection between simulated GW sources and galaxies in low redshift galaxy surveys 2MPZ and WISExSCOS (WISC).
We study this connection by creating a mock GW catalog, populating galaxy surveys  with binary black holes (BBHs) for different scenarios of the GW host-galaxy probability as a function of the galaxy stellar mass. We probe the observable consequences of this connection by exploring the spatial clustering of the GW sources in terms of the GW bias parameter. We consider a phenomenological broken power law model for the host-galaxy probability function, with a potential turnover $M_{\mathcal{K}}$ at high stellar mass ($10^{11}$ $M_{\odot}$ in the fiducial model) where the star formation efficiency begins to drop. We vary the parameters of the GW host-galaxy probability function and find that generically the GW bias increases as $M_{\mathcal{K}}$ increases (and gets suppressed as $M_{\mathcal{K}}$ decreases). The change in the GW bias parameter shows a maximum change of about $30\%$ for different scenarios explored in this work in comparison to the galaxy bias. Future measurements of the GW bias can help constrain $M_{\mathcal{K}}$ and the slopes of the host-galaxy probability function and thus offer insights into the underlying astrophysical processes.
}

\begin{document}
\maketitle

\flushbottom

\section{Introduction}
Following the first direct observation of gravitational waves (GWs) \cite{LIGOScientific:2016aoc} by the LIGO-Virgo-KAGRA (LVK) scientific collaboration \cite{KAGRA:2013rdx, LIGOScientific:2014pky, VIRGO:2014yos, PhysRevD.88.043007}, a new observational window has opened up for studying the Universe \cite{LIGOScientific:2018jsj, LIGOScientific:2020kqk, LIGOScientific:2021aug, KAGRA:2021duu}. Since then, GWs have allowed us to study astrophysics, cosmology, and fundamental physics from a completely new perspective \cite{Meszaros:2019xej, Berti:2022wzk, Adhikari:2022sve,LIGOScientific:2018mvr, LIGOScientific:2020ibl, LIGOScientific:2021djp}. For instance, binary mergers of compact objects are shedding new light on the ongoing $H_0$ tension in cosmology \cite{Schutz:1986gp, Oguri:2016dgk, LIGOScientific:2017adf, Chen:2017rfc, LIGOScientific:2018gmd, Mukherjee:2018ebj, DES:2019ccw, Mukherjee:2020hyn, Bera:2020jhx, Mukherjee:2022afz,LIGOScientific:2021aug,Afroz:2024joi}, and the first observations of the stochastic GW background from pulsar timing arrays are providing new insights in astrophysics and cosmology \cite{NANOGrav:2023gor, EPTA:2023sfo, Xu:2023wog, Sah:2024etc, Sah:2024oyg}.

There are many possible sources of GWs in the LVK frequency range, including different types of compact binary coalescing objects such as binary neutron stars, black hole neutron star binaries, and stellar binary black hole mergers (BBHs). Among these, BBHs are now mostly detected by the LVK network \cite{LIGOScientific:2018mvr, LIGOScientific:2020ibl, LIGOScientific:2021djp}. Understanding the properties of these binary systems requires comprehensive detailed models for their formation channels (see section 3.3 in \cite{Belczynski:2001uc} and section 3.1 in \cite{Neijssel2019}). To establish such a model, one needs to start from a population of stars and follow different possibilities of stellar evolution to the final stellar remnants \cite{Belczynski:2001uc}. These processes depend on the mass and metallicity of the stars, which in turn are functions of the mass, star formation rate, and metallicity of the host galaxies \citep{Belczynski:2001uc, Neijssel2019, OShaughnessy:2009szr, Dominik:2014yma, Mapelli:2017hqk, Giacobbo:2017qhh, Cao2018, Fishbach:2018edt, Santoliquido:2022kyu, Dominik:2012kk, Toffano:2019ekp, Artale:2019tfl, McCarthy2020,Afroz:2024fzp}. From stellar remnants to the final coalescence of the binary system, a variety of phenomena play a role, including stellar winds, initial masses and orbital geometry of the binary system, mass transfer, supernova kicks, and common envelope physics \cite{Belczynski:2001uc,Neijssel2019,Dominik:2012kk}. BBHs could arise from stellar remnants that were born together in a binary system, or they could arise through hierarchical mergers of smaller black holes \citep{2017PhRvD..95l4046G,2019PhRvL.123r1101Y,2020ApJ...893...35D}. Depending on the choice of model parameters, different possibilities exist for the formation channel of the binary compact objects. This complexity also reflects itself in the number of free input parameters in simulation codes for stellar evolution \cite{2011ApJS..192....3P}, galaxy evolution \cite{2014MNRAS.444.1518V,2023MNRAS.519.3154H}, and compact object binary formation \cite{Belczynski:2001uc,Neijssel2019,Dominik:2012kk,Santoliquido:2022kyu}.

Therefore, the final properties of these GW sources are the convolution of several physical quantities that are redshift-dependent and host-dependent from formation time to merger time. In particular, for BBHs, it is expected that their merger rates are influenced by the stellar mass, stellar metallicity, and star formation rate of the host galaxy \cite{Artale:2019tfl} and black hole (BH) masses may also have some dependencies on these properties \cite{2019ApJ...887...53F, 2022MNRAS.515.5495M, 2023MNRAS.523.4539K}. These global properties of the host galaxies are related to the formation and evolution of the galaxies' underlying dark matter halo.

Galaxy clustering measurements have been a rich source of information about the galaxy-halo connection (see \cite{Wechsler:2018pic} for a review), from the most basic information about host halo mass to the more subtle effects such as assembly bias and velocity bias \cite{Benson:1999mva, Sheth:1999mn, Cole2000, Peacock:2000qk, ValeOstriker04, Gao:2006qz, Croton:2006ys, Dalal08, Moster_2010, Tinker_2010, Behroozi_2013, Behroozi2019}. On large scales, clustering information is encoded in the bias parameter, which measures how much more galaxies are clustered compared to the underlying dark matter. In a similar way, the GW bias parameter relates the clustering of GW sources to the dark matter density, and naturally should depend on the formation history of the GW sources \cite{Scelfo:2018sny, Mukherjee:2019oma, Calore:2020bpd, Mukherjee:2020hyn,Diaz:2021pem, Libanore:2021jqv}. The dependence of GW bias parameter on redshift and spatial scales can therefore bring useful insight into the connection between stellar GW sources and their host galaxies.

Modeling the GW bias parameter at large cosmological scales using the relationship to stellar properties at the sub-parsec scale can be done using cosmological simulations together with BBH population synthesis models, but is challenging due to the range of scales involved in the problem (from sub-pc to Gpc). It is also a difficult problem to construct realistic simulations that can capture the actual cosmic evolution of stars and galaxies \citep[e.g.][]{Naab17,2023IAUS..373..283N,Sales:2022ich}. To avoid these difficulties, we use observed properties of galaxies and empirical galaxy characteristics to develop a phenomenological model for the GW-galaxy connection. In other words, we estimate the GW bias parameter by populating all-sky photometric galaxy surveys with synthetic GWs following a simple phenomenological relation for host-galaxy probability. We refer to this approach as the \textit{phenomenological approach}, as the observed galaxy properties are used to populate hosts of BBHs, instead of the \textit{simulation-based approach}, where GWs are populated based on semi-analytic models and combining population-synthesis simulations with synthetic galaxy catalogs from cosmological simulations \cite{Libanore21,2024MNRAS.530.1129P, Artale:2019tfl} or excursion set calculations \cite{Scelfo20}.

This paper thus investigates the potential of GW bias measurements to illuminate the GW-galaxy connection.
Here, we primarily focus on the dependence of the GW bias parameter on the stellar mass of host galaxies. This is the first logical step to take since stellar mass of galaxies is the primary galaxy property linked to the halo mass (and consequently linked to the large scale bias). While this approach cannot fully resolve uncertainties or parameter degeneracies in linking the observed GW bias to the underlying physics of the GW-galaxy connection, improving our understanding of the GW-galaxy connection is a valuable aim. This relation has important implications for astrophysical formation mechanisms of GW sources and for further contextualizing their use as cosmological probes.
We are also conducting follow-up research to incorporate additional galaxy characteristics that may influence GW bias behavior \cite{paper-II}.

Many other works have considered the observability of the GW-galaxy connection by studying their correlation \cite{Mukherjee:2021bmw} and also cross-correlations to study the GW bias parameter
\cite{Namikawa16, Mukherjee:2018ebj,Mukherjee:2019wcg, Mukherjee:2019oma, Calore:2020bpd, Mukherjee:2020hyn, Mukherjee:2020mha, Scelfo20, Libanore21, Diaz:2021pem, Libanore22, Gagnon23}.
Prior knowledge about the GW bias parameter can also be useful for cosmological model inference. Moreover, measurements of the GW bias parameter can differentiate between an astrophysical and a primordial origin for BBHs \cite{Raccanelli16,Scelfo18,Libanore23}; in this work we will hereafter only consider astrophysical origins for BBHs.

The structure of the paper is summarized as follows. In Section \ref{sec:bias} we will review the basics of the bias parameter for tracers and introduce our phenomenological approach to generate mock catalogs of BBH mergers from galaxy catalogs. Then, in Section \ref{sec:catalog} we will discuss the galaxy catalogs that were used in our study. In Section \ref{sec:modelling_gw}, we will describe how the mock catalogs of BBH mergers were generated. We present our prescription for calculating the 2D angular power spectra of BBH mergers and galaxies that were used to compute the BBH GW bias parameter in Section \ref{sec:power_spectrum}. The main results of this study will be presented and discussed in Section \ref{sec:results}. Finally, we make our concluding remarks in Section \ref{sec:conclusion}.

\section{Modeling Gravitational Wave bias parameter from BBHs based on observed galaxies} 
\label{sec:bias}

\subsection{GW bias parameter}
\label{sec:bias-def}
In cosmology, the bias parameter quantifies how the overdensity field of a tracer population follows the underlying matter overdensity field. In this work, we explore the properties of the gravitational wave (GW) bias parameter, $b_{GW}$, defined similarly to the well-known galaxy bias. That is, given the matter overdensity as a function of redshift $\delta(x,z)$, and similarly the overdensity of gravitational wave sources $\delta_{GW}(x,z)$, we define the GW bias to parameterise the operational relation $\delta_{GW}\equiv b_{GW}[\delta]$ (which in the linear approximation reduces to a proportionality constant). Note that in this work, we are only exploring the GW bias parameter for stellar BBH sources of GWs. However, in principle, it can refer to any kind of GW source, including binary neutron stars, neutron star-BHs, supermassive BHs, and primordial BHs.

Since we will be working with a photometric survey and the angular power spectrum, we will focus on the inference of the bias parameter in spherical harmonic space. To be more specific, we assume that $\delta_{GW}(\ell,z)$ depends on $\delta(\ell,z)$ as
\begin{equation}
     \delta_{GW}(\ell,z)=b_{GW}(\ell,z)\delta(\ell,z)\,.
\end{equation}

We measure $b_{GW}$ from the angular power spectrum of mock GW sources created from the observed galaxy catalog (described more fully in Section~\ref{sec:modelling_gw}). This measurement thus requires a theoretical model for the angular power spectrum, $C_\ell^{\textrm{Mod}}$, which we compute using the method described in \cite{2011PhRvD..84d3516C} (in particular, see Eq. (44) in that paper),
\begin{align}
    C_\ell^\mathrm{Mod}=
    4\pi \int d (\ln{k})~\mathcal{P}_\mathcal{R}(k) \left(\Delta_{\ell}(k)\right)^2 \,,
\label{eq:cl_the}
\end{align}
where $\mathcal{P}_\mathcal{R}(k)$ is the dimensionless power spectrum of the primordial curvature perturbation, and $\Delta_\ell(k)$ is the line-of-sight projection of the matter density, weighted by the galaxy redshift distribution $dN/dz$:
\begin{equation}
    \Delta_\ell(k) = \int \, dz \, \frac{dN}{dz} T_\delta (k, z) j_\ell(k \chi(z))\,,
\end{equation}
where $\chi$ is the comoving distance, $T_\delta(k,z)$ is the matter overdensity transfer function (which relates the linear matter power spectrum and the primordial curvature power spectrum $P_{\mathrm{mm}}(k) = (T_\delta(k,z))^2 \mathcal{P}_\mathcal{R}(k)$) and $j_\ell$ is a spherical Bessel function of degree $\ell$. We use the post-Limber calculation in the \verb|CAMBSOURCES| module \cite{Lewis:1999bs,2011PhRvD..84d3516C,Howlett:2012mh} to compute $C_\ell^{\textrm{Mod}}$, which neglects contributions from the unequal-time nonlinear power spectrum and rescales the linear transfer function to match the \verb|HALOFIT| nonlinear matter power spectrum \cite{Challinor05}, while still performing the 1D redshift integrals over the spherical Bessels rather than replacing them with the Limber approximation.

Injecting $dN/dz$ from the siren catalog and substituting the dark-matter power spectrum, the expected model angular power spectrum, $C_\ell^\mathrm{Mod}$, encodes the autocorrelation for a sample of the same number of sources if they were unbiased tracers of the dark matter distribution. We call this the ``model angular power spectrum'' as it is the projection of the theoretical 3D power spectrum onto the sky, using the observed redshift distribution.  In this work, we are interested in inferring the GW bias at fixed cosmology, and so we do not vary the cosmological parameters. We use the most recent values of the cosmological parameters from the Planck Collaboration result \cite{Planck:2018vyg} for the Flat $\Lambda$CDM (Lambda Cold Dark Matter) cosmological model, i.e. with $\Omega_{c}h^2 = 0.122$, $\Omega_{b}h^2 = 0.022$, $H_0 = 67.5$ km s$^{-1}$ Mpc$^{-1}$, $n_s = 0.965$, $A_s=2\times 10^{-9}$ calculating the power spectrum with \verb|CAMB| \cite{Lewis:1999bs,2011PhRvD..84d3516C}.

We then proceed by fitting a constant $b_{GW}$ within each redshift bin within a specified range of scales, chosen to ensure that we are within the range where linear bias is valid. We therefore fit the following model, $\hat{C}_\ell^{GWGW}$, to the observed power spectrum $C_\ell^{GWGW}$:
\begin{equation}
\label{eq:cl_ob}
    \hat{C}_\ell^{GWGW} = b_{GW}^2 C_\ell^{\textrm{Mod}}.
\end{equation}
We also check the scale-dependence of the ratio of GW and matter power spectrum to ensure that linear bias is a valid description on the scales that we are interested in, by testing an empirical linear scale dependence in $\ell$ for the bias (see Section~\ref{sec:bgw_ellbounds}). The GW bias parameter, similar to the galaxy bias, may have other non-local corrections, but that is beyond the scope of this work.

\subsection{Phenomenological approach: assigning mock BBHs to observed galaxies}
\label{subsec:method}
\begin{figure*}
\centering
\includegraphics[width=\hsize]{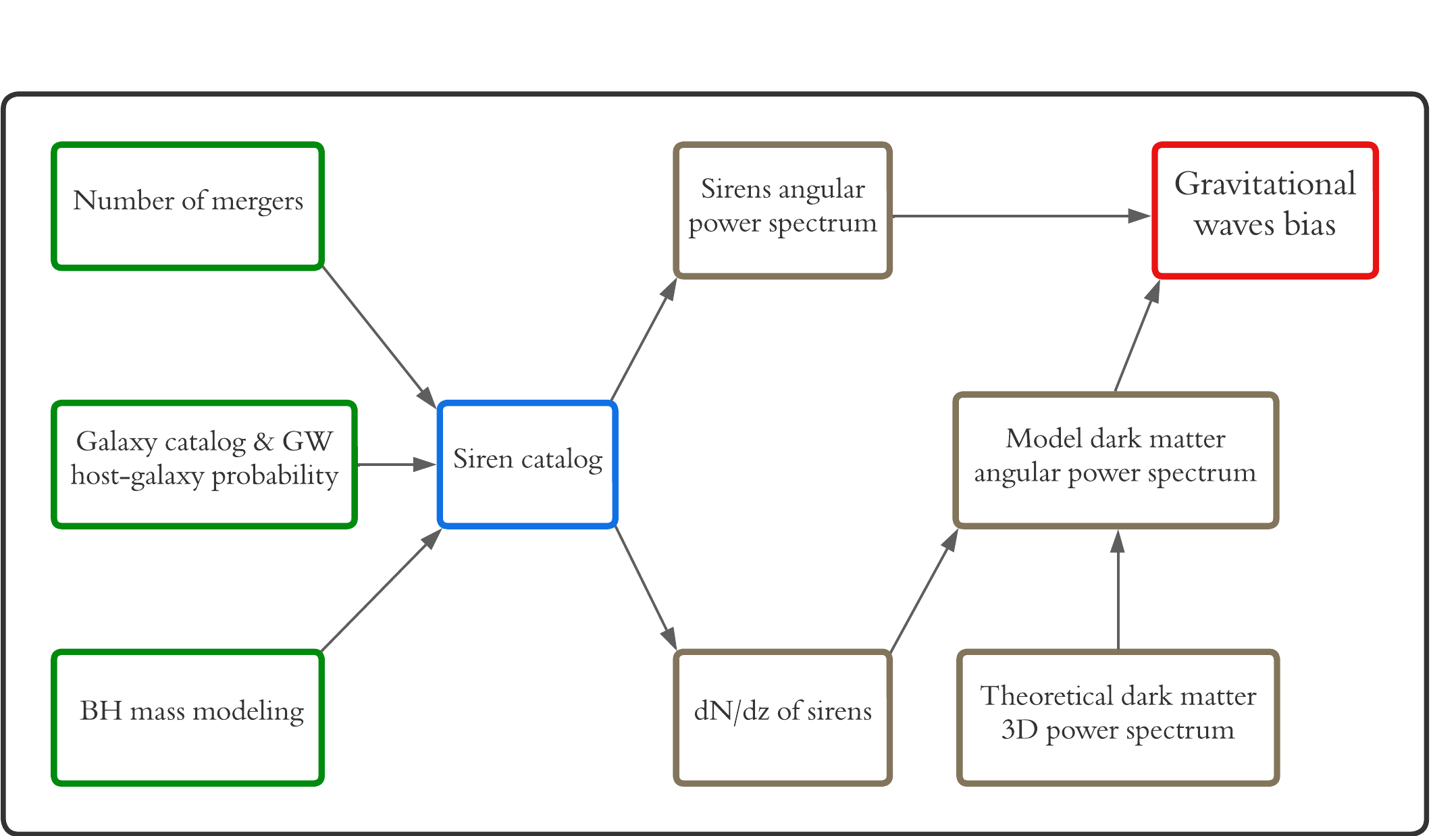}
\caption{Flowchart showing our procedure for modeling the GW bias by populating galaxy catalogs with mock BBHs. \label{fig:chart1}}
\end{figure*}

A schematic overview of our analysis is illustrated in Figure \ref{fig:chart1}. In order to estimate the GW bias parameter (the red box in Figure \ref{fig:chart1}), we require the angular power spectrum for the merging BBHs and the model dark matter angular power spectrum. To obtain the former, we generate a mock siren catalog (the blue box in Figure \ref{fig:chart1}) from existing observed galaxy catalogs.
Catalogs and their properties will be discussed in more detail in Section \ref{sec:catalog}. Generating the siren catalog requires three pieces of information (the green boxes in Figure \ref{fig:chart1}), (I) \textit{Number of mergers}: the expected number of sirens throughout cosmic history as a function of redshift, (II) \textit{Galaxy catalog and GW host-galaxy probability}: which together give the angular distribution of sirens in the sky, and (III) \textit{BH mass modeling}: which comes from the astrophysical properties of sirens and their hosts. In principle, all of these ingredients can be correlated with each other and thus require careful treatment when generating a working mock catalog of BBHs. However, as will become evident in our exploration of GW bias, only the host-galaxy probability plays a significant role in affecting clustering. Lastly, the model angular power spectrum is calculated from the redshift distribution of the siren catalog and the matter power spectrum using the method explained in Section \ref{sec:bias-def}.

The number of BBHs at a given redshift is expected to primarily depend on the star formation rate \cite{madau2014cosmic}. By considering a relation between the star formation rate, the BH formation, and the BH coalescences, we can estimate the BH merger rate, which will be discussed in detail in Section \ref{subsec:merger-rate}. Once the BBH merger rate as a function of redshift is determined, the merger rate can then be integrated over the observing time and the size of a redshift bin to determine the total number of expected sirens in that redshift bin. Although the number of BBHs in a given redshift bin can be calculated using the BBH merger rate, to evaluate GW bias, we also need the angular distribution of BBHs in the sky. We adopt a \textit{phenomenological survey-based} approach (or for short a \textit{phenomenological} approach), populating host galaxies with BBHs based on a phenomenological relation for the host-galaxy probability. Specifically, we assume that the probability of a given galaxy hosting a GW siren is a broken power-law function of the stellar mass. This is discussed in more detail in Section \ref{subsec:GW_Gal}.

There are several advantages of the phenomenological approach. First, galaxy formation is a difficult problem for simulations \citep[e.g.][]{Naab17,2023IAUS..373..283N,Sales:2022ich} and tying the BH properties directly to the observed stellar mass, star formation rate, and metallicity removes the need to marginalize over possible astrophysical models for sub-grid physics in galaxy simulations. On a related note, cosmological hydrodynamical simulations that can resolve properties of the gas in galaxies (and thus the star formation rate and metallicity) are limited to small volumes, usually $\sim10^{-3}$ Gpc$^3$, whereas we use surveys that span much larger volumes, $\sim1$ Gpc$^3$ for 2MPZ (2MASS Photometric Redshift catalog) and $\sim 10$ Gpc$^3$ for WISC (WISExSCOS Photometric Redshift catalog). Finally, our phenomenological model for the GW host-galaxy probability (i.e.\ the probability that a given galaxy hosts a BBH) allows flexibility in considering many different possibilities for the GW host-galaxy probability, rather than being tied to a single physical description of the galaxy-BBH connection. However, using simulations can be advantageous over using observations due to issues with the galaxy catalogs themselves, such as the completeness of a survey and observational errors (i.e., the fact that the galaxy survey does not contain all possible galaxies, but rather all galaxies above some fixed luminosity or mass threshold). In this work, we therefore always present our results in a way that minimizes the dependency on the uncertain details of the galaxy survey selection function.

The phenomenological approach requires an input galaxy catalog, either a photometric survey (requiring measurement of the angular power spectrum to infer the GW bias) or a spectroscopic survey (allowing us to use the 3D power spectrum instead). Although the 3D power spectrum using spectroscopic surveys can give a more precise measurement of clustering, and allows us to directly infer from the galaxy spectra the metallicity and the star formation rate, spectra are expensive to obtain and, therefore, spectroscopic catalogs are smaller and biased towards more massive and brighter galaxies. Furthermore, since most observations of GWs do not have a spectroscopic redshift measurement due to lack of an electromagnetic counterpart, the angular power spectrum is more closely tied to current observational prospects. While in this work we will use photometric galaxy surveys, we will explore the GW bias parameter using the 3D power spectrum and specifically using the SDSS-Main Galaxy Survey sample \cite{Ross_2015} in an accompanying paper \cite{paper-II}.

\section{Summary of the galaxy catalogs used in the analysis} 
\label{sec:catalog}

\subsection{2MPZ and WISC catalogs}
\begin{figure*}
\centering
\includegraphics[width=0.48\hsize]{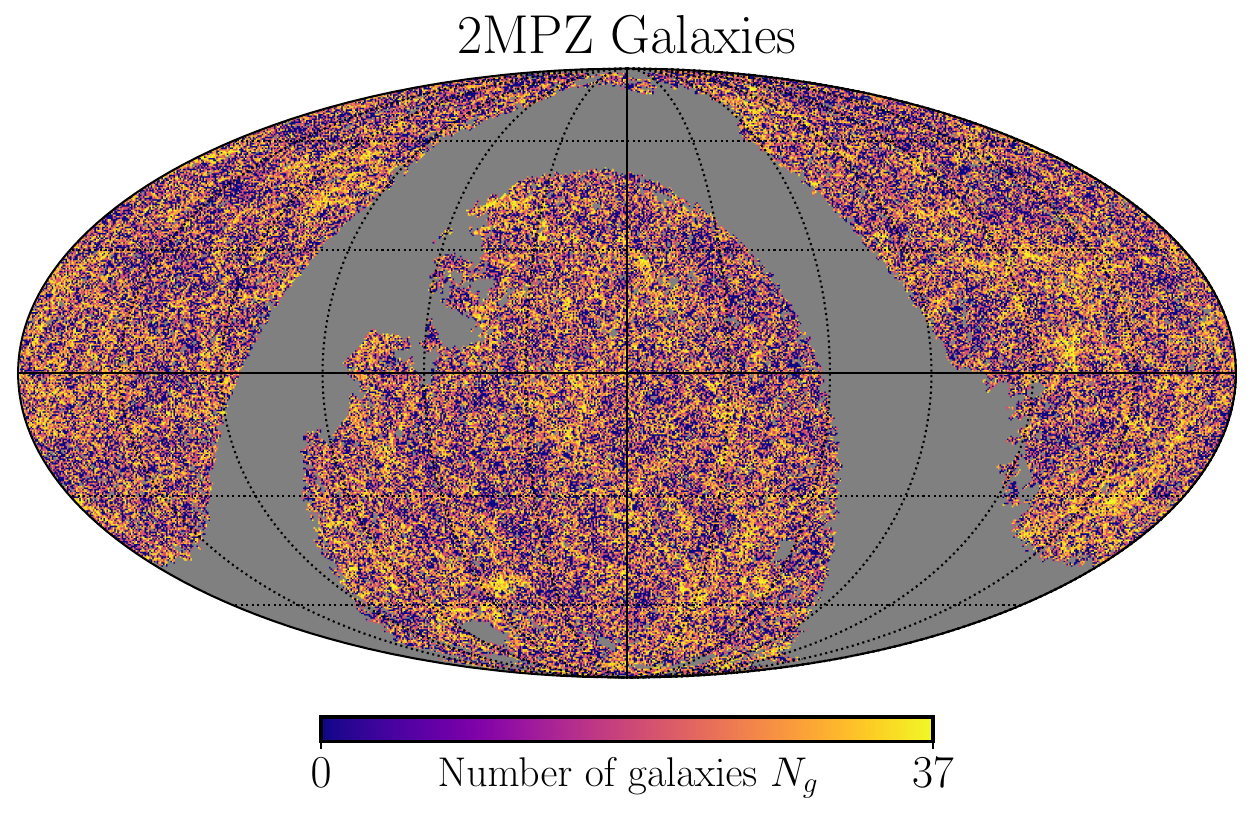}
\includegraphics[width=0.48\hsize]{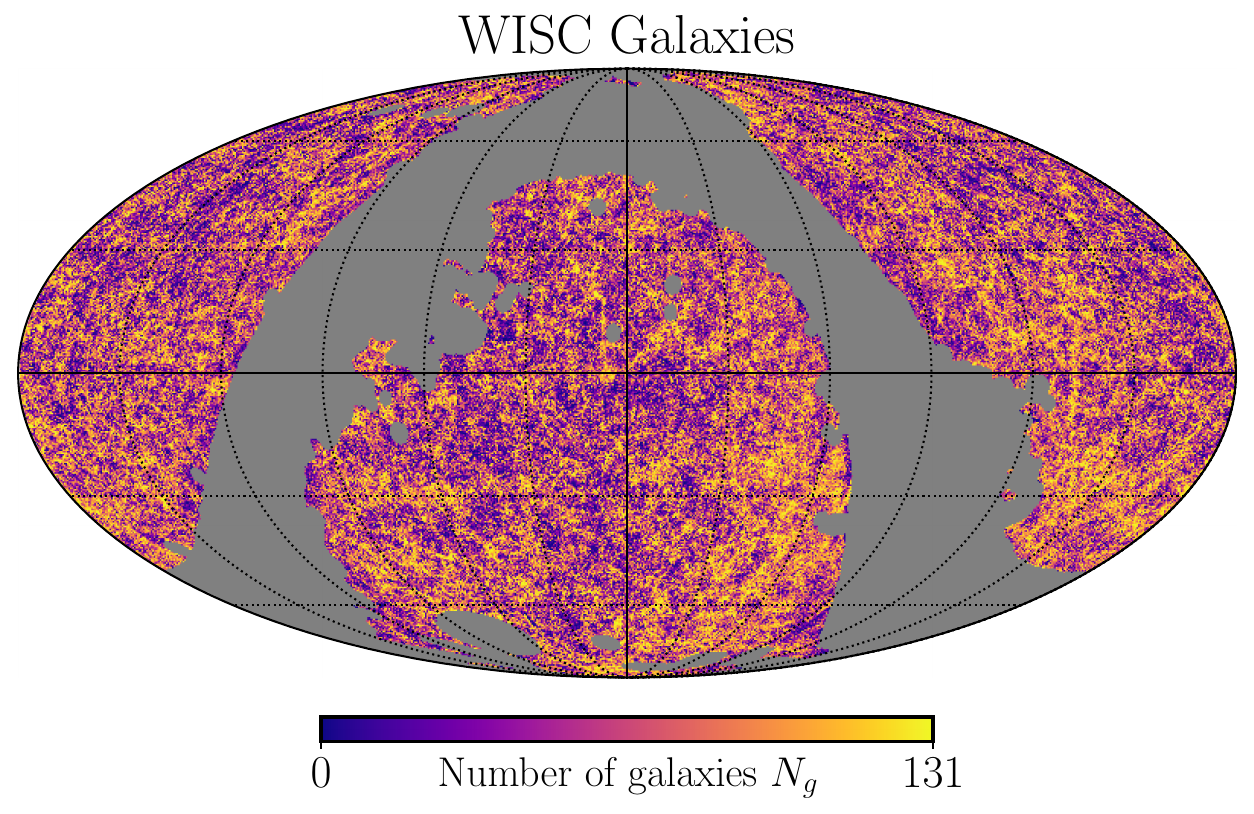}
\caption{2D sky maps of galaxies after masking, displayed in equatorial coordinates. The left panel shows the galaxies from 2MPZ, and the right panel shows the galaxies from WISC, both generated with $N_\mathrm{side} = 256$. Note that only WISC galaxies with $z\geq0.1$ are shown since we did not use galaxies from WISC with $z<0.1$ due to the higher stellar contamination rate. The gray regions are masked following the prescription in \cite{Mukherjee:2022afz}. }
\label{fig:galaxy_maps}
\end{figure*}

We populate mock sirens from two all-sky galaxy catalogs: 2MPZ and WISC (shown in Figure \ref{fig:galaxy_maps}), the details of which are described below. These galaxy catalogs have a variety of advantages in creating mock GW sources. Most importantly, they cover a large fraction of the sky and have a simple selection function. Moreover, 2MPZ and WISC have previously been analyzed as part of GLADE+ \cite{Dalya:2021ewn}, allowing for estimation of the stellar mass of each galaxy. However, only photometric redshifts are available for the majority of the galaxies in these surveys, restricting us to an analysis of the angular power spectrum. Also, without galaxy spectra, it is difficult to obtain accurate
measurements of star formation rate and metallicity. As a result, in this work, we only consider GW host-galaxy probabilities that depend on the stellar mass of the host galaxies, which is more easily measured for these samples. In our companion paper, we use the SDSS-MGS spectroscopic catalog instead, which provides a complementary view of GW clustering \cite{paper-II}.

\textbf{2MPZ catalog:} The 2MASS Photometric Redshift Catalog (2MPZ\footnote{\url{http://ssa.roe.ac.uk/TWOMPZ.html}}, \cite{Bilicki:2013sza})) is an all-sky photometric redshift galaxy catalog. It was built by cross-matching the 2 Micron All-Sky Survey Extended Source Catalog (2MASS XSC, \cite{2MASSXSC, 2MASS}), the All-Sky data release of the Wide-field Infrared Survey Explorer (WISE, \cite{WISE2010, WISEALLSKY}), and SuperCOSMOS \cite{Hambly:2001yj, Hambly:2001yi, Hambly:2001yh}. Then, the photometric redshifts were computed using the \verb|ANNz| artificial neural network package \cite{ANNz} trained on several overlapping redshift surveys, such as SDSS \cite{sdss9}, 6dFGS \cite{6df}, and 2dFGRS \cite{2DFGRS:2001zay}. The resulting catalog is comprised of almost 1 million galaxies, with a median $z= 0.07$.

\begin{figure}
\centering
\includegraphics[width=0.48\hsize]{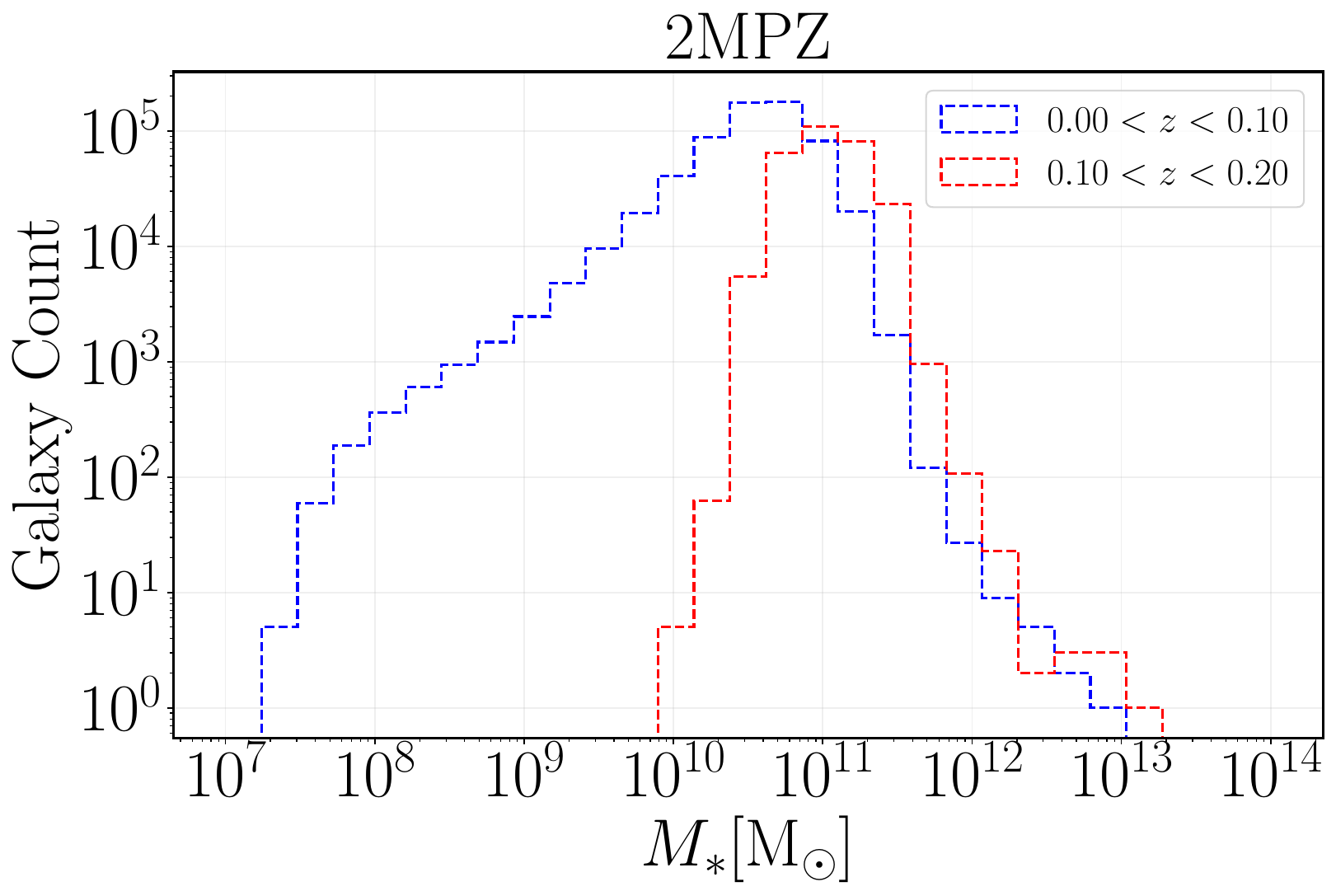}
\hspace{8pt}
\includegraphics[width=0.48\hsize]{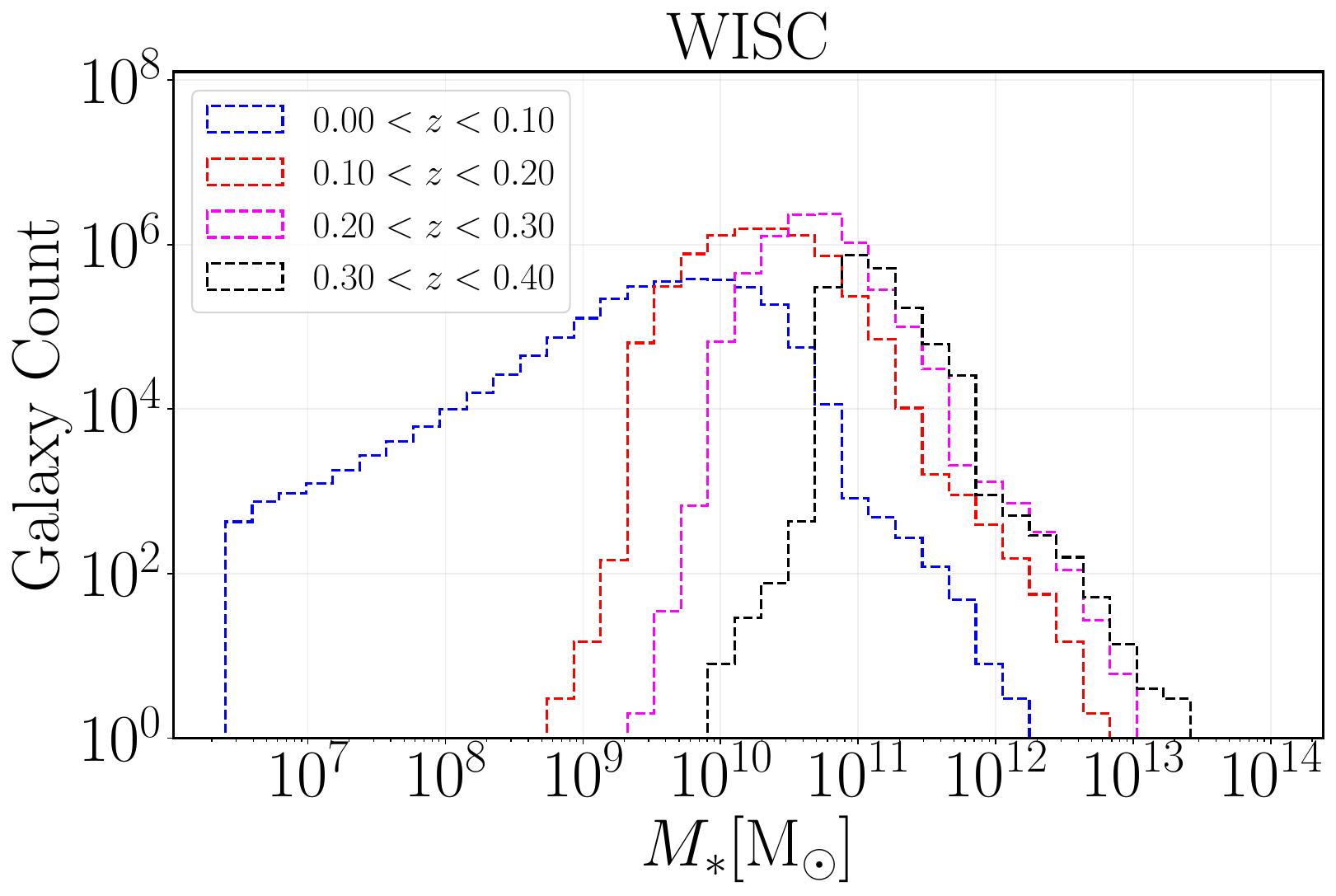}
\caption{Stellar mass distribution of the 2MPZ and WISC galaxy catalogs. This does not include masked galaxies as we do not use them in our simulation.}
\label{fig:gal_masses}
\end{figure}

\begin{figure}
\centering
\includegraphics[width=\hsize]{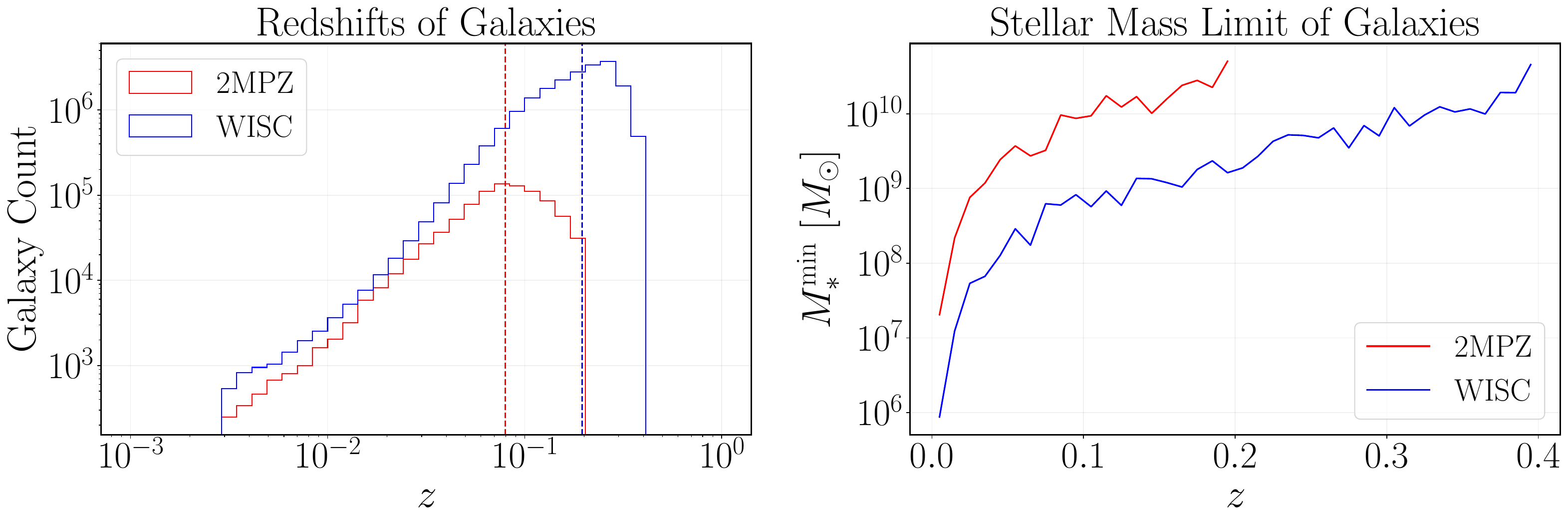}
\caption{\emph{Left panel:} Distribution of heliocentric redshifts for 2MPZ (red) and WISC (blue) in the range $z \in [0,0.2]$ for 2MPZ and $z \in [0,0.4]$ for WISC. The dashed vertical lines indicate median redshifts of these catalogs in these ranges, with $\Bar{z}_{\mathrm{2MPZ}} \approx 0.0795$ and $\Bar{z}_\mathrm{WISC} \approx 0.196$. \emph{Right panel:} Stellar mass limit for 2MPZ (red) and WISC (blue). The stellar mass limit is the lowest stellar mass of galaxies within redshift bins of $\Delta z = 0.01$. Note that 2MPZ only goes up to $z=0.2$ since we do not consider 2MPZ galaxies with $z>0.2$ for this work.}
\label{fig:gal_z_and_mass_lim}
\end{figure}

2MPZ includes all resolved sources in 2MASS XSC and is complete up to the magnitude band of $K_s = 13.9$ \cite{Jarrett04}. This translates to a redshift-dependent stellar mass cut, i.e. low-mass galaxies are only detected in the lowest redshift bins, as shown in Figure \ref{fig:gal_masses}. The stellar mass limit for 2MPZ galaxies ($i.e.$ the minimum stellar mass of a galaxy in a given redshift bin) can be seen in the right panel of Figure \ref{fig:gal_z_and_mass_lim}. This completeness limit is an unavoidable feature of using a galaxy catalog and its impact on our results will be discussed later. However, the simple selection of galaxies in the 2MPZ catalog guarantees that above the limiting magnitude of the survey, the completeness is high and it contains all types of galaxies\footnote{Indeed, the completeness for 2MASS galaxies for $K_s < 13.9$ is about 98.6\% \cite{Bell:2003cj}.}. This is critical for allowing us to produce a realistic mock catalog of GWs. Using a galaxy sample tailored for a particular galaxy population (for example, SDSS luminous red galaxies) is undesirable for creating a realistic mock GW catalog, since GWs may occur in any galaxy.

\textbf{WISC catalog:} We also use the WISExSCOS Photometric Redshift catalog (WISExSCOSPZ, or WISC for short\footnote{\url{http://ssa.roe.ac.uk/WISExSCOS.html}}, \cite{Bilicki:2016irk}) which was constructed by cross-matching the ALLWISE \cite{ALLWISE} data release of WISE and SuperCOSMOS Sky Survey (SCOS, \cite{Hambly:2001yj, Hambly:2001yi, Hambly:2001yh}) all-sky samples. \verb|ANNz| was once again used to derive the photometric redshifts of these samples, but with GAMA-II as the training data \cite{ GAMA, GAMAII}. This catalog is much larger and deeper than 2MPZ, containing about 18.5 million galaxies with a median of $z=0.2$.

WISC has a more complicated selection than 2MPZ, with selection limits in both the mid-infrared $W1$ band and the optical $B$ and $R$ bands, as well as a restriction to use only extended sources in optical SuperCOSMOS imaging \cite{Bilicki:2016irk}. They additionally apply a Galactic-latitude dependent infrared color cut to remove stars, which are difficult to remove from the $2^{\prime\prime}$ optical imaging alone\footnote{WISE has a $6^{\prime\prime}$ PSF and is therefore unsuitable for star-galaxy separation.}. These restrictions lead to $\sim 90\%$ completeness (i.e. the fraction of galaxies included in the catalog within the observational limit) in WISC at high Galactic latitudes ($|b| > 30^{\circ}$), with losses both from misclassification of compact galaxies as stars, and losses due to the color cut. However, despite these complications, WISC is advantageous as it is much larger than 2MPZ, leading to a lower stellar mass completeness limit (see the right panel of Figure \ref{fig:gal_z_and_mass_lim}).

As noted previously, Figure \ref{fig:galaxy_maps} illustrates the two galaxy catalogs, where the left panel presents the 2MPZ samples and the right panel shows the WISC samples. To generate the maps, we have used the Hierarchial Equal Area isoLatitude Pixelization (\verb|HEALPix|\footnote{\url{https://healpix.sourceforge.io}}, \cite{HEALPix}) library, employing the Python package \verb | HEALPy |\footnote{\url{https://healpy.readthedocs.io/en/latest/}} \cite{Zonca2019}. The gray-shaded regions in the figure represent masked areas, which were obtained using masks from \cite{Mukherjee:2022afz}, which followed the procedures in \cite{Balaguera-Antolinez:2017dpm} to mask 2MPZ and \cite{Xavier:2018owe} to mask WISC. Furthermore, for 2MPZ, the authors of \cite{Mukherjee:2022afz} manually excluded the region around LMC and SMC, as well as areas with $<85\%$ completeness in 2MPZ (determined by comparing the counts of 2MPZ sources and 2MASS XSC sources with $K_s < 13.9$). For WISC, the regions with high extinction $(E(B-V) > 0.1)$ and high stellar density were masked \cite{Mukherjee:2022afz}. These masks leave $\sim$ 68.5\% ($\sim$ 68.1\%) of the sky for 2MPZ (WISC). For further information on the generation of masks for 2MPZ and WISC, we refer interested readers to \cite{Mukherjee:2022afz}.

\subsection{Galaxy redshifts}
\label{sec:gal_z}
For 2MPZ, we used spectroscopic redshifts when available and photometric redshifts when not. Around 33\% of the 2MPZ sources have spectroscopic redshifts, nearly all from the overlapping areas of SDSS, 6dF, and 2dF, which provide redshifts for nearly all 2MPZ sources in their coverage areas. WISC provides only photometric redshifts; we use the photometric redshifts corrected for a hemispherical offset in the photometry. To convert the heliocentric redshifts to the frame of the cosmic microwave background (CMB), we use the standard formalism to convert redshifts between different frames, explained in Appendix \ref{sec:z_conversion}. We have only corrected for our relative motion to the CMB and have not corrected for the peculiar velocities of the galaxies themselves. The peculiar velocity corrections are small compared to the coarse redshift bins that we use ($\Delta z = 0.1$) and hence we do not include them.  Finally, galaxies with heliocentric redshifts $z < 0.003$ were excluded from our catalog due to the risk of stellar contamination \cite{Kettlety2018}. However, for WISC as will be described later, the findings suggested that contamination could be more significant, affecting our measurements to $z \lesssim 0.1$.
The redshift distribution is shown in the left panel of Figure \ref{fig:gal_z_and_mass_lim}.

\subsection{Determining the stellar masses of galaxies }
\label{sec:stellar_mass}
To estimate the stellar mass of the galaxies, we first classified galaxies as active or passive following the methodology described in \cite{Kettlety2018, Dalya:2021ewn}, with the former having active star formation and the latter without. To determine whether a galaxy was passive or active, we employed the $W2$ and $W3$ magnitudes, so that $W2 - W3 \leq 1.5$~ ($>1.5$) indicated active (passive) star formation. However, since 2MPZ and WISC did not explicitly provide $W3$ magnitudes in the catalogs, we first performed a one-to-one cross-matching search between the galaxies in these catalogs and those in ALLWISE \cite{ALLWISE} to extract the $W3$ magnitudes. Then, we corrected the magnitudes for dust extinction using the dust extinction coefficients for WISE filters of $A/E(B-V) = 0.184,\, 0.113$, and $0.0241$ for the $W1$, $W2$, and $W3$ bands, respectively \cite{Fitzpatrick1999} and $E(B-V)$ for each galaxy sample. Galaxies lacking a corresponding magnitude $W2$ or $W3$ from our cross-matching procedure were considered to be active, following the methodology in GLADE+ \cite{Dalya:2021ewn}.

After categorizing the galaxies, we computed the luminosity in the $W1$ band, using the $K$ correction given in \cite{Kettlety2018}
\begin{align}
    K = -7.1 \log_{10} (1+z), 
\label{eq:K_corr}
\end{align}
which arises from the fact that observed light from sources has different observed and emitted frequencies. Note that the redshift $z$ in Eq.~\eqref{eq:K_corr} is the heliocentric redshift of the galaxy. We then employed the mass-to-light in the $W1$-band ratio for passive galaxies (those with insignificant star formation) presented in \cite{Kettlety2018}
\begin{align}
    M_*/L_{W1} = 0.65 \pm 0.07, 
\label{eq:M_passive}
\end{align}
where $M_*$ is the stellar mass of the galaxy and $L_{W1}$ is its luminosity in the $W1$-band. According to \cite{Kettlety2018}, this straightforward method can effectively approximate the masses of galaxies with redshifts $z \leq 0.15$, and its results are comparable to those obtained using more sophisticated techniques. We applied this method to the entire catalog. We similarly estimated the stellar mass for active galaxies (those with significant star formation) using the mass-to-light ratio in the $W1$-band ratio provided for active galaxies in the same study
\begin{align}
    \log_{10}(M_*/L_{W1}) = -0.4 \pm 0.2. 
\label{eq:M_active}
\end{align}
The resulting distribution of stellar masses in various redshift bins can be seen in Figure \ref{fig:gal_masses}.

\section{Populating galaxies with BBHs} 
\label{sec:modelling_gw}
Provided with the galaxy catalogs, we now take the next step of making synthetic mock siren catalogs as outlined in Section \ref{subsec:method}, and schematically in Figure \ref{fig:chart1}. The following three subsections, respectively, explain the building blocks related to generating the binary black hole catalog (green boxes) in Figure \ref{fig:chart1}.

\subsection{The merger rate and $N(z)$ of BBHs}
\label{subsec:merger-rate}
The merger rate of BBHs \cite{Belczynski:2001uc, OShaughnessy:2009szr, Dominik:2012kk, Dominik:2014yma, Mapelli:2017hqk, Giacobbo:2017qhh, Cao2018, Fishbach:2018edt, Santoliquido:2022kyu, Mukherjee:2021bmw} is a complex process that is highly dependent on various astrophysical properties of the host galaxy of the BBH (which in turn determine the properties of the star population) and the delay time distribution of the BBH merger events. The delay time of binary mergers, $t_d=t_m - t_f$, is the time elapsed between the formation time of the progenitor stars $t_f$ in a binary system and the time of the merger event $t_m$. Numerous theoretical studies have investigated the behavior of the delay-time distribution \cite{OShaughnessy:2009szr, Banerjee2010, Dominik:2012kk, Toffano:2019ekp, Artale:2019tfl, McCarthy2020, mukherjee2021impact, Fishbach:2021mhp, Mukherjee:2021bmw, 2023MNRAS.523.4539K}, and its functional form, along with the necessary parameters to describe it, remain a topic of ongoing discussion. In this work, we consider a simple power-law delay-time distribution with a minimum cutoff expressed as
\begin{align}
    P_{D}(t_d) \propto \Theta(t_d - t_{d,\text{min}}) t_d^{-\kappa}.
\label{eq:ptd}
\end{align}
For formation histories from BBHs with initial uniform distribution in the log-space distribution of the binaries' separation, $\kappa\sim 1$ is expected \cite{OShaughnessy:2009szr, Dominik:2012kk, McCarthy2020}.
Assuming the BBH formation rate scales with the star formation rate and incorporating the delay time distribution, the merger-rate density of BBHs at a specific redshift can be written as
\begin{align}
    R_\text{GW}(z_m) = \mathcal{A}_0 \int_{z_m}^\infty P(t_d) \frac{d t_f}{dz_f} R_\text{SFR}(z_f) ~ dz_f, 
\label{eq:merger_rate}
\end{align}
where $z_m$ and $z_f$ denote the merger and formation redshifts, respectively, and $R_\text{SFR}(z_f)$ represents the redshift-dependent SFR density. Here, $\mathcal{A}_0$ denotes a normalization factor such that the local merger rate of gravitational wave sources at redshift $z_m=0$ matches the observed value $\mathcal{R}_\text{BBH} = 23.9^{+14.3}_{-8.6}$ Gpc$^{-3}$~yr$^{-1}$~ \cite{abbott2020gw190425}. In other words, $\mathcal{A}_0$ is given by
\begin{equation}
    \mathcal{A}_0= \frac{{\mathcal{R}_\text{BBH}}}{{ \int_{0}^\infty P(t_d) \frac{d t_f}{dz_f} R_\text{SFR}(z_f) ~ dz_f}} . 
\label{eq:A_0_normal}
\end{equation}
To estimate the quantity $R_\text{SFR}(z_f)$ in Eq.~\eqref{eq:merger_rate} and Eq.~\eqref{eq:A_0_normal} we take an analytical best-fit formula motivated by observational data \cite{madau2014cosmic} (the Madau-Dickinson SFR model), given by
\begin{align}
\label{eq:sfr}
    R_\text{SFR}(z) = 0.015 \frac{(1+z)^{2.7}}{1 + [(1+z)/2.9]^{5.6}} M_\odot \text{ yr}^{-1} \text{ Mpc}^{-3}.
\end{align}
Finally, $dt_{f} /dz_f$ in Eq.~\eqref{eq:merger_rate} is the standard change-of-coordinates Jacobian between time coordinate and redshift,
\begin{align}
    \frac{dt_f}{dz_f} = \frac{1}{H(z_f)} \frac{1}{(1+z_f)},
\end{align}
The SFR density $R_{\textrm{SFR}}(z)$ can vary with stellar metallicity, which in turn can impact the BBH population \cite{Chruslinska:2018hrb, Chruslinska:2022ovf}. However, we have not included this effect in this analysis and instead consider a universal Madau-Dickinson SFR model \cite{madau2014cosmic}. A possible way to explore that will be to study the correlation between the evolution of different BBHs mergers with the evolution of emission-line-galaxies \cite{Mukherjee:2021bmw}.

\begin{figure}
\centering
\includegraphics[width=0.98\hsize]{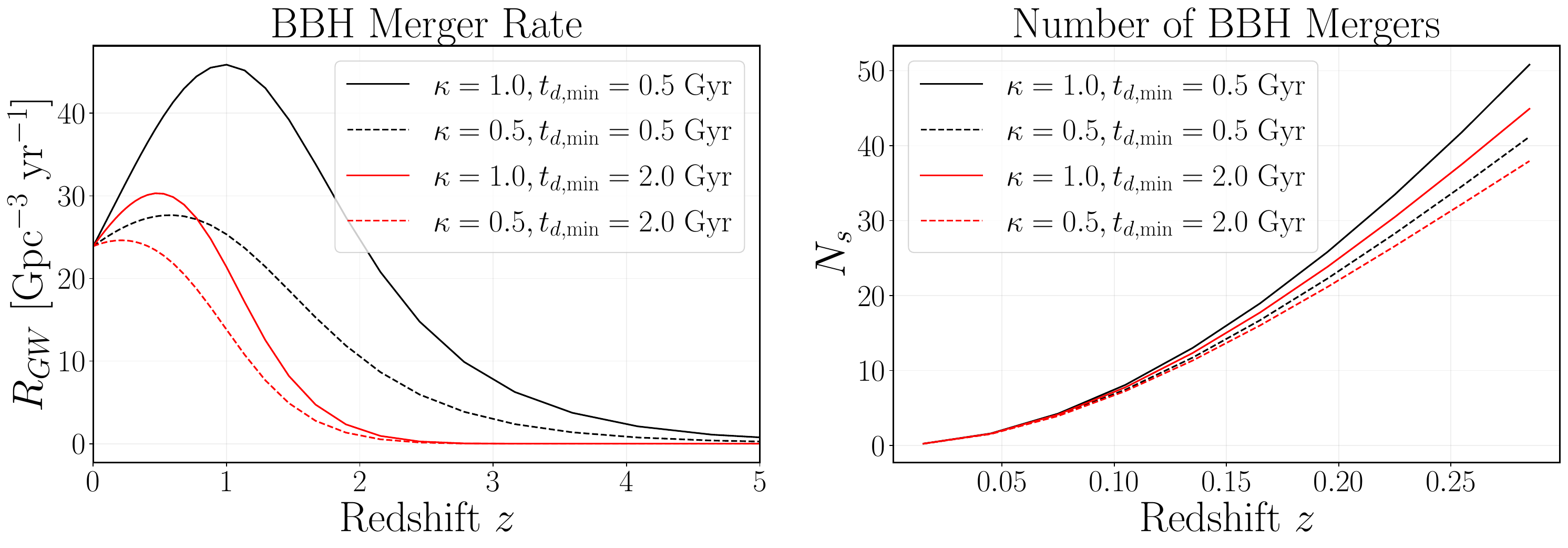}
\caption{\emph{Left panel}: BBH merger rate $R_\text{GW}(z)$ for two different minimum delay times $t_{d,\text{min}} = 0.5$ Gyr (\emph{black}) and $t_{d,\text{min}} = 2$ Gyr (\emph{red}). The solid curves correspond to the merger rates computed with the power law index $\kappa = 1.0$ while the dashed curves correspond to $\kappa = 0.5$. As discussed in the text, these curves have been normalized so that they coincide at redshift $z=0$ with a value of $\mathcal{R}_\text{BBH} = 23.9$ Gpc$^{-3}$ yr$^{-1}$ \cite{abbott2020gw190425}. \emph{Right panel}: Number of BBH mergers between redshifts $z=0$ and 0.3, computed using redshift bins of $\Delta z = 0.03$ and $T_\mathrm{obs} = 1$ yr. The colors and line styles of the curves are the same as in the left panel.}
\label{fig:merger_rate}
\end{figure}

We plotted the evolution of the BBH merger rate given in Eq.~\eqref{eq:merger_rate} using the delay time model in Eq.~\eqref{eq:ptd} for different values of the minimum delay time and values of $\kappa$ in the left panel of Figure \ref{fig:merger_rate}. As expected, a higher value of $t_{d,\mathrm{min}}$ results in a peak at lower redshifts, since on average merger events take longer to occur. Meanwhile, a higher value of $\kappa$ increases the steepness of the delay time distribution, resulting in a sharper fall-off from the peak.

Given the merger rate of BBHs, one can calculate the number of BBH mergers in a redshift bin by integrating the merger rate over the width of the bin. We write the number of observable mergers in a given redshift bin as
\begin{align}
    N_{s} (z_a) = \int_{z_a - \Delta z/2}^{z_{a}+ \Delta z/2} \frac{dV}{dz} \frac{R_{GW}(z)}{1+z} T_\mathrm{obs} ~ dz, 
\label{eq:N_GWcalculate}
\end{align}
where $z_a$ is the midpoint of each redshift bin labeled with index $a=1,\cdots, N_\mathrm{bins}$ (the total number of bins), and $\Delta z$ is the width of each redshift bin. In this expression, $T_\mathrm{obs}$ is a constant corresponding to the observing time of the GW detector. The number of mergers in redshift bins of size $\Delta z = 0.03$ from $z=0$ to $z=0.3$ is displayed in the right panel of Figure \ref{fig:merger_rate} for various choices of the parameters $\kappa$ and $t_{d,\mathrm{min}}$ in the delay time distribution, assuming $T_\mathrm{obs}=1$ yr. In this work, we do not examine the impact of variations of the parameters $\kappa$ and $t_{\textrm{d,min}}$ defined in Eq.~\eqref{eq:ptd} {on merger rate}, setting their fiducial values to $t_\mathrm{d,min}=0.5$ Gyr, $\kappa=1.0$, as the number of mergers does not directly change the clustering properties and these parameters are expected to have negligible effects on our results for GW bias. In other words, in our result the delay time parameters only influence the number of BH mergers but do not affect the GW-galaxy connection. However, as we discuss in the next subsection, the delay time distribution is expected to indirectly impact the host-galaxy probability. While our fiducial host-galaxy probability model assumes shorter minimum delay times ($t_{d, \rm min} < 0.5$ Gyr), we will also present variations in the host-galaxy probability appropriate for longer time delays (though parameterized through the shape of the host-galaxy probability function rather than explicitly via the delay time).

\subsection{Host-galaxy probability for BBHs}
\label{subsec:GW_Gal}
In the previous section, we provided the description for obtaining the number of BBH mergers at a given redshift. Here, we introduce our approach to modeling the probability distribution of host galaxies for BBH mergers, referred to as the "GW host-galaxy probability"\footnote{We use the following terms interchangeably throughout the paper: "host-galaxy probability", "host-galaxy probability function", and "GW host-galaxy probability".} (in agreement with the terminology used in \cite{Artale_2020}).

The host-galaxy probability can depend on different properties of the galaxy, such as stellar mass, metallicity, and SFR, and as shown in previous studies \cite{Artale:2019tfl,2023MNRAS.523.5719R,Salim07,Mannucci2010}, these properties are not independent of each other. However, the 2MPZ and WISC surveys contain only mass information; therefore, for the selection of host galaxies from the catalog, we only apply a conditional probability based on stellar mass, $P(GW|M_*)$. In the companion paper, we consider a more generic host-galaxy probability that also includes the dependencies on SFR and metallicity \cite{paper-II}.

From a practical perspective, in order to populate BBHs into the catalogs, we need to formulate $P(GW|M_*)$ that best encapsulates the physics from the formation to the merger of BBHs, but all of it solely manifested in terms of the stellar mass of the host galaxy at the time of the merger event. The stellar mass at merger time is influenced by other astrophysical properties at earlier times, such as the initial stellar mass, star formation history, and merger history of the galaxies. From a physics standpoint, one expects that the BBH merger rates are primarily linked to the star formation rate of the host galaxy during the period when the progenitor stars formed. The integral of the star formation rate then gives the stellar mass (up to a correction for mass loss as stars evolve). This leads to the expectation that the merger rate per galaxy will be positively correlated with the stellar mass at later times, which is also consistent with the results of population synthesis simulations \cite{Artale:2019tfl, Artale_2020, Santoliquido:2022kyu}.

The stellar mass--BBH merger rate relationship may not be completely monotonic, as it is also influenced by changes in the galaxy properties during the time between binary formation and merger. Specifically, high mass galaxies form stars rapidly at early times before their star formation is quenched \citep[e.g.][]{Peng10}, leading to the well-known bimodality between star-forming and quiescent observed at late times (see Figure 16 in \cite{Salim07}, showing that the division is around $M_*\sim 10^{11} M_\odot$).
Hence the relationship between BBH host-galaxy probability and late-time stellar mass depends on the two timescales in the problem, the delay time and the quenching time. If the delay time is shorter than the quenching timescale, one expects a suppression in the number of BBHs at high stellar mass, because BBHs that formed at early times when the SFR was high would have already merged. Inferring from the empirical relation between SFR and stellar mass \cite{Salim07}, we expect that the suppression in the BBH host-galaxy probability function would also occur around $10^{11} M_\odot$. Current observational constraints on the delay time distribution are weak and seem to indicate that the minimum delay time is less than $5$ Gyrs \citep{2023MNRAS.523.4539K}. On the other hand, some population synthesis simulations \citep[e.g.][]{Artale:2019tfl,Toffano:2019ekp,Mapelli_2018} prefer longer delay times for $z<1$, leading to a host-galaxy probability function with no suppression and even some amplification at higher stellar masses at $z \sim 0.1$ \citep[e.g. Figure 1 in ][]{Artale:2019tfl}.

\begin{figure}
\centering
\includegraphics[width=0.5\hsize]{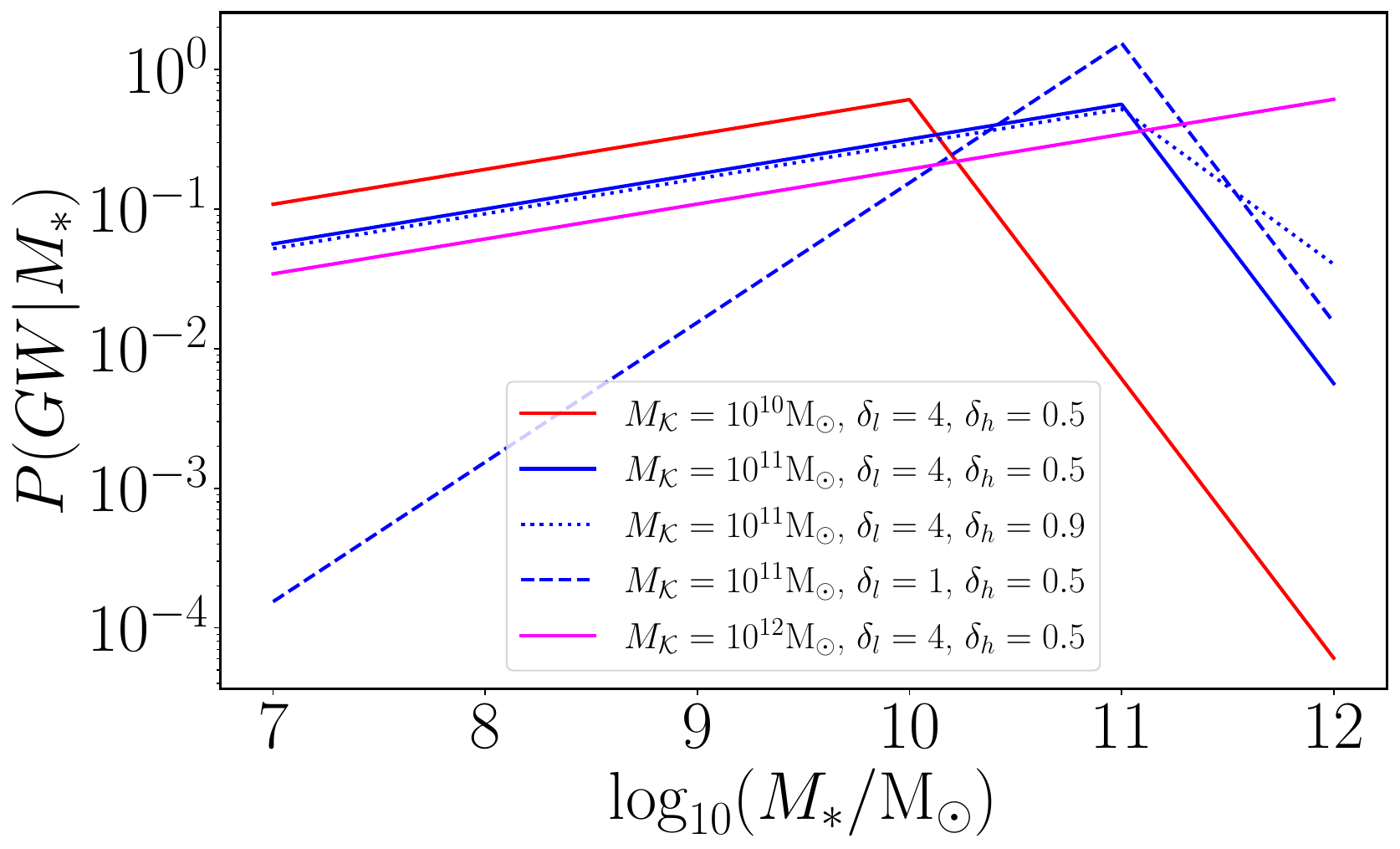}
\caption{Stellar mass-dependent BBH host-galaxy probability, Eq.~\eqref{eq:selection}, for different choice of model parameters. The probability of a galaxy hosting binary BH peaks at the pivot mass $M_\mathcal{K}$ and declines afterwards, following the power laws relations above and below the pivot mass. The host-galaxy probabilities are normalized to all integrate to unity. The solid blue line corresponds to our fiducial model.}
\label{fig:mzr}
\end{figure}

Consequently, considering all these factors, we utilize a broken power law to represent the GW host-galaxy probability function, where the break in the power law is designed to reflect the effect of suppressed SFR history in high-mass galaxies. In this framework, one can also easily study the possibility of removing the suppression at higher masses if considering longer delay times \citep[e.g.][]{Toffano:2019ekp,Cao2018,Artale:2019tfl,2023MNRAS.523.5719R}, by pushing the break scale to higher values (recovering a monotonic host-galaxy probability function in the limit of an infinite pivot mass). We parameterize the host-galaxy probability function described above\footnote{Practically, we use \texttt{numpy.random.choice} package for selecting the hosts with probabilities coming from the host-galaxy probability introduced here.} in the following way
\begin{align}
\label{eq:selection}
    P_{\rm host}(\rm GW|M_*)=
    \left\{\begin{array}{ c l } A_{M_*}10^{(\log_{10}(M_*)-\log_{10}(M_\mathcal{K}))/\delta_l} &\text{for}\,\, 7 \leq \log_{10}(M_*) \leq \log_{10}(M_\mathcal{K}),\nonumber\\
    A_{M_*}10^{(\log_{10}(M_\mathcal{K})-\log_{10}(M_*))/\delta_h} \,\, &\text{for} \,\, \log_{10}(M_*) \geq \log_{10}(M_\mathcal{K}),
    \end{array}
    \right.\\
\end{align}
where $1/\delta_l$ and $1/\delta_h$ control the power of the growing and decaying regime of the host-galaxy probability function, $M_\mathcal{K}$ is the pivot mass that divides the broken power laws, and $A_{M_*}$ is the normalization factor. For our fiducial model (Figure \ref{fig:mzr}), we set the values of the power indices as $\delta_l=4$ assuming a mild growth (positive slope of $\sim 0.25$), $\delta_h= 0.5$, a steep fall off in the suppression region (negative slope of $\sim -2$), and $M_\mathcal{K}= 10^{11} M_\odot$ for the pivot mass. As described earlier, this fiducial model is motivated by a scenario where the delay time is shorter than the quenching time, and thus binaries in high-mass galaxies have already merged.

In our full analysis {on the GW bias discussed in Section \ref{sec:results}}, we will vary these parameters ($\delta_l, \delta_h$ and $M_\mathcal{K}$), exploring scenarios with milder suppression $\delta_h \geq 1$, as well as those closer to the scenario obtained in \cite{Artale:2019tfl} with faster growth and no suppression, by setting $\delta_l\sim 1$ and taking the limit of $M_\mathcal{K}\sim 10^{12} \rm M_\odot$. Overall, we investigate a wide range of $\delta_l, \delta_h$ and $M_\mathcal{K}$ to explore the impact of these parameters and understand limiting cases. We investigate $\delta_l \in [0.5,10]$, $\delta_h \in [0.1,5]$, and $M_\mathcal{K} \in [10^{9} \rm M_\odot,10^{12} \rm M_\odot]$. These parameters are roughly chosen for consistency with host-galaxy probabilities presented in past works based on simulations \cite{Cao2018, Artale:2019tfl, Santoliquido:2022kyu}. To compare with \cite{Artale:2019tfl}, their parameter $\alpha_1\sim 0.8$ at $z \leq 1$ translates into $1/\delta_l$ for us, yielding $ \delta_l\sim 1.2$. Similarly \cite{Santoliquido:2022kyu} estimate the parameter $b$ in their models based on MZR and FMR relations to be $\sim 0.47$ and $\sim 0.58$ (see Table 2) which translates to $\delta_l\sim 2.1$ and $\delta_l\sim 1.72$. Last, the host-galaxy probability of \cite{Cao2018} can be inferred by comparing the stellar mass distributions of BBHs from their Figure 2 (derived from populating the Millenium-II simulation with semi-analytic models for galaxies and BHs) to the stellar mass function of \cite{Guo2011} for these simulations. For BBHs with delay times of 2 Gyr or less, this corresponds to $\delta_l \sim 1$ and $\delta_h \sim 0.5$--2; for very long delay times of 12.7 Gyr, the \cite{Cao2018} host-galaxy probability has $\delta_l \sim 0.5$ with no break at high mass.

\subsection{BH mass distribution} \label{subsec:bh_masses}
In the previous subsection, we outlined our method for selecting host galaxies for BBH mergers. The next step involves sampling the masses of the BBHs. We follow a phenomenological model fitted to the BH mass distribution from the GWTC-3 LVK catalog \cite{2023MNRAS.523.4539K}. This model assumes that BHs follow the same trend as stellar populations described by the Kroupa initial mass function (Kroupa IMF) \cite{2001MNRAS.322..231K}, i.e. a power-law distribution for BH masses subject to additional minimum and maximum mass cut-offs. The lower limit is imposed since we know that below some mass scale, stellar BHs do not form. In addition to the minimum mass cutoff, there is an upper bound for the allowed range of BH mass due to pair instability supernovae, called $m_{\rm PISN}$ \cite{2019ApJ...887...53F, 2022MNRAS.515.5495M, 2023MNRAS.523.4539K}. This instability is expected for stars surpassing a specific mass limit, where the core generates electron-positron pairs, leading to shock waves and consequent instability, establishing a maximum permissible mass for the resulting BH. This distribution can be expressed as
\begin{equation}
\label{eq:kroupaIM}
 P(m)\propto
     \begin{cases}
         0 & m < m_{\mathrm{min}}, \\
         m^{-\alpha} & m_{\mathrm{min}}\leq m \leq m_{\rm PISN}\\
         0 & m> m_{\rm PISN}\,,
     \end{cases}
\end{equation}
where $m$ is the BH mass and we set $m_{\mathrm{min}}=5 M_\odot$ \citep{2006csxs.book..157M}, and $m_{\rm PISN}=45M_\odot$ \citep{2019ApJ...887...53F}. $\alpha$ is a free parameter describing the power-law index of the mass distribution which is currently weakly constrained from the observations of GW sources \cite{KAGRA:2021duu, LIGOScientific:2020kqk, 2023MNRAS.523.4539K}. We set $\alpha=2.3$, the same as the power-law index of the Kroupa IMF \cite{2001MNRAS.322..231K} in the analysis. Though the parameters that control the BBHs mass distribution are impacted by the stellar metallicity of the parent stars \cite{Belczynski:2001uc}, it will not impact the large scale ($\gtrsim 1$ $h^{-1}$ Mpc scale) clustering of the GW sources and hence the GW bias parameter is not influenced by this choice of mass modeling. In other words, with this model, the GW bias, by construction, will be independent of the BH masses.

\section{Calculating the angular power spectra for sirens and galaxies} 
\label{sec:power_spectrum}

\subsection{Power spectrum measurement}
\label{sec:cl}
Given the setup of the mock GW catalog in the previous sections, and as mentioned in Section \ref{sec:bias}, we wish to compare the 2-D angular power spectra of sirens and galaxies and measure the bias by comparing to the angular matter power spectrum (Eq.~\ref{eq:cl_ob}). To do so, we will generate overdensity maps $\delta_X$ in redshift bins for both galaxies and sirens and compute the angular power spectrum (pseudo)-$C_\ell$ using \verb|anafast| \cite{Gorski_2005}. Given a catalog of either the galaxies or mock sirens, we produced a density map of $\{\delta_{X,i}^a\}$, by computing the local overdensity in each pixel with label $i$ and within the redshift bin labeled $a$ such that
\begin{align}
    \delta_{X, i}^a = \frac{N_{X, i}^a}{\Bar{N}_{X}^a } - 1,
\end{align}
where $X = g \,(GW)$ for galaxies (sirens). $N_{X,i}^a$ is the number of the tracer $X$ in the pixel $i$ and the redshift bin $a$. $\Bar{N}_{X}^a$ is the average number of galaxies or sirens in the redshift bin computed from $\Bar{N}_{X}^a = N_{X}^a / n_\mathrm{pix}$, where $N_{X}^a$ and $n_\mathrm{pix}$ are respectively the total number of objects and the number of pixels in that redshift bin. We then measure $C_\ell$ as the spherical harmonic transform of the density field, using \verb|anafast|. We divide by the pixel window function\footnote{\url{https://healpy.readthedocs.io/en/latest/generated/healpy.sphtfunc.pixwin.html}} for the value of $N_\mathrm{side}$ that we use when creating the overdensity maps, $N_\mathrm{side} = 512$.
Since galaxies or sirens are only obtained in locations where the mask is nonzero, we need to correct the power spectrum for mask convolution. On scales smaller than the characteristic scale of the mask, this is equivalent to dividing the power spectrum by the sky fraction of the mask \citep{2004MNRAS.348..885E}, $f_{\textrm{sky}}$. We adopt this approximation since we always work on sufficiently small scales where it is valid ($\ell > 20$). Finally, the shot noise is subtracted from the power spectrum; the shot noise itself is computed as the inverse of the number density observed given by $n_X^a=\frac{N_X^a}{4\pi f_{\textrm{sky}}}$.

We smooth the power spectrum by bin-averaging with width $\Delta \ell = 10$. For estimating the bias uncertainty, we use the error for a Poisson-sampled Gaussian random field \cite{Knox95}
\begin{align}
\label{eq:sigma_cl}
    \sigma_{C_{\ell}^{XX}}=\sqrt{\frac{2(\overline{C_{\ell}^{XX}+1/n_{X}})^2}{(2l+1)\Delta l f_{\textrm{sky}}}},
\end{align}
where $\overline{C_{\ell}^{XX}+1/n_{X}}$ means first the shot noise is added back and then we calculate the average. $\ell$ and $\Delta \ell$ are the center and width of the bin. Also note that we have dropped the index $a$ in Eq.~\eqref{eq:sigma_cl} to make the equation less crowded, but the calculations are done for each redshift bin separately. An example of the galaxy and GW power spectra is shown in Appendix~\ref{sec:random} in Figure~\ref{fig:random_sel_pow}, along with their shot noise, error bars, and $\Delta \ell = 10$ binned power spectra.

Note, here we provide computable theoretical models of GW bias by establishing a mapping between GW sources and galaxies. Therefore, since our primary aim is to explore the phenomenological model describing GW bias dependence on galaxy stellar mass, we consider only Poisson noise from a limited number of sources. While this work focuses on theoretical modeling, any work aiming for a realistic forecast of GW bias measurement from observations will require independent frameworks that incorporate GW selection effects and instrumental noise to infer the GW bias. In particular, for the redshift range considered here, future GW observations are expected to have a selection function close to 1 (see e.g. Figure 6 in \cite{Leandro_2022}), suggesting a nearly complete GW catalog that mitigates the impact of detector-related systematics on inferences of the GW-host galaxy probability.

We measure the angular power spectrum using relatively broad redshift bins. For the 2MPZ catalog, we opt for two equal bins within the range $z\in [0,0.2]$, while for WISC, we choose three equal bins within the range $z\in [0.1,0.4]$. The range $z\in [0,0.1]$ for WISC is excluded due to the anticipated stellar contamination in the dataset \cite{Bilicki:2016irk,Xavier:2018owe}.

The number of mergers in a redshift bin can be calculated using Eq.~\eqref{eq:N_GWcalculate}, with $z_a$ representing the center of the bin. Having chosen $\kappa=1.0$, and $t_{d,\text{min}} = 0.5$ Gyr, determining the total number of mergers requires specifying the observational time, $T_\mathrm{obs}$ in Eq.~\eqref{eq:N_GWcalculate}. As we aim to estimate the theoretical GW bias parameter that links galaxies and BBHs, it is crucial to have a substantial number of binaries to minimize the error bars, thereby allowing us to effectively study the behavior of the theoretical GW bias parameter. A realistic observational time, spanning only a few years, is not a viable choice for our purposes of modeling the bias, as it would yield an insufficient number of sources in the siren catalog, resulting in noise-dominated bias calculations. Therefore, in our computations, we assign a considerably higher value to $T_{\textrm{obs}}$.

However, there is still an unavoidable practical statistical limit for $T_{\textrm{obs}}$. This upper bound is dictated by the fact that there are finite numbers of galaxies at different stellar masses in the catalogs. When we select hosts for sirens from the galaxy catalog, we do so without replacement, to avoid any galaxy being selected in the catalog multiple times. If $T_{\textrm{obs}}$ is so high that all galaxies within a certain mass range are selected as hosts, then hosts will subsequently be selected from other mass ranges, misrepresenting the properties of the hosts away from the desired mass distribution set by the selection criteria. We could instead choose to select host galaxies with replacement, but this is undesirable, as it can distort the shot noise and small-scale clustering properties, due to some galaxies being included in the sample multiple times. Hence, $T_{\textrm{obs}}$ must be within a sweet spot of being large enough, not to make the clustering measurement shot noise-dominated, but not so large as to select all galaxies with a particular stellar mass.

To quantify this, we applied the host-galaxy probability function to the survey several times while each time increasing the observational time, i.e.\ selecting more and more galaxies. The maximum observational time was chosen so that the mean stellar mass for the GW hosts changed by $<5\%$. Our results are insensitive to the exact value of the observational time chosen, near this threshold. This allows us to choose $T_{\textrm{obs}}$ for analyzing the GW bias parameter with minimal noise, as permitted by our galaxy catalogs. Following this process, the fractions $\{0.0091,0.13\}$ and $\{0.0047, 0.012,0.090\}$ of galaxies were selected as hosts in the corresponding redshift bins (in ascending order) of the two equal bin widths in the range $z\in [0,0.2]$ for 2MPZ and the three equal bin widths in the range $z\in [0.1,0.4]$ for WISC, respectively.

Finally, since we anticipate the bias to be approximately scale independent on sufficiently large scales \cite{2004ApJ...601....1W}, we will fit a constant bias within a specific range of $\ell$ in $[\ell_{\mathrm{min}},\ell_{\mathrm{max}}]$. The value of $\ell_{\mathrm{max}}$ is determined based on the maximum value of the 3D comoving wavenumber $k_{\mathrm{max}} = 0.4$ $h$ Mpc$^{-1}$, translating into $\ell_{\mathrm{max}} \approx \chi(z_a) k_{\mathrm{max}}$, with $\chi(z_a)$ being the comoving distance to the middle of the redshift bin $z_a$, calculated using our reference cosmology \cite{Planck:2018vyg}. For $z_a = \{0.05,0.15,0.25,0.35\}$, this yields $\ell_{\mathrm{max}}=\{60,180,290,400\}$. As we will explain in Section \ref{sec:bgw_ellbounds}, we also verified the robustness of our bias calculation against this assumption by varying $k_{\textrm{max}}$ to higher values and checking a simple model for scale-dependent bias. In Appendix \ref{sec:l_min} we also discuss our strategy for choosing the best values for $\ell_{\mathrm{min}}$, avoiding large scales where contamination in the catalogs leads to spurious additional clustering. We start with $\ell_{\rm min} = 20$, and if the linear bias model does not fit the data well, we increase $\ell_{\rm min}$ until we find a good fit. We use $\ell_{\rm min} = 20$ for 2MPZ and $\ell_{\rm min} = [30, 50, 60]$ for the three WISC bins.
The higher values of $\ell_{\textrm{min}}$ for WISC are due to stellar contamination in the WISC sample, which adds additional power at large scales.

\subsection{Model angular power spectrum}
\label{sec:bias_measure}
The estimation of the GW bias requires calculation of the model angular power spectrum $C^{\rm Mod}_\ell$, in Eq.~\eqref{eq:cl_the}. This process is described in Section~\ref{sec:bias-def} and requires accurate knowledge of the redshift distribution of galaxies and sirens. We therefore take into account the redshift uncertainties in the photometric surveys. We discuss the modeling of the photometric redshift uncertainty and how it is propagated to the inference of the redshift distribution $dN/dz$ in the following subsection.

\subsubsection{Photometric redshift modeling}
\label{sec:photoz_modelling}
To estimate the actual redshift distribution of the tracers, given their measured photometric redshift distributions, we convolve the photometric redshift distribution with a Gaussian or modified Lorentzian function denoted $P_{\mathrm{conv}}(\delta z)$ as per \cite{2018MNRAS.481.1133P,2018MNRAS.476.1050B}. This can be represented by the equation
\begin{align}
\label{eq:zconv}
     \frac{dN}{dz_\mathrm{conv}}(z_{\textrm{conv}}) =\int \frac{dN}{dz_\mathrm{photo}} P_{\mathrm{conv}}(\delta z) \, dz_\mathrm{photo} \,,
\end{align}
where, $z_\mathrm{photo}$ represents the photometric redshift, $z_\mathrm{conv}$ represents the convolved redshift,\footnote{In the literature, $z_\mathrm{spec}$ typically refers to the convolved redshift, which is meant to reflect the true spectroscopic redshift distribution. However, in our case, some 2MPZ galaxies have both photometric and spectroscopic redshifts, whereas others have only photometric redshifts. We therefore save $z_\mathrm{spec}$ for the spectroscopic redshift of galaxies with both photometric and spectroscopic redshifts.} and $\delta z\equiv z_\mathrm{conv}-z_\mathrm{photo}$. For 2MPZ, as shown in \cite{2018MNRAS.476.1050B}, $\frac{dN}{dz_\mathrm{conv}}$ can be estimated from the photometric redshift distribution by convolving with a redshift-dependent Gaussian. The method given in \cite{2018MNRAS.476.1050B} uses 2MPZ galaxies with both photometric and spectroscopic redshifts to determine the scatter of the observed spectroscopic redshift, $\sigma(z)$, in particularly narrow photometric redshift bins and finds
\begin{equation}
    \sigma(z) = 0.034 \tanh{(-20.78 z^2 +7.76 z + 0.05)}.
\label{eqn:sigmaz}
\end{equation}
This quantity is later used to estimate the spectroscopic redshift distribution of all the galaxies in the survey. In this analysis, Ref.~\cite{2018MNRAS.476.1050B} find that a Gaussian PDF centred at that particular photometric redshift provides a good fit for $\frac{dN}{dz_\mathrm{conv}}$ compared to the spectroscopic redshift $\frac{dN}{dz_\mathrm{spec}}$ (see Figure 6 in \cite{2018MNRAS.476.1050B}). That is, the photometric redshifts have some scatter but are overall unbiased relative to the spectroscopic redshifts. Although this is justified for the entire 2MPZ catalog, in our work we measure the clustering of mock siren catalogs, which, depending on the GW host-galaxy probability, have different stellar mass distributions from 2MPZ as a whole. Given that the photometric redshift scatter can depend on the stellar mass (fainter galaxies with lower mass will likely have worse photometric redshifts), then their photometric redshift distribution may be biased depending on the mass selection. Therefore, assuming unbiased photometric redshifts and using Eq.~\eqref{eqn:sigmaz} for mock siren catalogs with different GW host-galaxy probabilities may be incorrect.

\begin{figure}
\centering
\includegraphics[width=0.48\hsize]{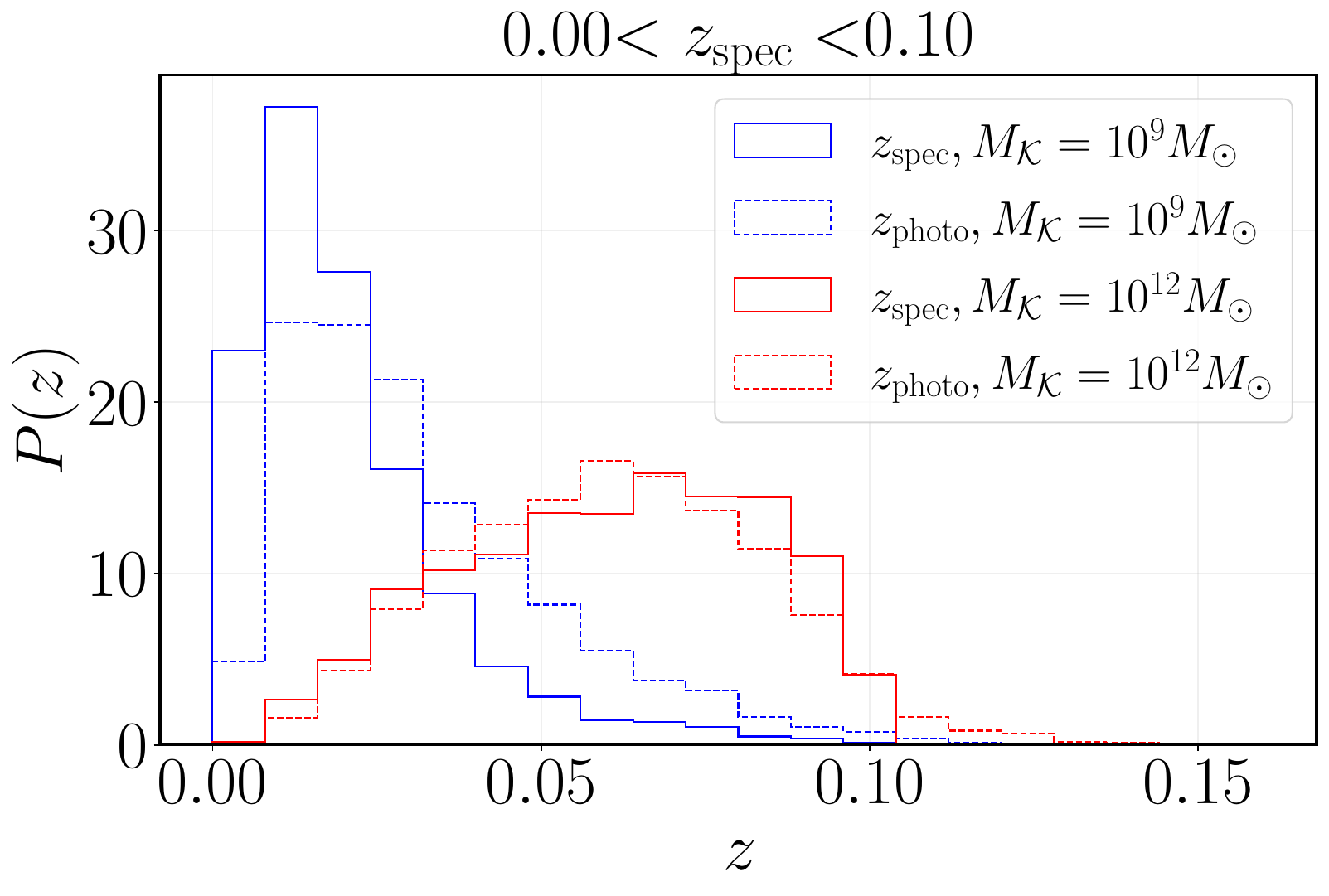}
\includegraphics[width=0.48\hsize]{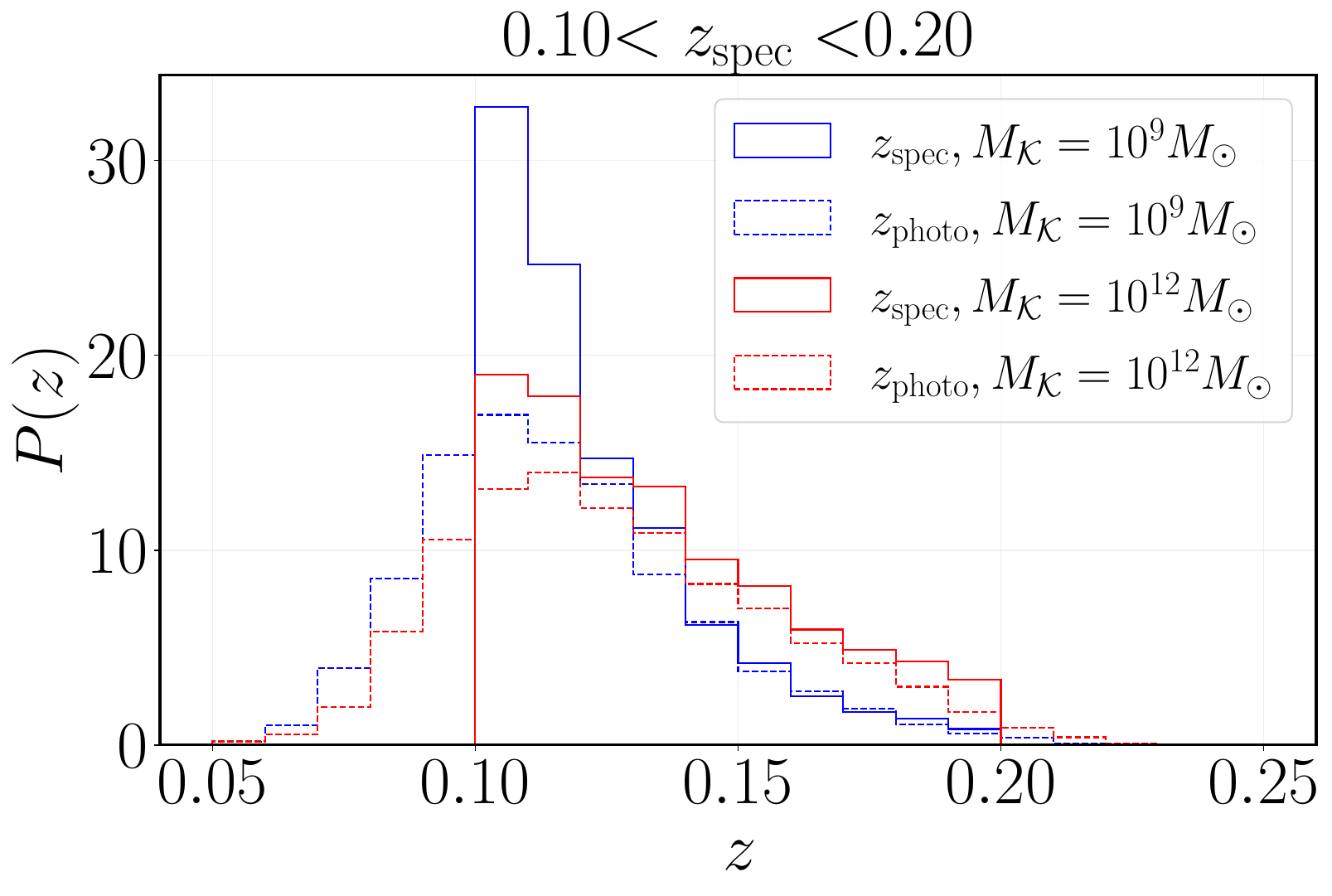}
\caption{The distribution of photometric and spectroscopic redshifts for the subset of the 2MPZ catalog that includes both photometric and spectroscopic redshifts. This subset then is divided into two equal redshift bins within the range $z_\mathrm{spec}\in[0,0.2]$. For each bin, we apply the host-galaxy probability function at the lowest and highest values of $M_\mathcal{K}$, which respectively prioritize low-mass and high-mass galaxies. Here, the values for $\delta_l$ and $\delta_h$ are set at $4$ and $0.5$, respectively. This plot shows that the scatter between photometric and spectroscopic redshifts is similar for both values of $M_{\mathcal{K}}$, but the low-$M_{\mathcal{K}}$ sirens show a substantial bias in their photometric redshifts, which overestimate the spectroscopic redshifts.}
\label{fig:z_examp}
\end{figure}

As an example, we extract the subset of the 2MPZ catalog that includes both photometric and spectroscopic redshifts and divide it into two equal redshift bins within the range $z_\mathrm{spec}\in[0,0.2]$. For each bin, we apply the host-galaxy probability function at the lowest and highest values of $M_\mathcal{K}$, which respectively prioritize low-mass and high-mass galaxies (keeping $\delta_l$ and $\delta_h$ fixed at their fiducial values). We then plot the resulting distributions of photometric and spectroscopic redshifts. The results, shown in Figure \ref{fig:z_examp}, indicate that the choice of $M_\mathcal{K}$ results in a different photometric redshift distribution in comparison to the spectroscopic redshift distribution, particularly in the first bin with $0 < z_{\textrm{spec}} < 0.1$. Although the photometric and spectroscopic redshift distributions are more similar for $M_\mathcal{K} = 10^{12}$ $M_\odot$, the photometric redshift distribution is shifted to higher values than the spectroscopic redshift distribution for $M_\mathcal{K} = 10^9$ $M_\odot$.

To address this potential issue, we allow the redshift distribution for sirens to differ compared to the entire galaxy population in 2MPZ and to vary depending on the host-galaxy probabilities used. To be specific, we repeat the analysis given in \cite{2018MNRAS.476.1050B} for siren hosts obtained in different runs of the host-galaxy probability, that is, different values of $M_\mathcal{K}$, $\delta_l$ and $\delta_h$. We fit Gaussians with two free parameters, the scatter $\sigma(z)$ and an offset parameter $z_{\textrm{shift}}\equiv z_\mathrm{photo}-\overline{z_\mathrm{spec}}$, in narrow photometric redshift bins of width 0.02. In other words, we allow the centre of the photometric redshift distribution to deviate from the mean spectroscopic redshift within each narrow bin by $z_{\textrm{shift}}$. Within each bin in photometric redshift, indexed with $i$, we fit the function,
\begin{align}
\label{eq:P_conv_2mpz}
    P_{\textrm{conv}, i}(\delta z)
    &\propto \exp\left(-\frac{1}{2} \left(\frac{z_{\textrm{conv}} - \overline{z_{\mathrm{spec},i}}}{\sigma_i} \right)^2\right)\nonumber\\
    &\propto \exp\left(-\frac{1}{2} \left(\frac{\delta z + \overline{z_{\mathrm{shift},i}}}{\sigma_i} \right)^2\right).
\end{align}

\begin{figure}
\centering
\includegraphics[width=0.48\hsize]{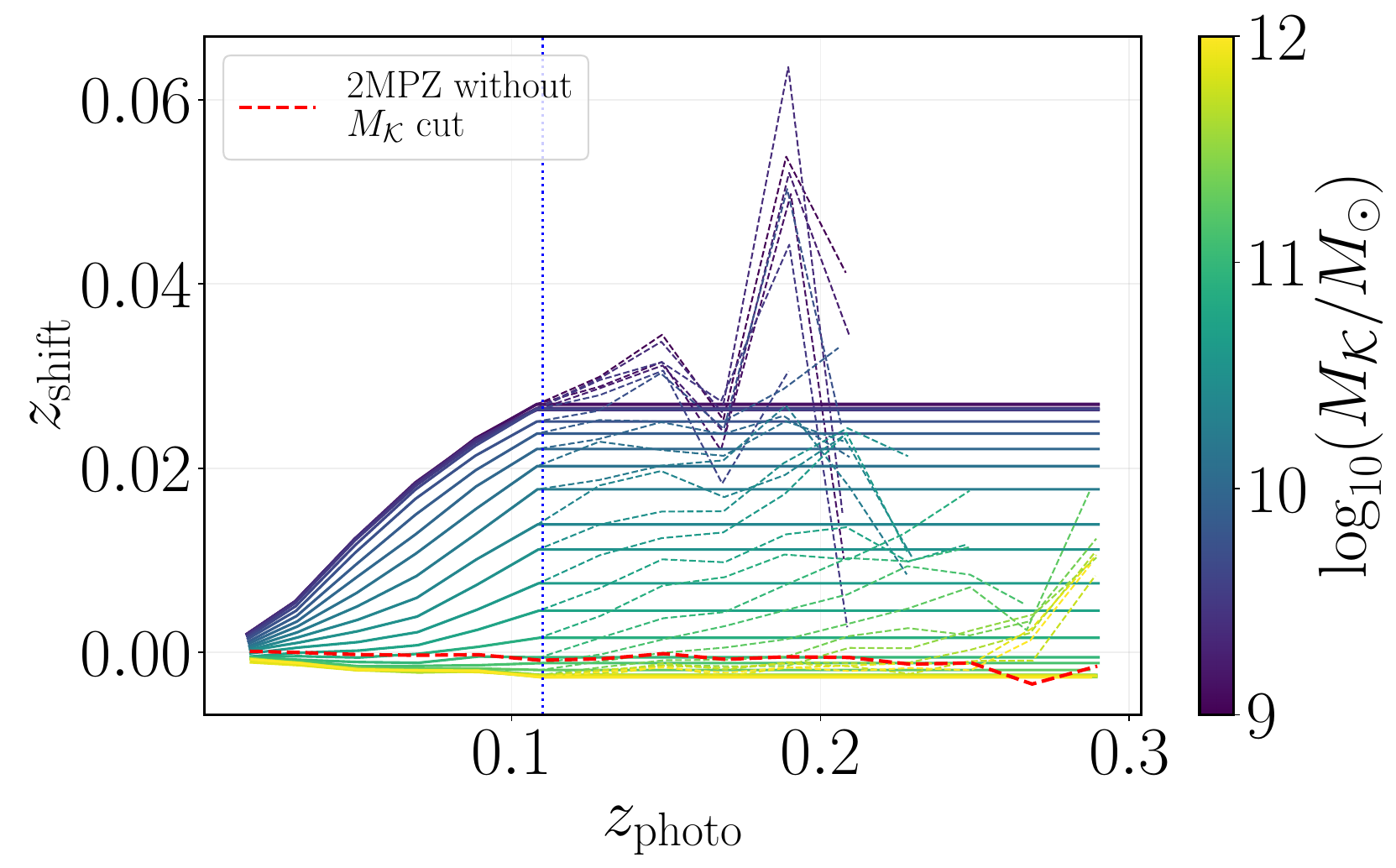}
\includegraphics[width=0.48\hsize]{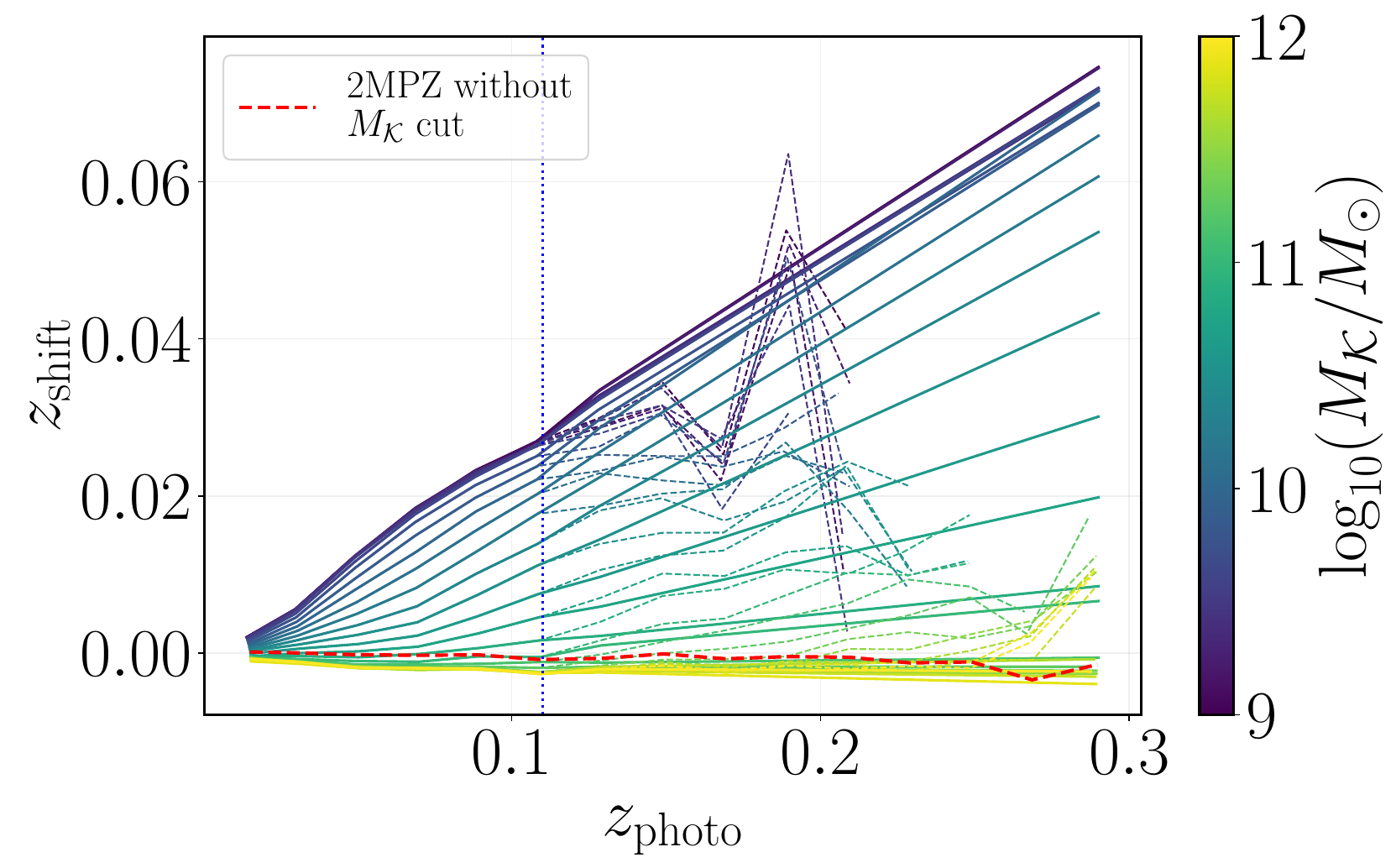}
\caption{The offset between the spectroscopic and photometric redshift distribution, i.e.\ $z_\mathrm{shift}=z_\mathrm{photo}-\overline{z_\mathrm{spec}}$, for sirens of various $M_{\mathcal{K}}$, setting $\delta_l=4$ and $\delta_h=0.5$), in narrow photometric redshift bins ($\sim 0.02$), as generated from the 2MPZ catalog. This quantity is used in Eq.~\eqref{eq:P_conv_2mpz}. The dashed lines represent the actual data. From $z \simeq 0.11$ onward, there are not enough hosts (in the siren catalog), the data fluctuate, and we approximate it with (I) a constant (left panel) or (II) a linear extrapolation (right panel), shown by solid lines. The red dashed line ($z_\mathrm{shift}\approx 0$) corresponds to the result for the entire 2MPZ catalog.}
\label{fig:2mpz-offset}
\end{figure}

We find that there is significant deviation from a zero value of $z_{\textrm{shift}}$ for different values of $M_\mathcal{K}$, $\delta_l$ and $\delta_h$. For example, Figure \ref{fig:2mpz-offset} illustrates how $z_{\textrm{shift}}$ changes with respect to $M_\mathcal{K}$ at $\delta_l=4$ and $\delta_h=0.5$, where the dashed lines represent the actual data. Note that after $z \simeq 0.11$, there are relatively few hosts in the siren catalog and therefore the errors on $z_{\textrm{shift}}$ are large. We tested a few methods to determine $z_{\textrm{shift}}$ in this region: (I) a constant or (II) a linear extrapolation. The difference between the two has a minor impact on the final bias calculation. For fiducial values of $\delta_l$, and $\delta_h$, we found the constant $z_{\textrm{shift}}$  generally matches the data better so we applied that extrapolation. The red dashed line in Figure \ref{fig:2mpz-offset} shows $z_\mathrm{shift}$ for the entire 2MPZ catalog. It follows zero closely, which matches the previous assumption for the entire catalog and the findings of \cite{2018MNRAS.476.1050B}. We can see that, as $M_\mathcal{K}$ decreases, the offset becomes larger, indicating that the photometric redshift measurement is worse for lower mass galaxies. While constant extrapolation works well for fiducial values of $\delta_l$, and $\delta_h$, for other configurations the fit may or may not remain valid. In principle, it is possible to determine the optimal fit for each individual case, but the sparse number of galaxies at $z > 0.11$ led us to simplify this by directly using the actual data with an upper limit imposed on $z_{\textrm{shift}}$. That is, we generate the corresponding dashed lines of Figure \ref{fig:2mpz-offset} for other configurations of $\delta_l$ or $\delta_h$, but to avoid producing spurious data, we impose the condition of $|z_\mathrm{shift}|<0.03$.

Next, we investigate the effect of the host-galaxy probability on the scatter. We find that the $\sigma(z)$ formulation of Eq.~\eqref{eqn:sigmaz} is still a good fit, as illustrated in Figure \ref{fig:2mpz_sigma} for different choices of the host-galaxy probability. This is further discussed in Appendix \ref{sec:z-conv} and the impact of choosing different convolution schemes on GW bias is presented in Figure \ref{fig:b_with_diff_conv}.

\begin{figure}
\centering
\includegraphics[width=0.48\hsize]{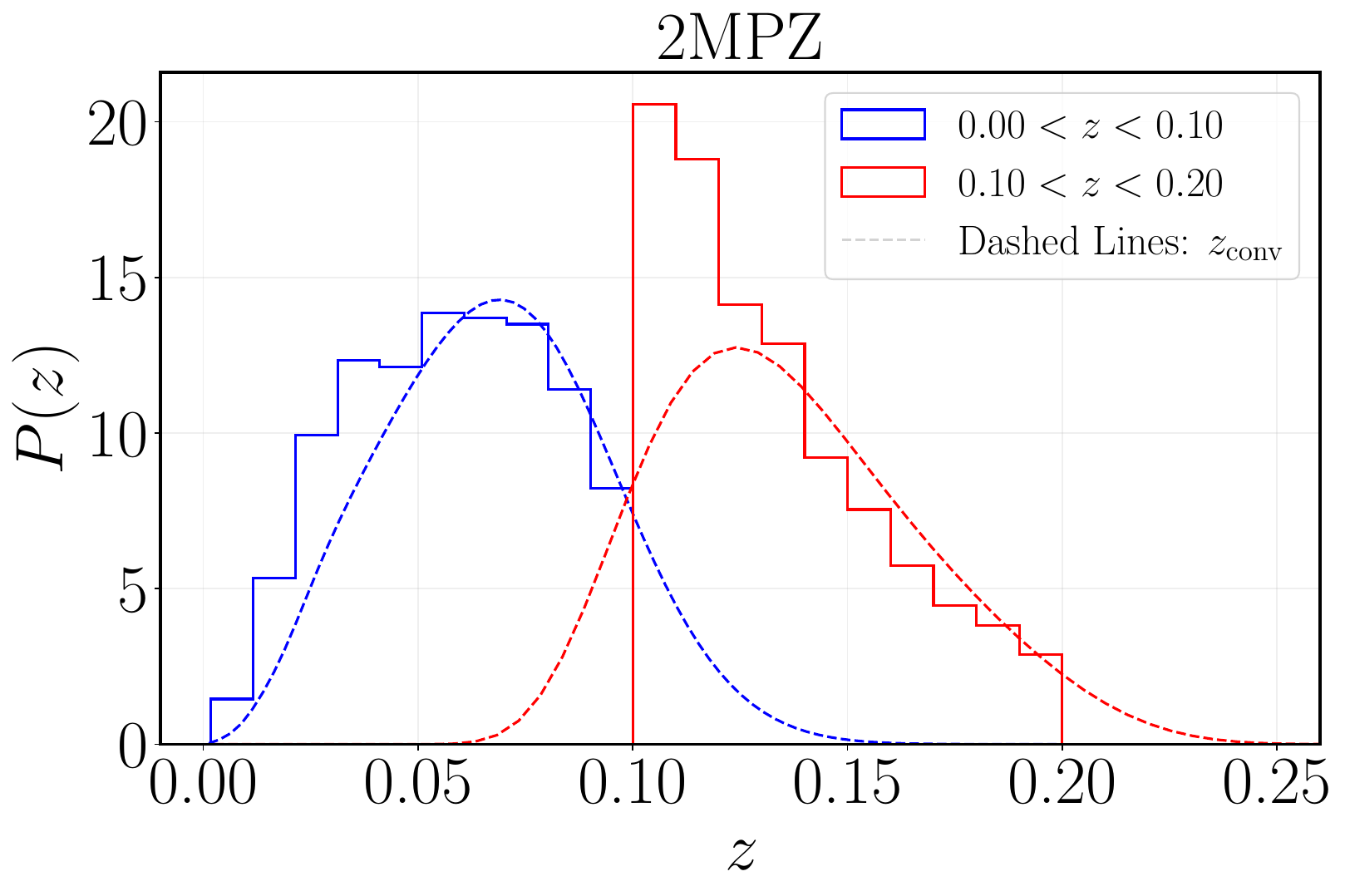}
\includegraphics[width=0.48\hsize]{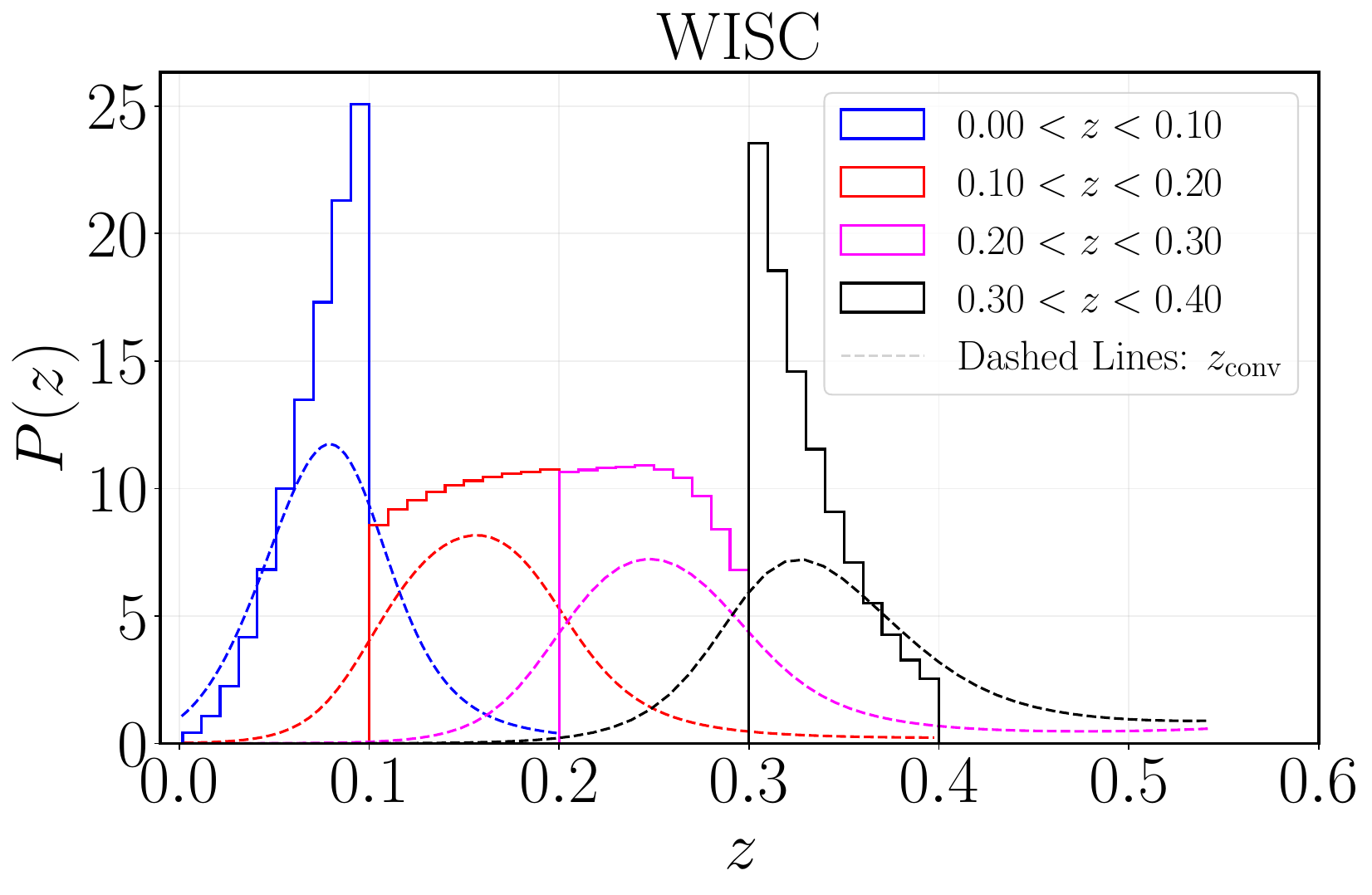}
\caption{
Comparison between the photometric redshift distribution (solid) and the convolved redshift distribution (dashed), for the entire galaxy samples of 2MPZ and WISC. The latter is taken into account in the calculation of the model angular power spectrum. This figure shows only galaxies that do not have spectroscopic redshifts.}
\label{fig:dNdz}
\end{figure}

To summarize our results for 2MPZ, the left panel of Figure \ref{fig:dNdz} shows the original distribution of photometric redshifts in 2MPZ alongside the convolved redshift distribution derived using this approach.

Meanwhile, for WISC, we convolve the photometric redshift distribution with a generalized Lorentzian \cite{2018MNRAS.481.1133P},
\begin{equation}
    P_{\mathrm{conv}}(\delta z) \propto \left(1 + \frac{\delta z ^2}{2 a(z) s(z)^2}\right)^{-a(z)},
\end{equation}
where $a(z) = -4z + 3$ and $s(z) = 0.04 z + 0.02$. Figure \ref{fig:dNdz} (right) shows the photometric redshift convolution for the galaxy catalogs in their corresponding redshift bins. We cannot conduct the same detailed tests for WISC as for 2MPZ, since we do not easily have access to the corresponding spectroscopic redshift catalogs. Moreover, due to the higher mean redshifts of the WISC bin, the impact of changes in the redshift distribution is expected to be smaller. Hence, we consider only the simple case where the redshift convolution is the same for all the host-galaxy probabilities considered.

\section{Results for gravitational wave (GW) bias and discussion}
\label{sec:results}
To confirm the validity of our analysis pipeline, we first measured the GW bias for a uniform host-galaxy probability (see Appendix \ref{sec:random}). Once we showed that the GW bias agrees with the galaxy bias in that limit as expected, we used the GW host-galaxy probability and investigated how the GW bias depends on the parameters of the host-galaxy probability function, redshift, and angular scale.

In this section, we present our results for the distinct siren catalogs associated with 2MPZ and WISC, analyzed in their respective redshift bins as outlined in Section \ref{sec:power_spectrum}. There is one common redshift bin between the two galaxy catalogs used here $z \in [0.1,0.2]$. Within that shared redshift bin, the results for the GW bias agree on the overall trend for varying different parameters; however, the overall amplitude is different. This is due to different completeness levels in the two catalogs: WISC is more complete for lower masses compared to 2MPZ, and consequently yields a lower GW bias, consistent with the lower galaxy bias in WISC.

In the following, we discuss this and provide more details on how the GW bias is influenced by various parameters, namely (I) variation in relation to the parameter $M_\mathcal{K}$, (II) dependence on the parameters $\delta_l$ and $\delta_h$, (III) changes in redshift, and (IV) variation with respect to the choice of $\ell_{\rm max}$.

\subsection{Variation in pivot mass $M_\mathcal{K}$}
\label{subsec:vary_pivot_mass}
\begin{figure}
\centering
\includegraphics[width=0.48\hsize]{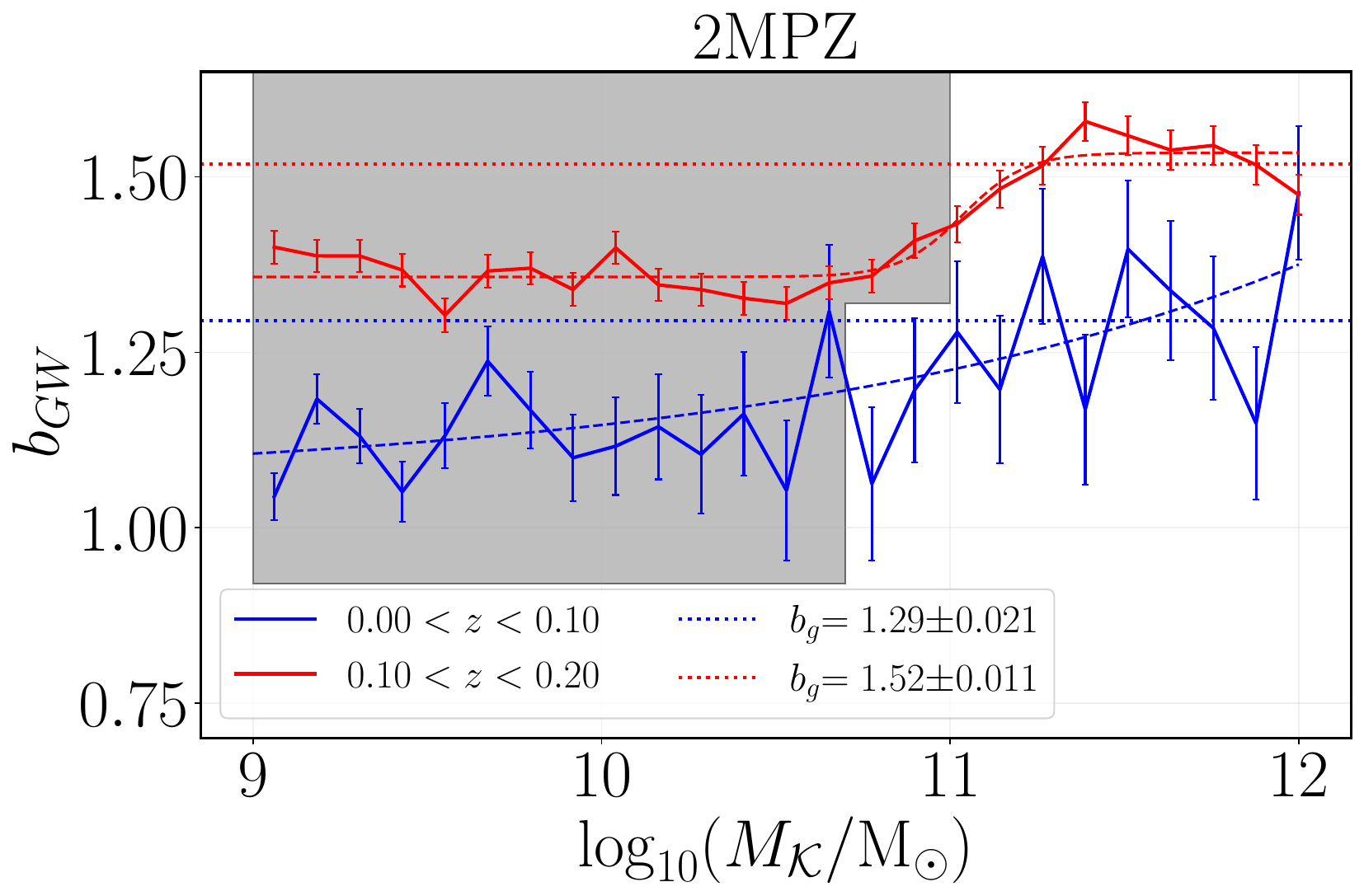}
\hspace{8pt}
\includegraphics[width=0.48\hsize]{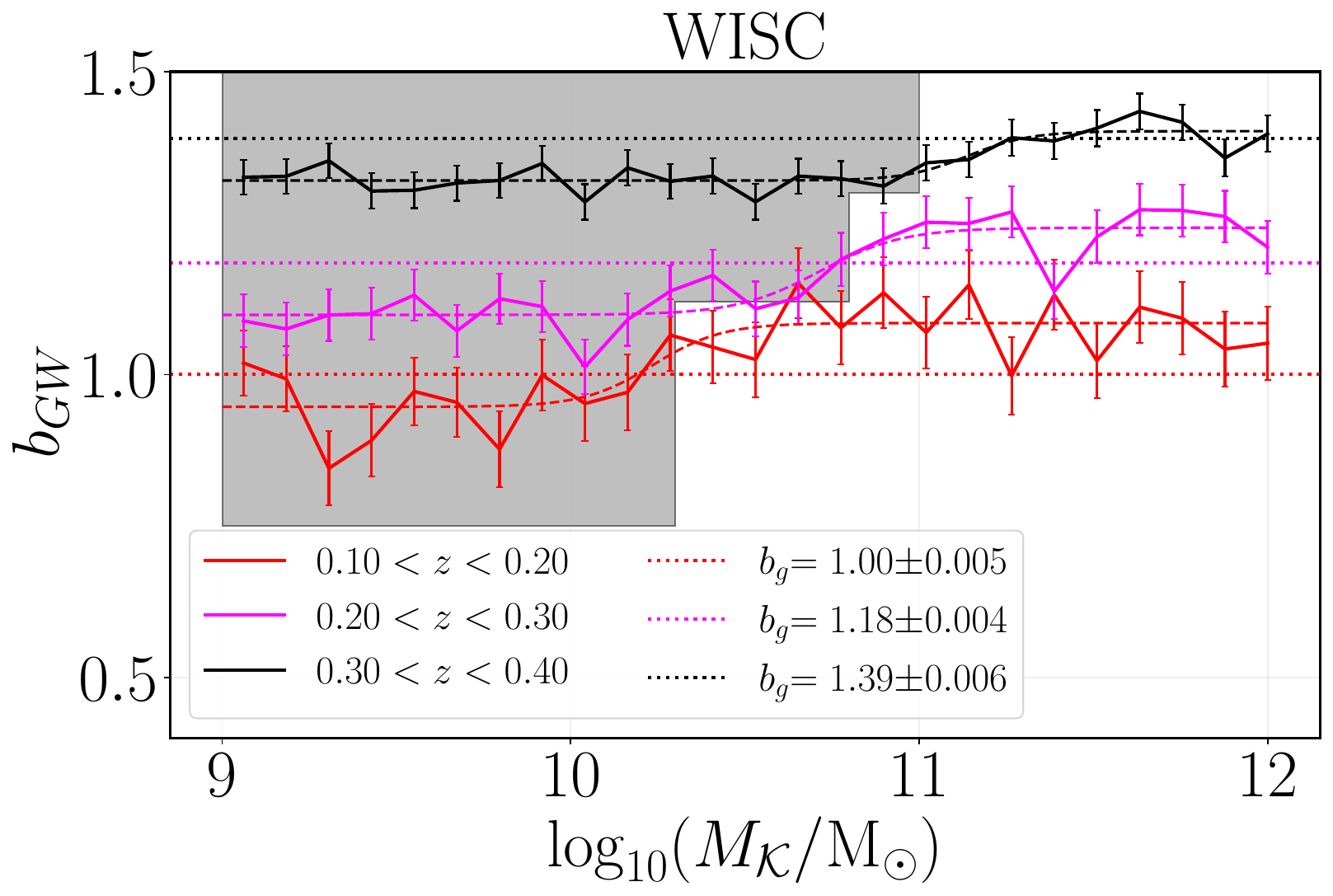}
\caption{Bias for varying the pivot mass $M_\mathcal{K}$ in the GW host-galaxy probability for 2MPZ and WISC catalogs. In this case the low and high-mass slopes are fixed by setting $\delta_l=4$, and $\delta_h=0.5$. The gray areas show roughly the range that galaxy catalogs are incomplete. The dotted lines represent the galaxy bias. The dashed lines show the fitted logistic function introduced in Eq.~\eqref{eq:logistic}. The fitted parameters are shown in Table \ref{tab:log}. Even though we show a fitted function for the first redshift bin in 2MPZ (blue curve in the left panel), the fit is not an accurate description for the entire range of $M_\mathcal{K}$ and the data points probe only the slope-changing part of the logistic function. The bars indicate $1\sigma$ error.}
\label{fig:m_k}
\end{figure}

We start with investigating how the GW bias changes in response to the host-galaxy probability of Eq.~\eqref{eq:selection} for different values of $M_\mathcal{K}$. The plots in Figure \ref{fig:m_k} show $b_{GW}$ as a function of $\log(M_\mathcal{K})$, while $\delta_l$ and $\delta_h$ are kept fixed at their fiducial values. We observe similar trends in both catalogs and across all redshift bins, that is, monotonically increasing, flatter towards the endpoints and a steeper slope at some midpoint, with some exceptions in the first redshift bin for 2MPZ, which will be discussed separately. To capture the observed behavior of the GW bias versus $\log (M_\mathcal{K})$, we fit a logistic function to the data points in Figure \ref{fig:m_k} as follows
\begin{align}
\label{eq:logistic}
    b_{GW}=\frac{A}{1+\exp{\left(-R(\log_{10}(M_\mathcal{K}/ \mathrm{M}_\mathrm{mid})\right)}}+b_\mathrm{min}\,,
\end{align}
where $A$, $R$, $M_\mathrm{mid}$, and $b_\mathrm{min}$ are fitting constants whose values are shown in Table \ref{tab:log}. $b_\mathrm{min}$ represents the asymptotic minimum value of $b_{GW}$ as $M_{\mathcal{K}}$ approaches its lower end. $M_\mathrm{mid}$ represents the point where the slope is steepest, and $AR/4$ is the value of the derivative at that point. $A+b_\mathrm{min}$ represents the maximum value of the GW bias, and thus $A$ represents how much the GW bias changes in the range of $M_{\mathcal{K}}\ll M_\mathrm{mid}$ to $M_{\mathcal{K}}\gg M_\mathrm{mid}$. While in Figure \ref{fig:m_k}, we also display a fitted function for the first redshift bin of 2MPZ (blue curve in the left panel of Figure \ref{fig:m_k}), in this case the best-fit function does not flatten at endpoints of $M_{\mathcal{K}}$, and the fit is not accurate in the tails, since the data points probe only the slope-changing part of the logistic function. Therefore, we do not report the fitted parameters for that bin in Table \ref{tab:log}.

\begin{table}[]
\centering
\scriptsize
\begin{tabular}{|ccccccc|}
\hline
Catalog& $z$ & $A$ & $R$& $b_\mathrm{min}$ & $\log_{10}(M_\mathrm{mid}/M_\odot)$ & \rule{0pt}{2.5ex} $\log_{10}(\overline{M_g}/M_\odot)$\\ \hline\hline
2MPZ & $0.1<z<0.2$ & $0.18 \pm 0.01$ & $11.0 \pm 4.4$ & $1.36 \pm 0.01$ & $11.0 \pm 0.04$ & $11.1$ \\ \hline
WISC & $0.1<z<0.2$ & $0.14 \pm 0.03$ & $10.1 \pm 11.8$ & $0.95 \pm 0.02$ & $10.2 \pm 0.1$ & $10.4$\\ \hline
WISC & $0.2<z<0.3$ & $0.14 \pm 0.02$ & $8.4 \pm 6.3$ & $1.10 \pm 0.01$ & $10.7 \pm 0.1$ & $10.8$\\ \hline
WISC & $0.3<z<0.4$ & $0.08 \pm 0.02$ & $11.2 \pm 11.4$& $1.32 \pm 0.01$ & $11.2 \pm 0.1$ & $11.1$\\ \hline
\end{tabular}
\caption{Parameters of Eq.~\eqref{eq:logistic} fitted to the data points in Figure \ref{fig:m_k}. $A$, $R$, $M_\mathrm{mid}$, and $b_\mathrm{min}$ are the fitting coefficients and constants. $b_\mathrm{min}$ represents the asymptotic minimum value of $b_{GW}$ as $M_{\mathcal{K}}$ approaches its lower end. $A$ represents how much GW bias is changed in the whole range of $M_{\mathcal{K}}$, in other words, $A+b_\mathrm{min}$ represents the maximum value of GW bias. $M_\mathrm{mid}$ represents the point where the slope is steepest, and $AR/4$ is the derivative at that point. The last column corresponds to the horizontal dashed lines of Figure \ref{fig:host_mass_dist}, i.e. the mean mass of all the galaxies in the bin.}
\label{tab:log}
\end{table}

\begin{figure}
\centering
\includegraphics[width=0.48\hsize]{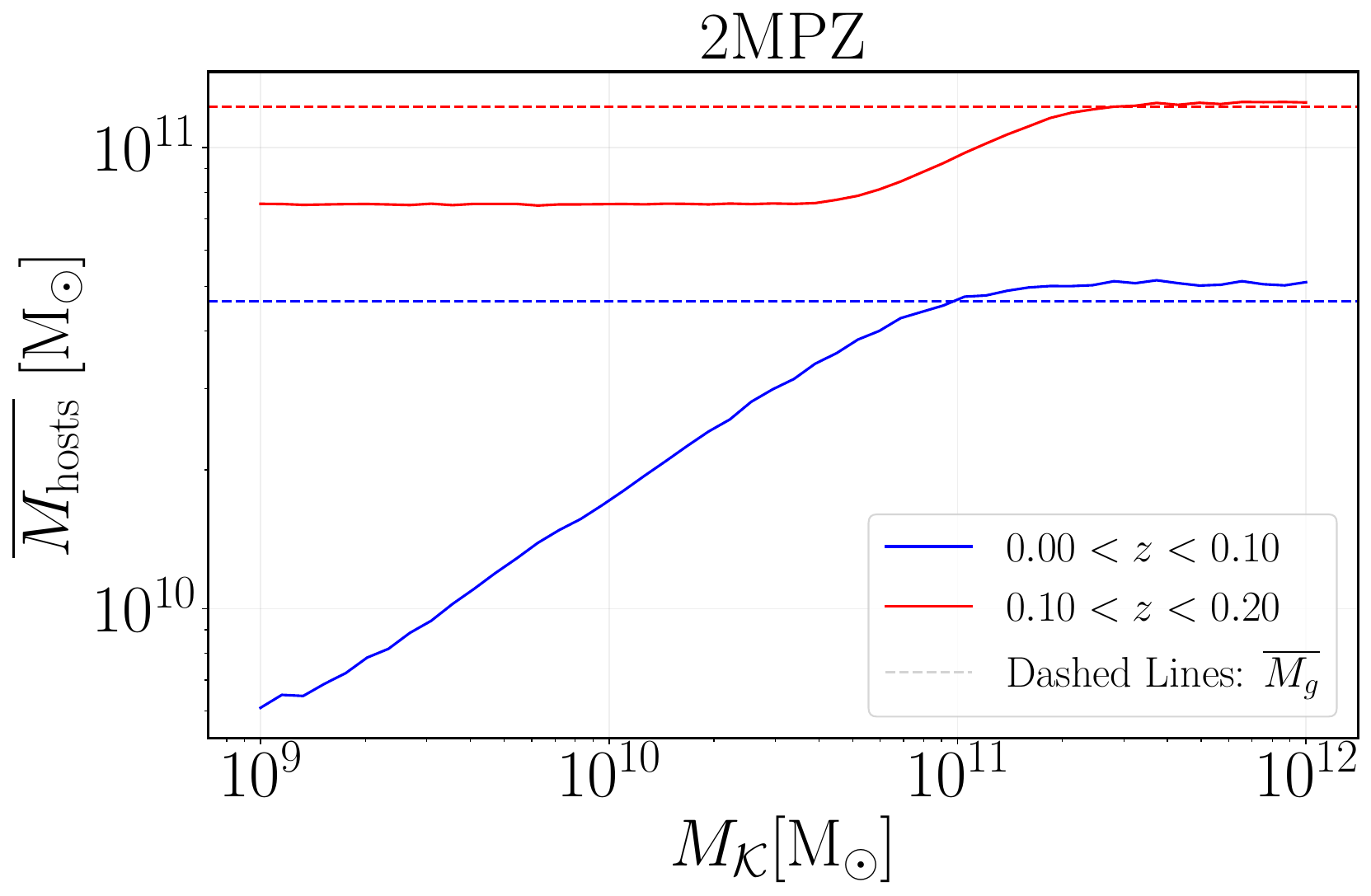}
\hspace{8pt}
\includegraphics[width=0.48\hsize]{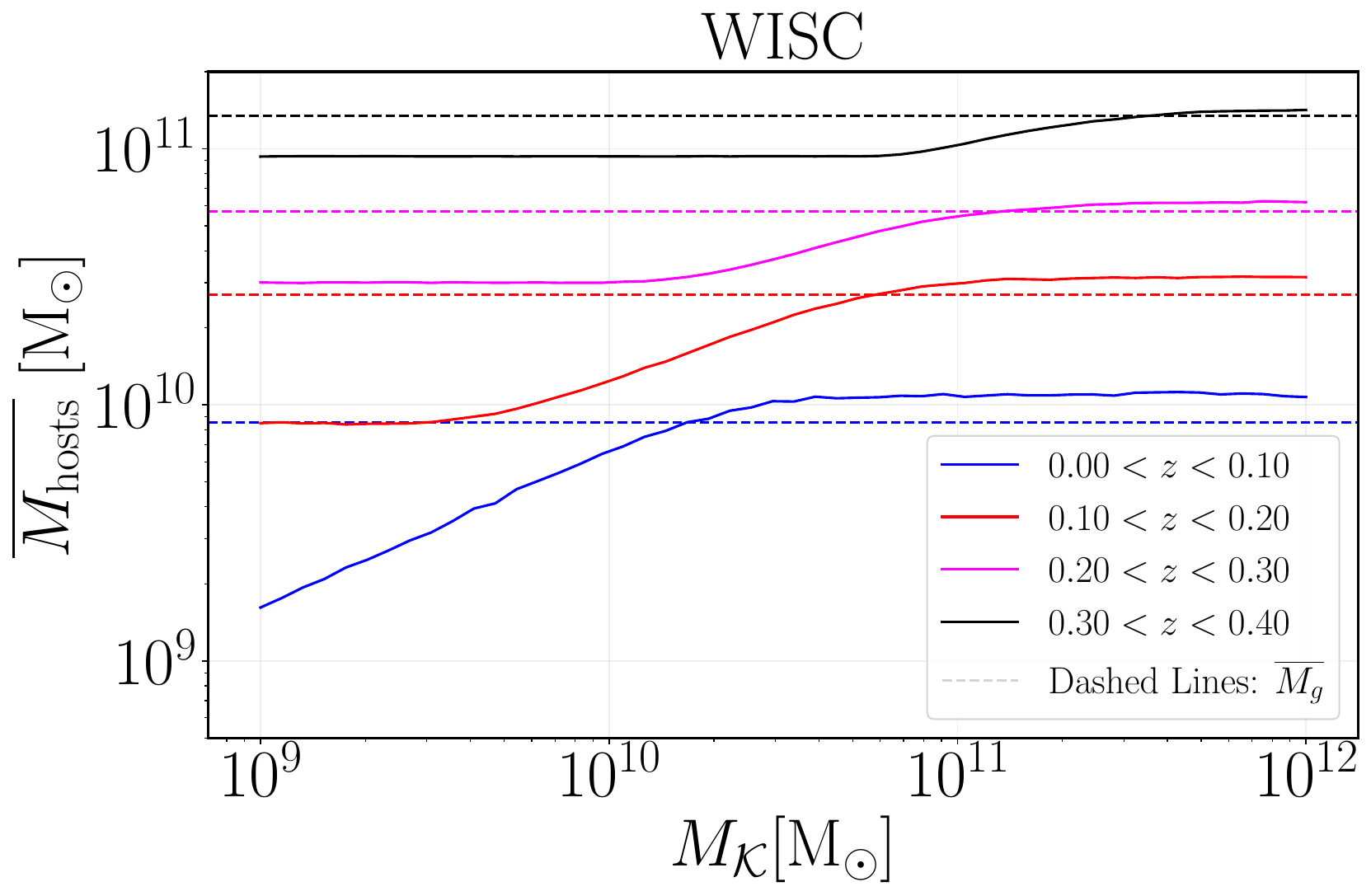}
\caption{Mean stellar mass of host galaxies for different $M_\mathcal{K}$ values. In this case $\delta_l=4$, and $\delta_h=0.5$. The dashed lines show the mean mass of all galaxies (without any $M_\mathcal{K}$ cut) within that redshift bin. Even though we are not modeling the GW bias in WISC for $0<z<0.1$ (the blue curve in the right panel), we show the host mass distribution to demonstrate similar behavior for 2MPZ and WISC in this particular redshift bin.}
\label{fig:host_mass_dist}
\end{figure}

To better understand the trends in Figure \ref{fig:m_k} we examine the variation of the GW host stellar mass distribution with respect to $M_\mathcal{K}$ and compare it to the stellar mass distribution of galaxies in 2MPZ and WISC. Figure \ref{fig:host_mass_dist} shows the mean stellar mass of all the galaxies, $\overline{M_g}$, as well as the host galaxies, $\overline{M_\text{hosts}}$, at different values of $M_\mathcal{K}$. This plot arises by the combination of two effects, (i) GW host-galaxy probability\footnote{As mentioned in Section \ref{subsec:GW_Gal}, our fiducial model (illustrated with a solid blue line in Figure \ref{fig:mzr}) exhibits two phases: a gentle increase followed by a steep decline just past $M_\mathcal{K}$.} and (ii) stellar mass distribution of the galaxies in the catalog. When $M_\mathcal{K}> \overline{M_g}$, the increasing part of the host-galaxy probability function becomes more relevant causing the mean stellar mass of hosts, $\overline{M_\text{hosts}}$, to be around or slightly higher than the mean stellar mass of galaxies, $ \overline{M_g}$ (Figure \ref{fig:host_mass_dist}). As $ M_\mathcal{K}$ increases further above $ \overline{M_g}$, $\overline{M_\text{hosts}}$ rises marginally since most of the galaxies correspond to masses less than $M_{\mathcal{K}}$ and the increase rate (for $\overline{M_\text{hosts}}$ vs. $M_{\mathcal{K}}$) is set by the shallow slope of the growing regime of the host-galaxy probability function. As a result of the increase in $\overline{M_\text{hosts}}$ with higher $M_{\mathcal{K}}$, since galaxy bias is related to stellar mass \citep{Moster_2010,2010ApJ...724..878T,2010ApJ...717..379B}, there is a gradual rise in the GW bias, making $b_{GW}\gtrsim b_g$ (Figure \ref{fig:m_k}). Conversely, when $M_\mathcal{K} < \overline{M_g}$, the decreasing part of the host-galaxy probability function is more relevant, and the sharp drop-off causes the host-galaxy probability to strongly favor low-mass galaxies. Consequently, as $M_{\mathcal{K}}$ decreases, $\overline{M_\text{hosts}}$ also decreases rapidly (see Figure \ref{fig:host_mass_dist}). As a result, $b_{GW}$ decreases below $b_g$ (see Figure \ref{fig:m_k}). Naturally, the transition between the two phases from the $M_{\mathcal{K}} <  \overline{M_g}$ regime to the $M_{\mathcal{K}} >  \overline{M_g}$ regime occurs near $M_{\mathcal{K}} \sim  \overline{M_g}$. In fact, the values of $M_\mathrm{mid}$, where the slope is steepest for each bin, align well with the values of $\overline{M_g}$ in the corresponding bin (see Table \ref{tab:log}). When $M_{\mathcal{K}}$ drops into the very low-mass regime, where the catalog is incomplete and there are very few galaxies, further lowering $M_{\mathcal{K}}$ makes no difference, and $\overline{M_\text{hosts}}$
(and therefore $b_{GW}$) saturates.

The behavior of the GW bias in the tails of $ M_\mathcal{K}$ range can also be understood in terms of the
Schechter function \cite{Schechter76,Cole01} with a cut-off at low masses due to the incompleteness of the catalogs. If $M_{\mathcal{K}}$ is greater than the characteristic stellar mass $\hat{M}_{\star}$ of the Schechter function then further increasing $M_{\mathcal{K}}$ prefers more sirens in high-mass galaxies. But since there are exponentially few galaxies with stellar mass higher than $\hat{M}_\star$, this will have minimal impact on the mean stellar mass of the siren catalog and thus the GW bias. Hence both the GW bias and the mean stellar mass flatten out for $M_{\mathcal{K}}$ above the characteristic mass of the Schechter function $\hat{M}_{\star}$. Likewise, if $M_{\mathcal{K}}$ drops below the stellar mass where the catalogs become incomplete, since there are so few galaxies at these low masses, further decreasing $M_{\mathcal{K}}$ will have little impact on the mean stellar mass and thus the GW bias approaches a constant value. Due to the above reason, we find that the asymptotic value of the GW bias at low $M_\mathcal{K}$ is highly impacted by the completeness of the galaxy catalogs. In Figure \ref{fig:m_k}, we roughly indicate the incomplete regime of the galaxy catalogs in different redshift bins with the shaded gray area, taking the completeness threshold to start around the peak of the mass distribution in Figure \ref{fig:gal_masses}, motivated by the fact that the Schechter function is a monotonically decreasing function of mass.

Another important feature noticeable from Figure \ref{fig:m_k} is the difference in the variation of the GW bias with $M_\mathcal{K}$ between the first redshift bin (blue curve in the left panel of Figure \ref{fig:m_k}), in comparison to the other redshift bins. Firstly, note that it has relatively larger error bars, since there are fewer mergers in that redshift bin. Secondly, the stellar mass distribution of the galaxies within this low redshift bin is more complete at the lower-end of the stellar mass in comparison to a higher redshift bin. We will discuss in more detail the variation with redshift in Sec. \ref{sec:redshift}.

\begin{figure}
\centering
\includegraphics[width=0.48\hsize]{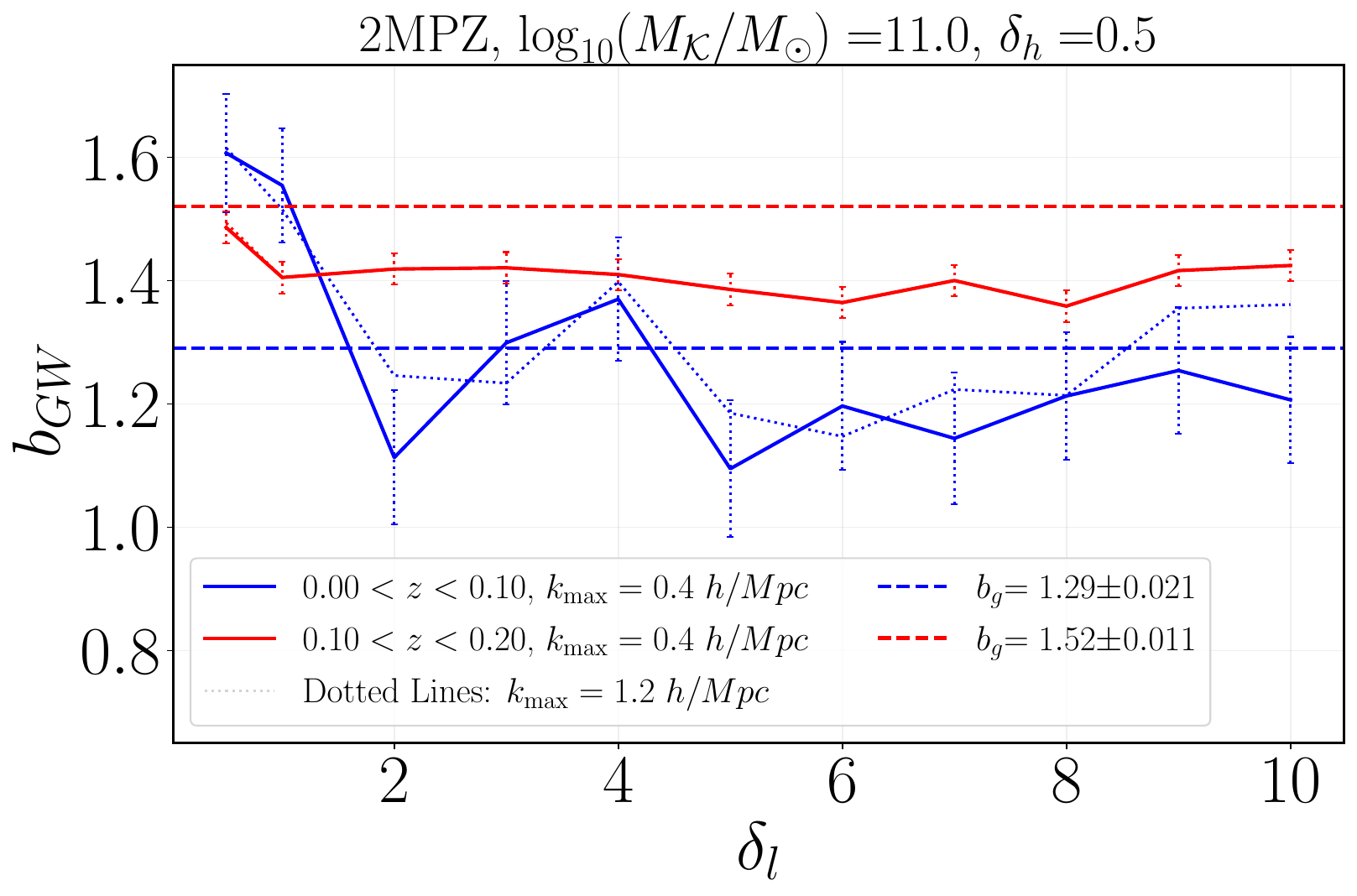}
\hspace{8pt}
\includegraphics[width=0.48\hsize]{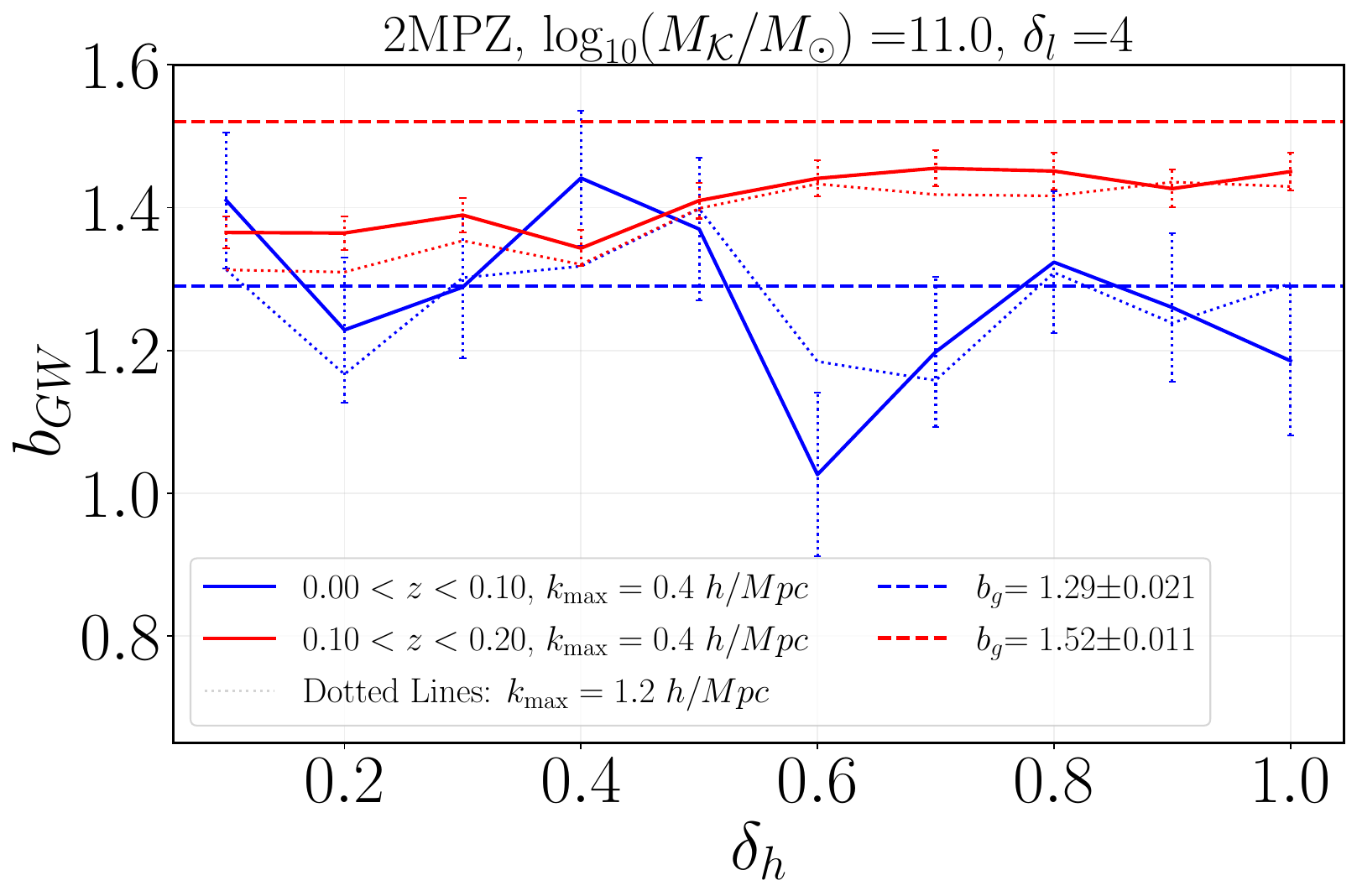}
\includegraphics[width=0.48\hsize]{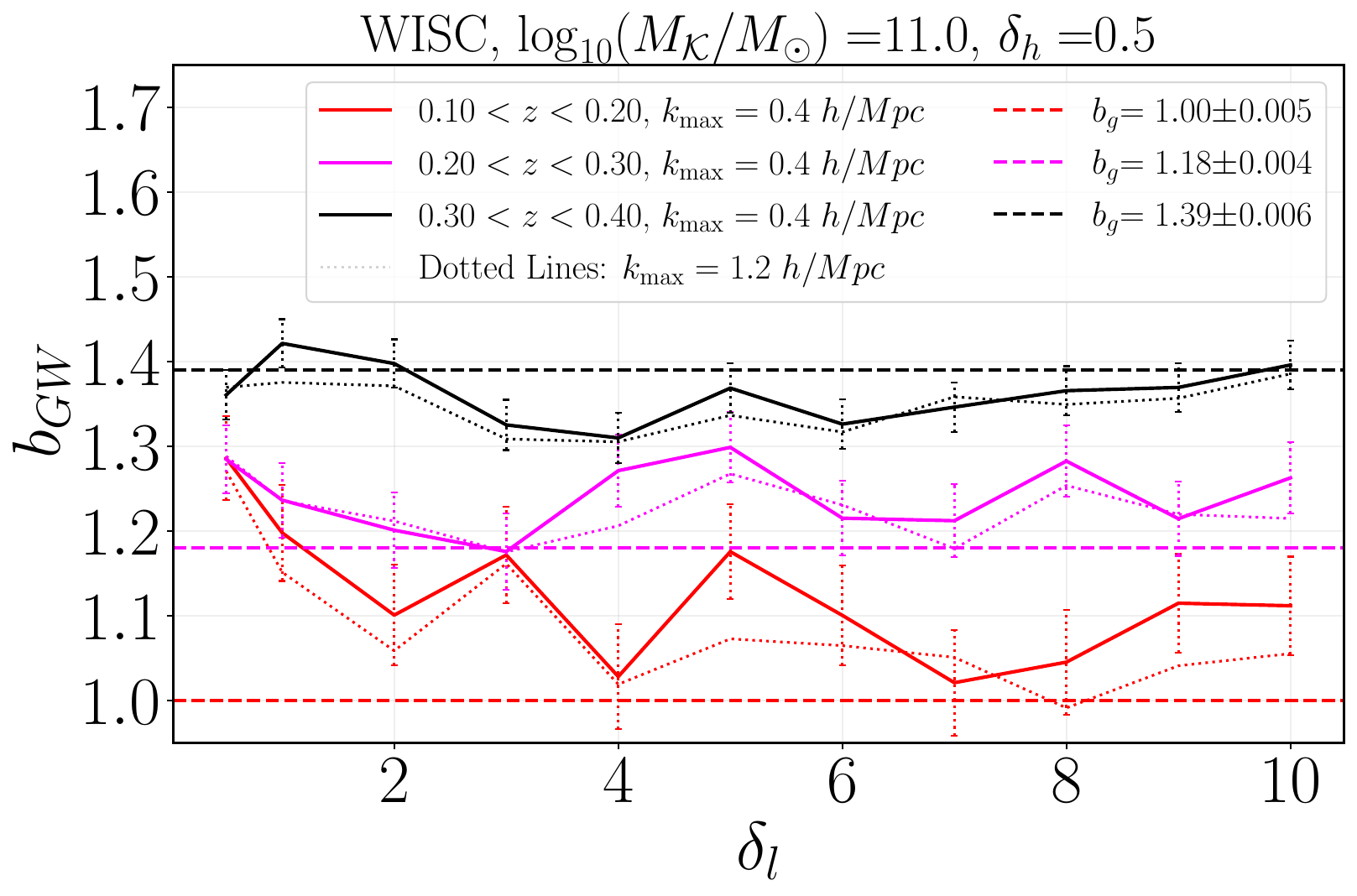}
\hspace{8pt}
\includegraphics[width=0.48\hsize]{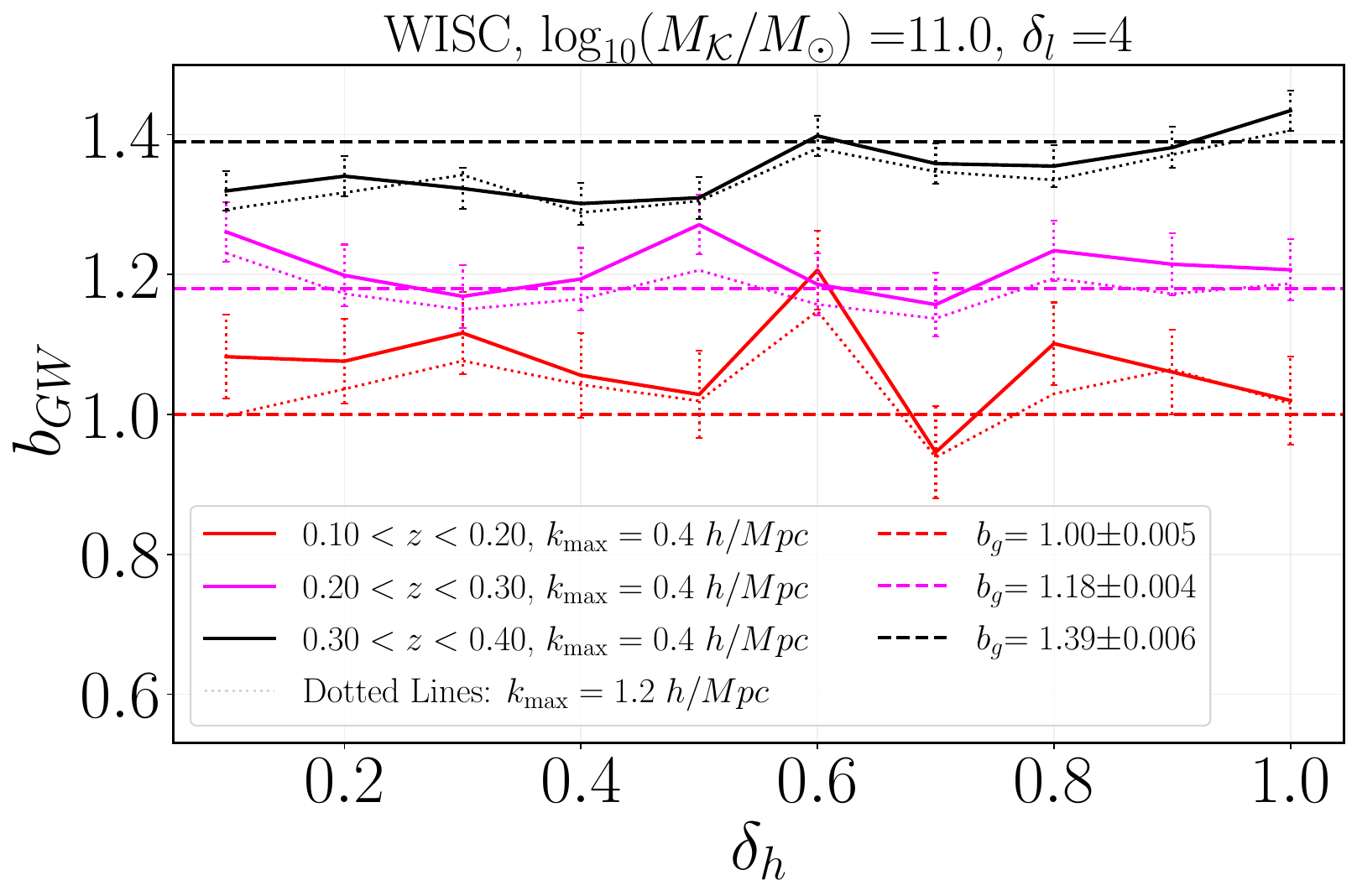}
\caption{Bias for varying $\delta_l$ (left panel) and $\delta_h$ (right panel) in the GW host-galaxy probability function setting $M_\mathcal{K}=10^{11}~M_\odot$. GW bias is weakly dependent on both $\delta_l$ and $\delta_h$ for this particular choice of $M_\mathcal{K}$, which is close to the mean stellar mass of the galaxies in different bins. Nevertheless, the left (right) panels illustrate that in lower (higher) redshift bins, where $\overline{M_g}$ is below (above) $10^{11}~M_\odot$, there is a noticeable increase (decrease) in GW bias as we increase the growth (suppression) rate by adopting $\delta_l \lesssim 1$ ($\delta_h \lesssim 1$). The dashed lines represent the galaxy bias. The dotted lines represent the same results for a different choice of $k_\mathrm{max}$ which confirms the scale independence to higher $\ell$ values. The bars are $1\sigma$ error. For the simultaneous varying all three parameters of the host-galaxy probability function $\delta_l$, $\delta_h$, and $M_{\mathcal{K}}$, see Appendix \ref{sec:3d}.}
\label{fig:deltas}
\end{figure}

\subsection{Variation in slopes $\delta_l$ and $\delta_h$}
\label{sec: bGW-deltal-deltah}
In the last sub-section, the fiducial model is chosen with $\delta_l=4$ and $\delta_h=0.5$. Here we consider the effect of varying $\delta_l$ and $\delta_h$ on the GW bias while keeping $M_\mathcal{K}=10^{11}~M_\odot$ (shown in Figure \ref{fig:deltas}).
For this particular choice of $M_\mathcal{K}=10^{11}~M_\odot$, the GW bias is weakly dependent on both $\delta_l$ and $\delta_h$. The reason is that $M_\mathcal{K}=10^{11}~M_\odot$ is close to the peak of the catalog mass function (Figure \ref{fig:gal_masses}) for all the redshift bins. Therefore, changing the slopes of the host-galaxy probability function only alters the selection probability of galaxies at the low- and high-mass ends, which have competing effects on the GW bias. This only leads to a noticeable impact on GW bias in cases where there are a significant number of galaxies in the low or high-mass tails, e.g. WISC at $0.1<z<0.2$ (for $\delta_l$, from the long tail of galaxies to low mass in this bin) or $0.3<z<0.4$ (for $\delta_h$, from the long tail of galaxies to high mass in this bin). However,  this may not hold for a different choice of $M_\mathcal{K}$, especially if the delay times are long so the pivot mass is at $M_\mathcal{K} > 10^{11}$ $M_\odot$ or there is no suppression in the host-galaxy probability function at all.

\begin{figure*}
\centering
\includegraphics[width=0.49\hsize]{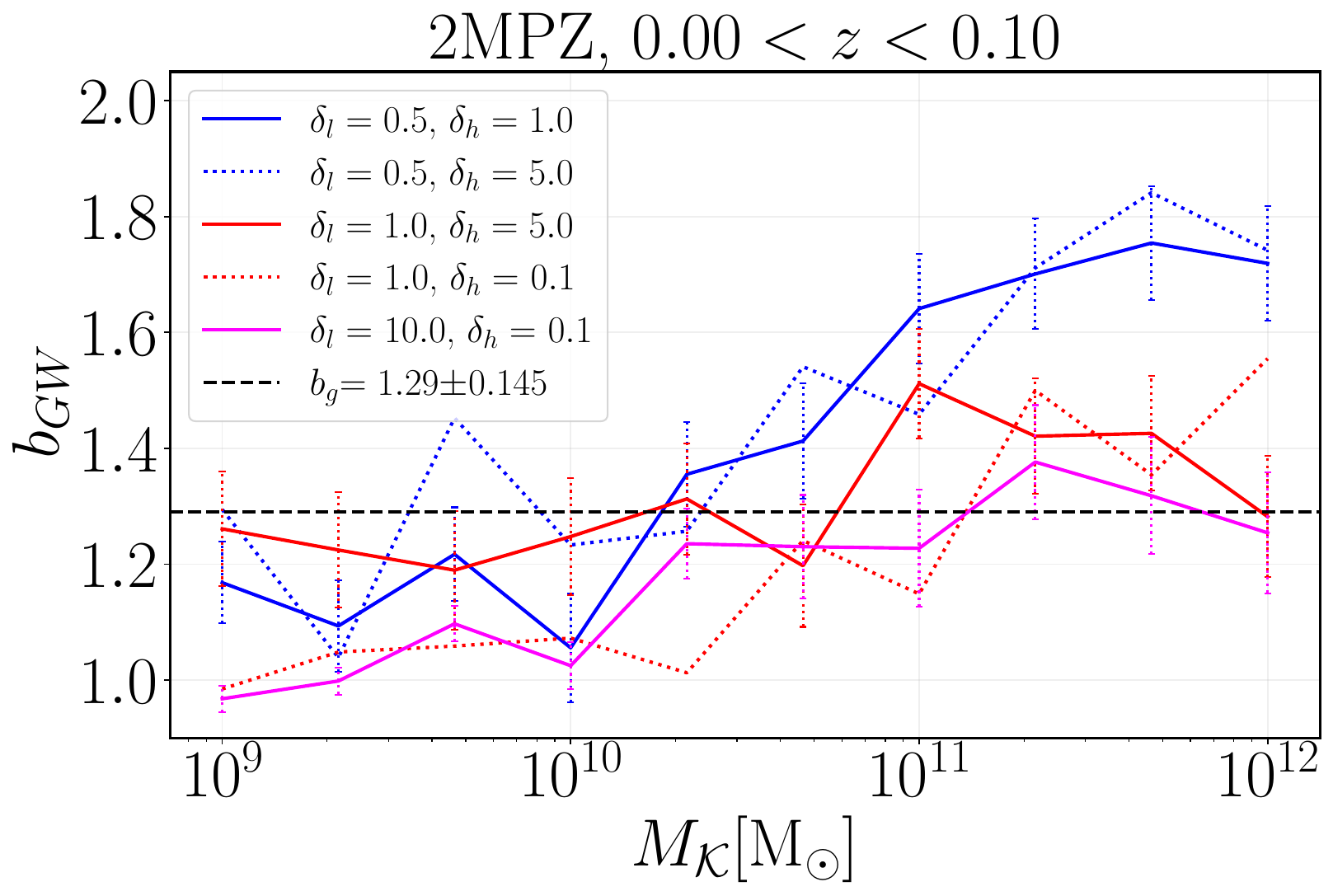}
\includegraphics[width=0.49\hsize]{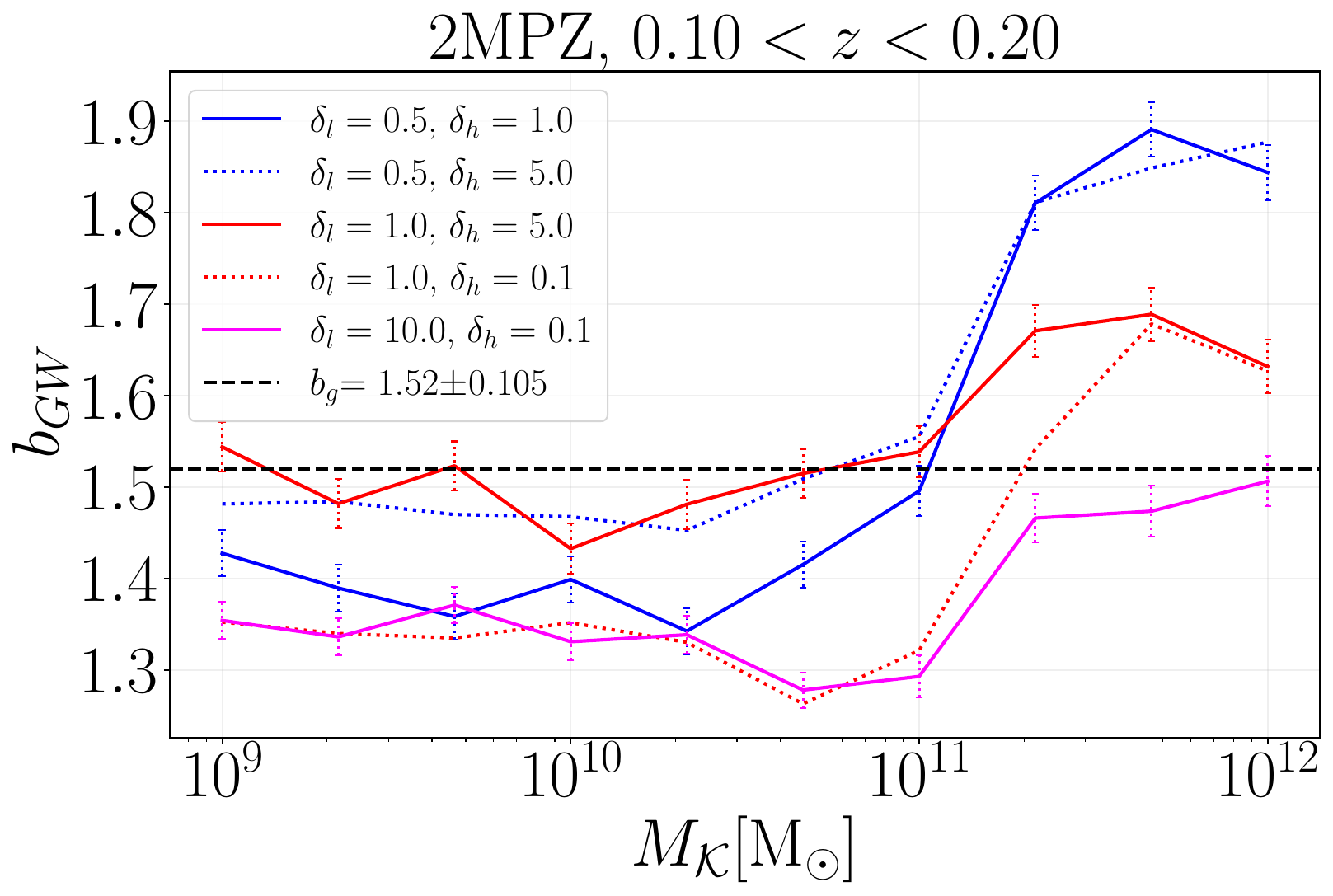}
\includegraphics[width=0.49\hsize]{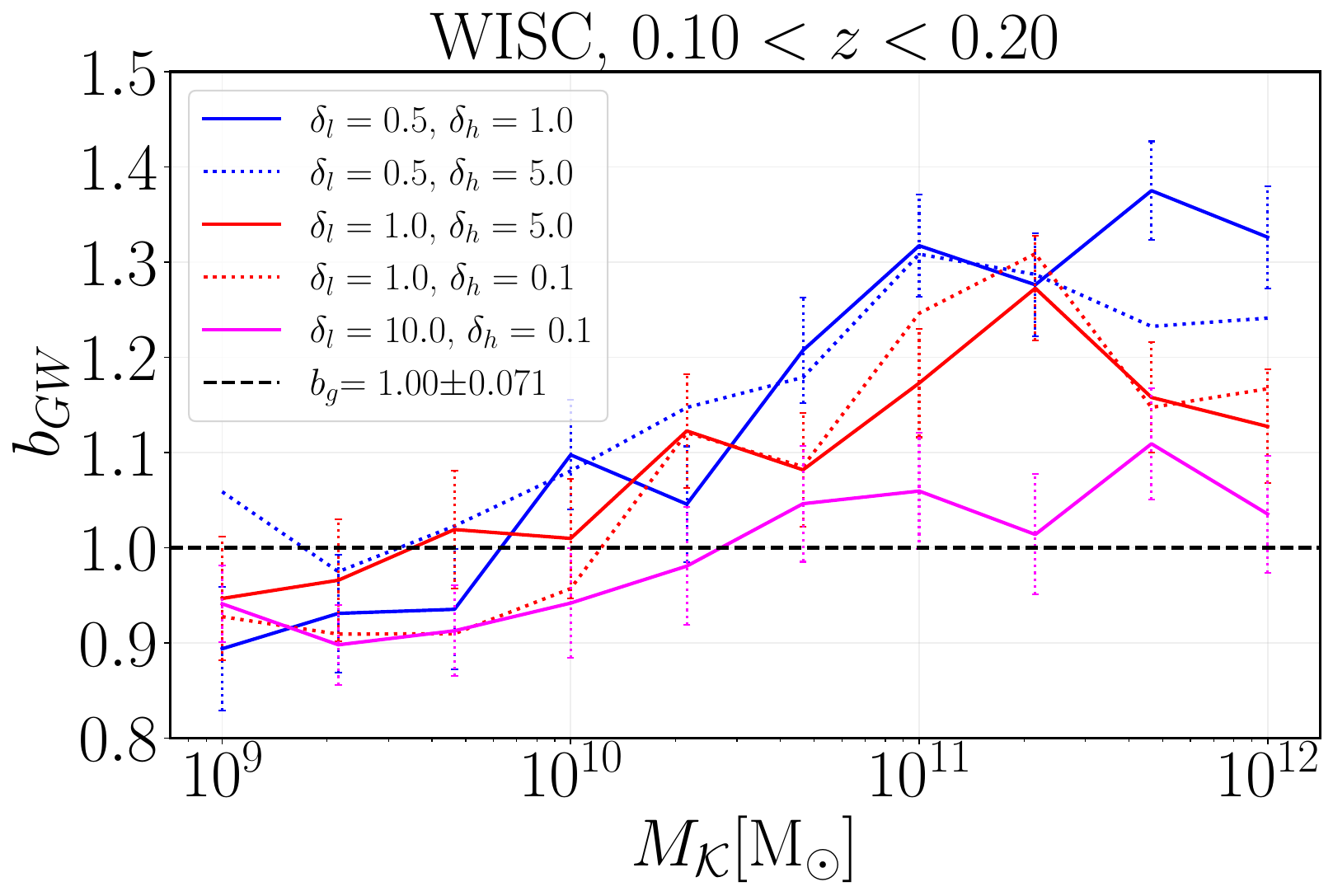}
\includegraphics[width=0.49\hsize]{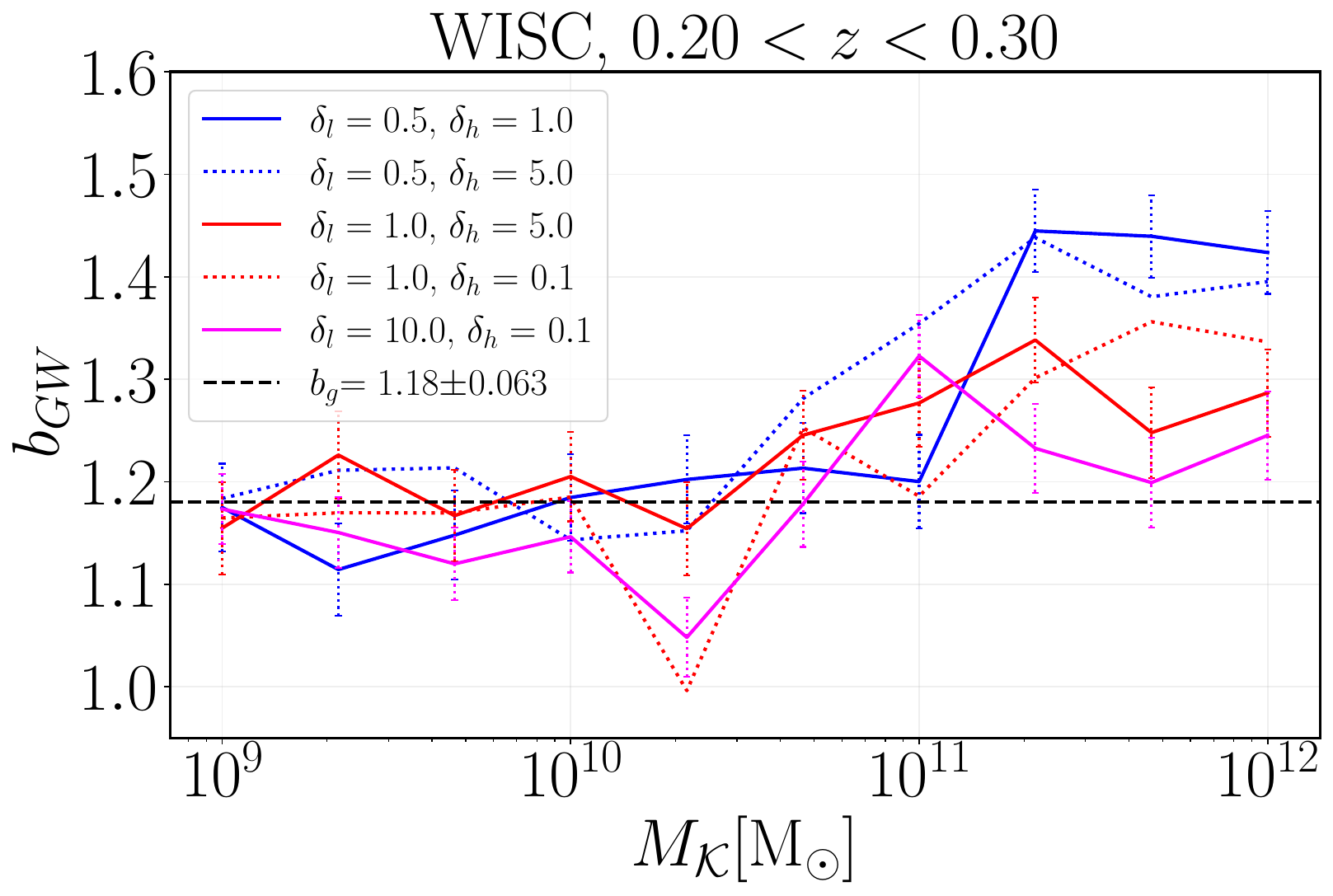}
\includegraphics[width=0.49\hsize]{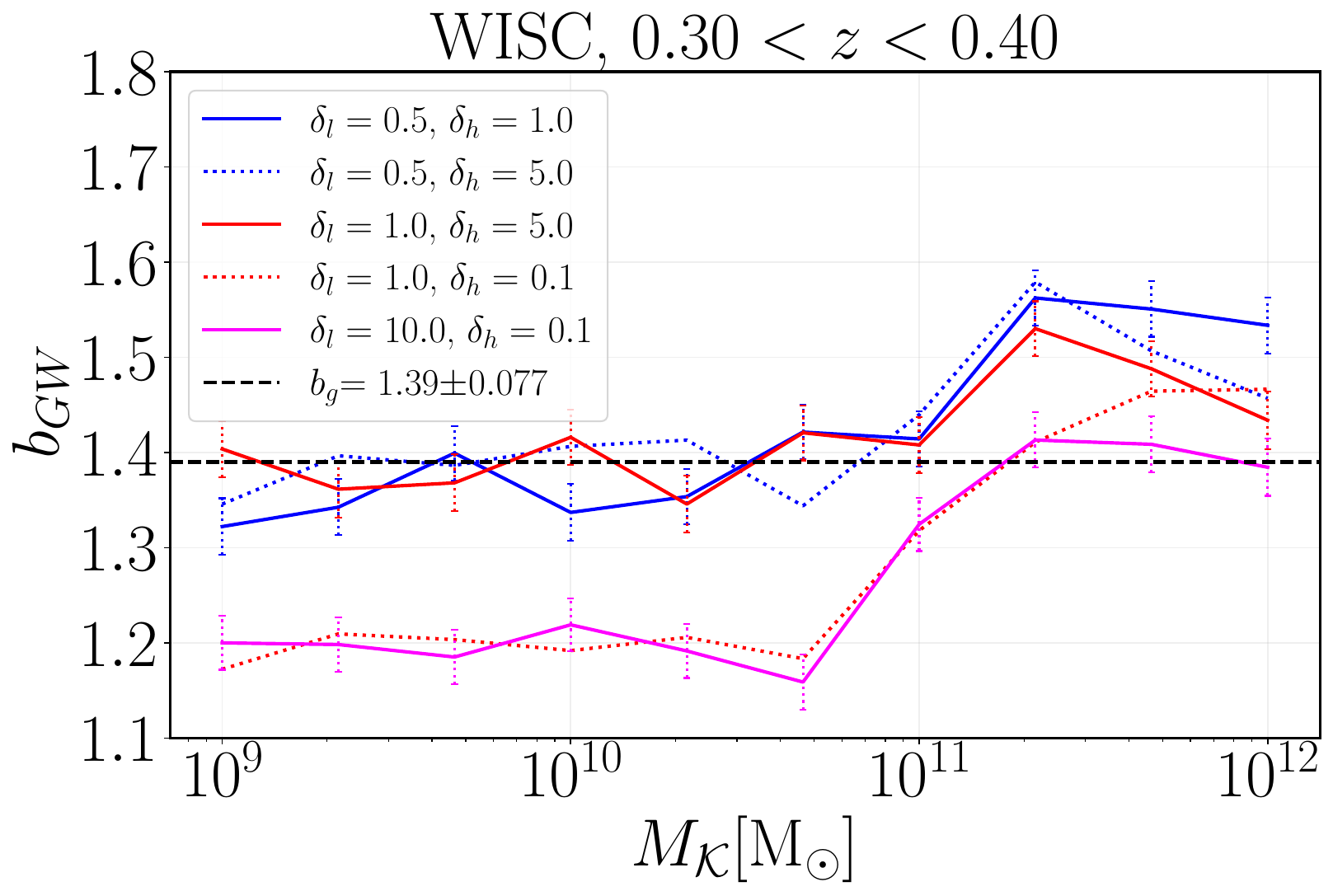}
\caption{
Comparisons of GW bias relative to $M_\mathcal{K}$ for 2MPZ and WISC under different $\delta_l$ and $\delta_h$ choices. These choices include very strong $\delta_l=0.5$ with a very clear impact at high values of $M_\mathcal{K}$ to almost flat growth $\delta_l=10$ resulting in $b_{GW}\sim b_g$ in that limit. It also includes a mild suppression $\delta_h=5$ and a strong suppression $\delta_h=0.1$. The solid red curve in the limit of $M_\mathcal{K} \to 10^{12}~M_\odot$ depicts scenarios that align with the simulations of \cite{Artale:2019tfl}, that is, a rise with $\delta_l\sim 1$ followed by minimal suppression. We also observe greater fluctuations in the first redshift bin of 2MPZ. The bars indicate $1\sigma$ error. To avoid cluttering the figure, we removed the error bars from the dotted lines. However, the errors are approximately the same as those for the corresponding solid lines.}
\label{fig:2mpz_deltas}
\end{figure*}

Therefore, we performed a more complete variation of the parameters of host-galaxy probability function, allowing all three parameters to vary simultaneously. We display examples of GW bias as a function of $M_\mathcal{K}$ with different choices of $\delta_l$ and $\delta_h$ in Figure \ref{fig:2mpz_deltas} for both 2MPZ and WISC. These examples range from a linear increasing slope $\delta_l=1$ showing a significant effect on GW bias at high $M_\mathcal{K}$ values (blue lines) to an almost flat growth phase with $\delta_l=10$ (magenta), where $b_{GW}\sim b_g$. We also considered one case where the increasing phase is steeper than linear growth ($\delta_l=0.5$) to investigate how the values of $\delta_l<1$ can enhance the GW bias. Similarly, the high stellar mass slope in these examples ranges from a moderate suppression at $\delta_h=5$ to a strong suppression at $\delta_h=0.1$. Specifically, for cases with close to linear rise in the host-galaxy probability function with stellar mass, consistent with the simulation results in Ref.~\cite{Artale:2019tfl} ($\delta_l\sim 1$), refer to the red curves in those figures, with the solid (dotted) red curves corresponding to an insignificant (significant) suppression at high mass. In all bins except for the first redshift bin of 2MPZ where the GW bias fluctuates more, as $M_\mathcal{K}\to 10^{12}~M_\odot$ and passes the Schechter upper limit $\hat{M}_{\star}$, the decreasing part of the host-galaxy probability function becomes irrelevant and the solid red line and the dotted line converge (as they should) to the same value (within the error bar). We also show
a 3D plot of the GW bias vs. $\delta_l$, $\delta_h$, and $M_\mathcal{K}$ in Appendix \ref{sec:3d} to illustrate the variation across the range of these parameters.

In summary, depending on the value of $M_\mathcal{K}$, changing the slopes of the host-galaxy probability function can impact the GW bias. In particular, for $\delta_l\sim 1$  and $\delta_l\sim 0.5$, there can be, respectively,  up to $\sim 10\%$ and $\sim 30\%$ enhancement on GW bias relative to galaxy bias at high $M_{\mathcal{K}}$ (e.g. $0.1 < z<0.2 $ for both WISC and 2MPZ in Figure \ref{fig:2mpz_deltas}. On the other hand, at low $M_{\mathcal{K}}$, the bias is suppressed by $\sim$10\% for small values of $\delta_h$, corresponding to a steeply falling slope of the host-galaxy probability function at low mass. We find that WISC at $0.2 < z < 0.3$ is an exception to this trend, driven by anomalously high bias in the galaxy catalog at $M_* < 10^{10} M_{\odot}$. This increase is unphysical and may come from contamination in the galaxy catalog or problems with the photometric redshift (both of which occur more often for faint, low-mass galaxies).

\begin{figure}
\centering
\includegraphics[width=0.49\hsize]{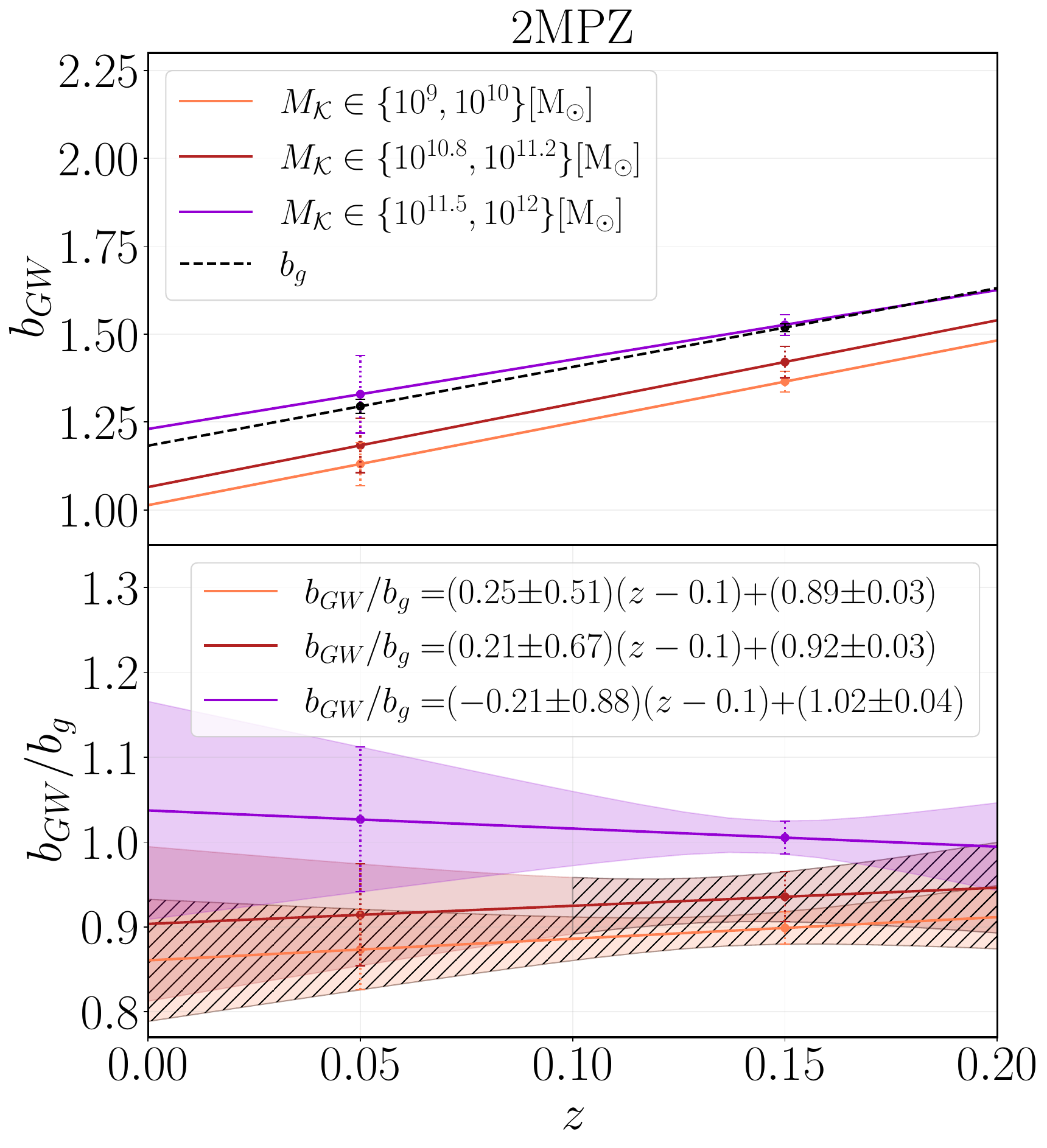}
\includegraphics[width=0.49\hsize]{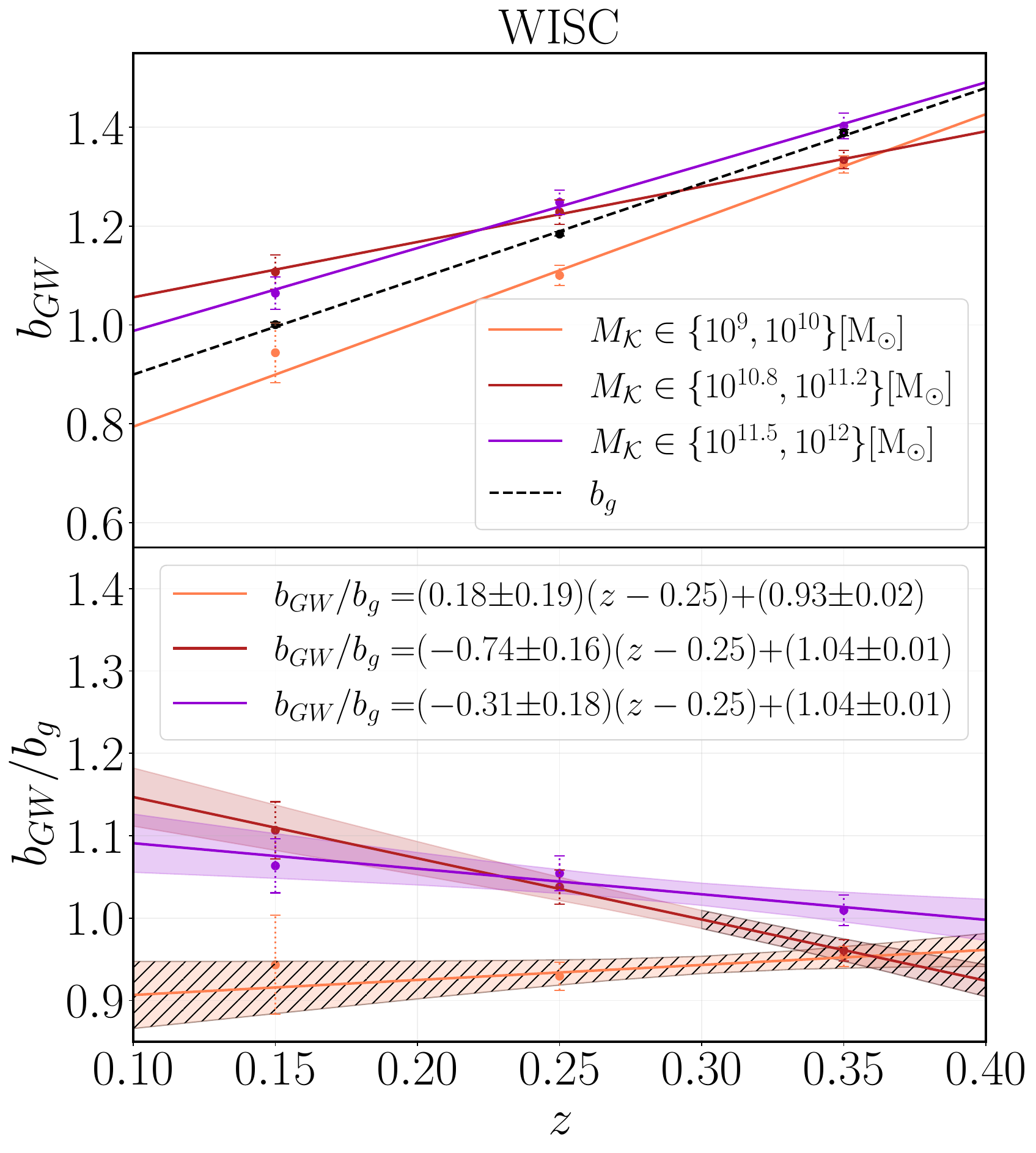}
\caption{\emph{Upper panels:} The redshift dependence of GW bias for three broad ranges in the pivot mass $M_{\mathcal{K}}$; (I) average of $8$ data points for different $M_\mathcal{K}$ values with $M_\mathcal{K}\in [10^{9},10^{10}] M_\odot$, (II) the fiducial model represented by average of $4$ data points for different $M_\mathcal{K}$ values with $M_\mathcal{K}\in [10^{10.8},10^{11.2}] M_\odot$, and (III) average of $5$ data points for different $M_\mathcal{K}$ values with $M_\mathcal{K}\in [10^{11.5},10^{12}] M_\odot$. The galaxy bias is also plotted in the black dashed line. \emph{Lower panels:} the same information plotted as the ratio of GW bias to galaxy bias. The hatched bands indicate the data points affected by the galaxy catalog's incompleteness. The bars are $1\sigma$ error.}
\label{fig:z_dep}
\end{figure}

\subsection{Variation with redshift}
\label{sec:redshift}
As is evident in Figure \ref{fig:gal_masses} and Figure \ref{fig:m_k}, the completeness is different for the different surveys and in different redshift bins. This leads to the trend of increasing $b_{GW}$ with increasing redshift in Figure \ref{fig:m_k}. This is because the galaxy catalogs become increasingly incomplete at higher redshifts, favoring higher-mass galaxies and thus more biased galaxies (greater values of $b_g$). Likewise, at $0.1 < z < 0.2$, where we can compare the results between 2MPZ and WISC, we find lower bias values for galaxies and sirens in WISC, which has a lower completeness limit ($\log_{10}{M_*}$ $\sim10$ vs.\ $\sim10.5$ for 2MPZ). As a result, the most relevant measurement within each redshift bin is not the absolute value of $b_{GW}$ but rather the comparison between $b_{GW}$ and the galaxy bias $b_{g}$. That is, we cannot reliably compare the GW clustering to matter clustering ($b = 1$) because our mock GWs occupy a biased subset of galaxies, and thus dark matter halos. Therefore, it is more meaningful to always check whether the host-galaxy probability causes the GWs to be more or less biased with respect to the underlying galaxy bias inferred from a catalog.

The redshift dependence of the GW bias for our fiducial values of $\delta_l=4$ and $\delta_h=0.5$, is plotted in Figure \ref{fig:z_dep}. We compare three broad ranges in pivot mass $M_\mathcal{K}$, aggregating the results within each range of $M_{\mathcal{K}}$ to reduce errors. We consider the average bias across (I) $8$ different values of $M_\mathcal{K}$ with $M_\mathcal{K}\in [10^{9},10^{10}] M_\odot$, (II) 4 different $M_{\mathcal{K}}$ values with $M_\mathcal{K}\in [10^{10.8},10^{11.2}] M_\odot$, and (III) $5$ different values of $M_\mathcal{K}$ with $M_\mathcal{K}\in [10^{11.5},10^{12}] M_\odot$.  The error bars are calculated by evaluating the standard deviation of the $b_{GW}$ data points within the corresponding $M_{\mathcal{K}}$ range. This allows us to study the redshift dependence of the GW bias for host-galaxy probabilities favoring low-mass or high-mass galaxies.

The redshift dependence of GW bias inherits the trend in the redshift dependence of galaxy bias, i.e.\ GW bias increases with redshift, and this fact stands for all choices of host-galaxy probabilities. As discussed above, this is mostly an observational effect, since as redshift is increased, the surveys are biased to observe massive galaxies more than low mass galaxies due to the flux limit. The two bottom panels of Figure \ref{fig:z_dep} show the ratio of GW bias over the galaxy bias ($b_{GW}/b_g$) for both 2MPZ and WISC along with 1$\sigma$ errorbar on the fit. The plot indicates that the variation in the ratio $b_{GW}/b_g$ with redshift is less than $15\%$ up to $z=0.4$ considered in this analysis for different choices of $M_\mathcal{K}$. The ratio shows a value less than one primarily for those scenarios that are impacted by the galaxy catalog incompleteness (lower $M_\mathcal{K}$ values), shown by striped lines in Figure \ref{fig:z_dep}. This mild redshift evolution for different scenarios of GW source and host galaxy connection is an interesting signature that can be explored in the future with more observations.

\subsection{Variation in $\ell_{\rm max}$} 
\label{sec:bgw_ellbounds}
\begin{figure}
\centering
\includegraphics[width=0.49\hsize]{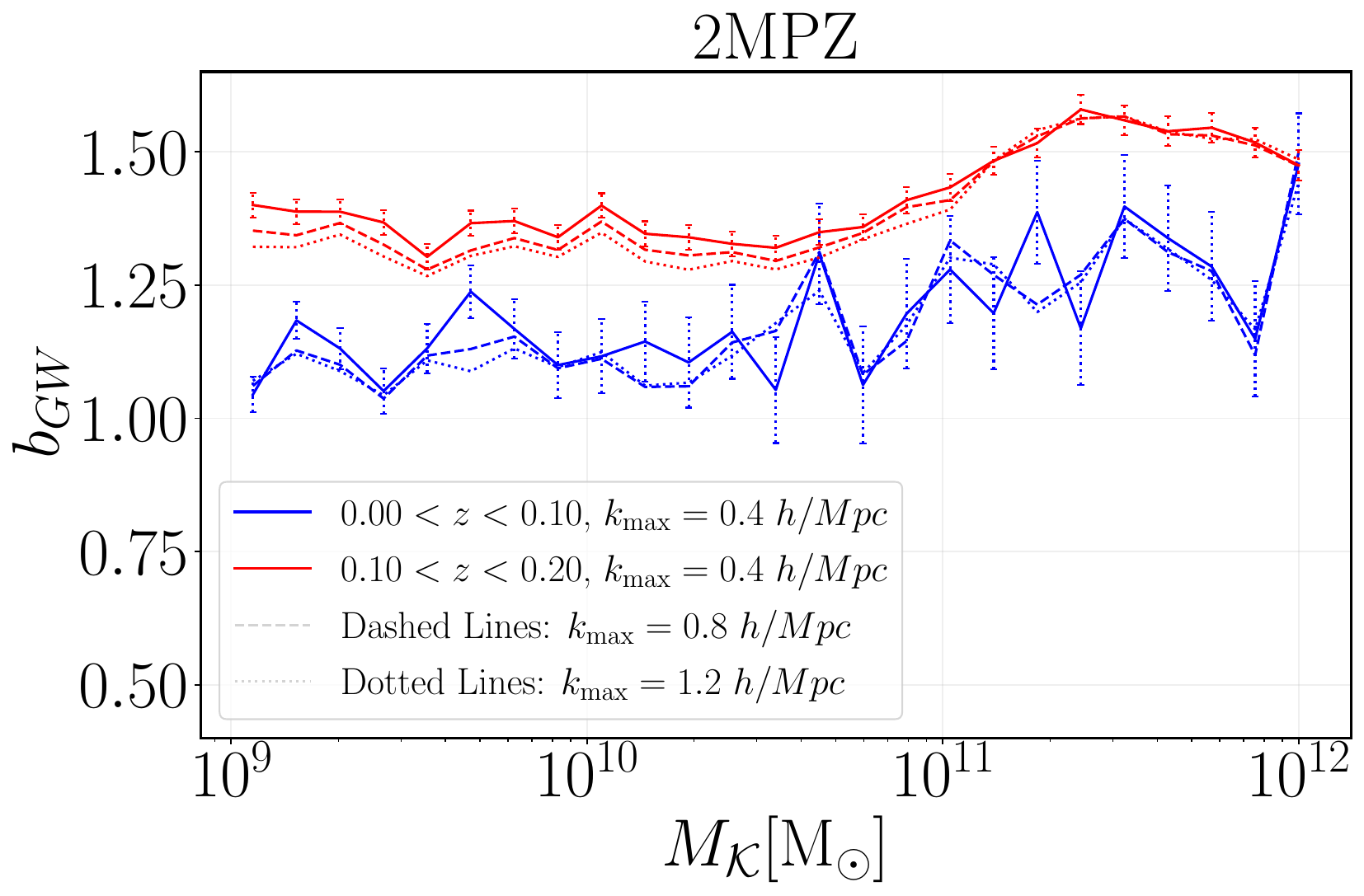}
\includegraphics[width=0.49\hsize]{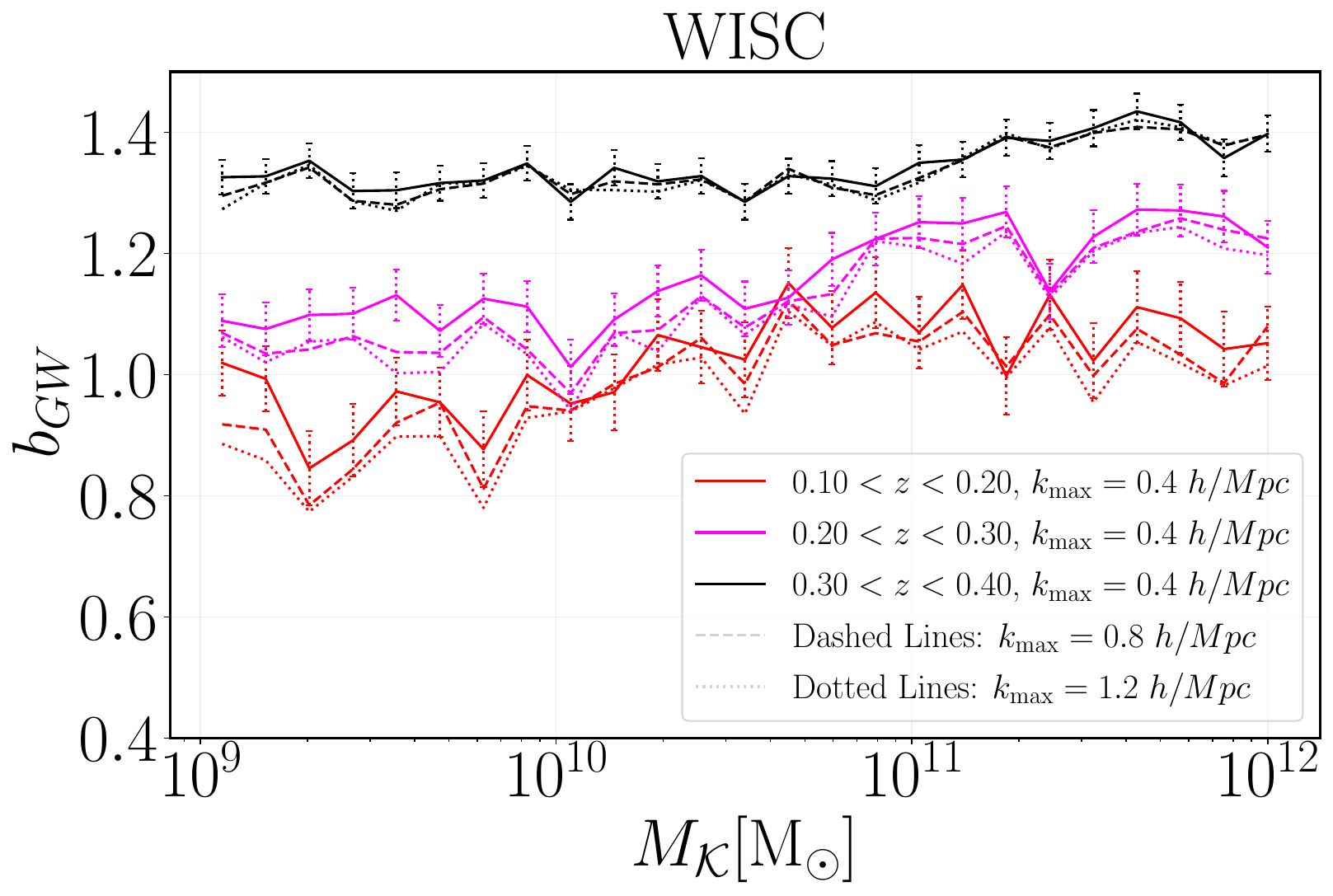}
\caption{Same as Fig.~\ref{fig:m_k} but for different values of $k_{\mathrm{max}}$ (different line styles), showing that our results are robust to the scales used. In this case $\delta_l=4$, and $\delta_h=0.5$. The bars are $1\sigma$ error.}
\label{fig:kmax}
\end{figure}

We investigated extending our calculations to smaller scales (that is, larger $k_{\textrm{max}}$ and $\ell_{\textrm{max}}$) for both catalogs. In Figure~\ref{fig:kmax}, we show the GW bias versus $M_\mathcal{K}$ for different choices of $k_\mathrm{max}$. This shows that our results are robust to changing $k_\mathrm{max}$: using smaller scales does not noticeably change the trends in bias vs.\ $M_\mathcal{K}$, and Figure~\ref{fig:kmax} also shows that the linear bias provides a good fit even with higher $k_{\textrm{max}}$. We also explored the possibility of $b_{GW}(\ell)$ with a linear scale dependence. A linear function in multipoles ($\ell$) did not fit much better than the constant bias. We noticed a slight preference for a decreasing bias at high $\ell$; however, it was not statistically significant. We conclude that our results are robust to the choice of the minimum scale, $k_{\textrm{max}}$.

\subsection{Comparison with previous studies}
Our results can be compared to other results in the literature studying the GW bias in simulations \cite{Scelfo20,Libanore21,Libanore22,2024MNRAS.530.1129P}. These works have found a strong evolution of $b_{GW}$ with redshift but have not considered how $b_{GW}$ depends on the GW host-galaxy probability and the physics relating BBHs to their host galaxies. We reiterate that our study is most effective in examining the sensitivity of the ratio between $b_{GW}$ and $b_g$ to the parameters of the host-galaxy probability function, given the incompleteness of the galaxy catalogs. We find $b_{GW}$ is within $15\%$ of $b_g$ for our fiducial model at $0 < z < 0.4$ (see red data points in lower panel of Figure \ref{fig:z_dep}), in good agreement with \cite{2024MNRAS.530.1129P} who find $b_{GW}$ = 1 at low redshift.
In contrast, \cite{Libanore21} finds a higher value of $b_{GW} \sim 2$ at low redshift using a galaxy catalog from the EAGLE simulation and using a HOD-based approach. They also find a higher galaxy bias of $b_{g} \sim 1.5$, implying $b_{GW}$ is $\sim 30\% $ higher than $b_g$. In scenarios with longer delay times and a moderately rising host-galaxy probability phase ($\delta_l\sim 1$), we find higher GW bias values. However, in our host-galaxy probability models, achieving a $30\% $ enhancement of GW bias would require a rapid growth rate, around $1/\delta_l \gtrsim 2$ (refer to the blue lines in Figure \ref{fig:2mpz_deltas}. The comparison of our results and those from simulations shows how this ``phenomenological'' approach can complement the ``simulation-based'' approach of directly populating GWs into cosmological simulations. However, completeness of galaxy catalogs and photometric redshift uncertainties are potential hurdles in the ``phenomenological'' approach.

\section{Conclusion} 
\label{sec:conclusion}
In this work, we have studied the dependence of the GW bias parameter for BBHs on the GW host-galaxy probability, using angular power spectrum measurements of mock sirens populating galaxy catalogs.
We generated mock catalogs of BBHs using physically motivated relations between GWs and the properties of the host galaxies, populating the 2MPZ and WISE all-sky galaxy catalogs. In particular, we examined an astrophysically inspired phenomenological model for the GW host-galaxy probability, characterized by a broken power law in stellar mass that rises at low mass and falls at high mass. The presence of a turnover point in the host-galaxy probability function $M_{\mathcal{K}}$ is motivated by the expected transition scale between star-forming and quiescent galaxies, which is predicted to be around $M_{\star} \sim 10^{11}$ $M_\odot$. Essentially, our reasoning is that the BBH merger rate is expected to follow the SFR with a certain non-zero delay time. If the delay time is short compared to the quenching timescale of the galaxies, high-mass galaxies whose star formation ceased in the past should have a smaller probability of hosting a GW merger, leading to a drop in the host-galaxy probability at high mass. On the other hand, if the delay time is long, then we would expect the host-galaxy probability to trace the stellar mass and increase at high mass. We explored these physical differences by varying the pivot mass and slopes in our model.

We observed that increasing $M_{\mathcal{K}}$ and $1/\delta_l$ increases the GW bias. For the fiducial values of $\delta_l=4$ (mild increase rate) and $\delta_h=0.5$ (strong suppression rate), we observed that increasing $M_{\mathcal{K}}$ increases the bias, but only modestly. The GW bias at $M_{\mathcal{K}} = 10^{12}$ $M_\odot$ exceeds the galaxy bias by at most $\sim 5\%$ and exceeds the galaxy bias at $M_{\mathcal{K}} = 10^{11}$ $M_\odot$ by $\sim 10\%$ (see Figure \ref{fig:m_k}). Assuming that the host-galaxy probability increases at a higher rate, as predicted by certain population synthesis models ($\delta_l \sim 1$) \cite{Artale:2019tfl}, our findings indicate that the GW bias can surpass the galaxy bias by approximately 10\% when $M_{\mathcal{K}} = 10^{12}$ $M_\odot$, as illustrated in Figures \ref{fig:2mpz_deltas}. It is important to note that the model we have described for the GW host-galaxy probability is quite simple, largely due to limitations imposed by the photometric galaxy catalogs, which lack direct measurements of the star formation rate and galaxy metallicity. We also explore more intricate host-galaxy probabilities that take these factors into account in our companion paper \cite{paper-II}.

We also note that since the hosts of the BBHs were taken as a sub-sample of the 2MPZ and WISC galaxy catalogs, the incompleteness of the galaxy catalogs themselves is also reflected in the mock BBH catalog. Hence, our calculation of the GW bias parameter has some systemic error depending on the incompleteness level of the input galaxy catalog. We address this issue by mainly comparing the GW bias to the galaxy bias of the primary galaxy catalog (which is similarly affected by incompleteness). However, for the host-galaxy probabilities with $M_{\mathcal{K}}$ greater than the catalog completeness limit,  the bias values are less influenced by incompleteness so we believe the result is more robust in this regime. Our ``phenomenological'' approach, populating real galaxy catalogs, is complementary to ``simulation-based'' approaches, allowing more flexibility for the host-galaxy probability; probing a larger volume than hydrodynamical simulations; and avoiding dependence on uncertain galaxy formation physics. However, the phenomenological approach generally suffers from the incompleteness of the galaxy catalogs, observational errors, and less depth in redshift space, out to $z \sim 0.3$ compared to simulations spanning $0 < z < 6$.

To summarize, this method is the first of its kind, using a phenomenological approach with a physics driven host-galaxy probability to connect observed low-redshift galaxy catalogs to GW sources in terms of the GW bias parameter. This provides a framework that can be easily tested with future observations. The inference of the GW bias parameter provides a complementary probe to explore the astrophysical population of BBHs. 
 We acknowledge that it is likely that with a single large-scale clustering measurement, uncertainties and parameter degeneracies would make it difficult to infer more than one or two effective parameters describing the GW-host galaxy probability (i.e.\ break mass and slope).
However, this would still provide valuable insight into the GW-galaxy connection, and this information could be enhanced by smaller-scale clustering measurements or higher-point measurements of the small-scale GW-galaxy cross-correlation. We leave these avenues to future work.

Insight into the GW bias parameter would be valuable for inference of cosmological parameters, such as the Hubble constant and dark energy equation of state using BBH-galaxy cross-correlation \cite{Mukherjee:2020hyn, Diaz:2021pem}. In these studies, it is crucial to include the variation of GW bias and marginalize over its uncertainties; hence a better understanding of the GW bias parameter could strengthen the analyses. Similarly to the GW bias parameter for BBHs, it is possible to extend this technique for calculating the GW bias parameter for other GW sources such as binary neutron stars and neutron star black hole systems. The future holds even greater potential for this approach to modeling the GW bias, as galaxy catalogs from DESI \cite{dey2019overview}, Euclid \cite{collaboration2022euclid}, Vera Rubin Observatory \cite{ivezic2019lsst}, and Roman Space Telescope \cite{Dore:2019pld} will illuminate the GW-galaxy connection up to high redshifts.

\section*{Acknowledgements}
The authors thank the Wide Field Astronomy Unit (WFAU) for publicly providing the 2MPZ and WISC data. This work used Astropy:\footnote{\url{http://www.astropy.org}} a community-developed core Python package and an ecosystem of tools and resources for astronomy \cite{astropy:2013, astropy:2018, astropy:2022}.  The computational aspects of this work were performed at the computing clusters at the Perimeter Institute (PI) for Theoretical Physics and the Center for Advanced Computing (CAC) at Queen's University, supported in part by the Canada Foundation for Innovation. AD, GG and DSH research is supported by Discovery Grant from the Natural Science and Engineering Research Council of Canada (NSERC). AD, GG, DSH, and AK are also supported in part by the Perimeter Institute for Theoretical Physics. Research at Perimeter Institute is supported by the Government of Canada through the Department of Innovation, Science and Economic Development Canada and by the Province of Ontario through the Ministry of Colleges and Universities. JLK is supported by the NSERC Canada Graduate Scholarship - Doctoral (CGS-D) and the Arthur B. McDonald Canadian Astroparticle Physics Research Institute.  AK was supported as a CITA National Fellow by the Natural Sciences and Engineering Research Council of Canada (NSERC), funding reference \#DIS-2022-568580.
The work of SM is a part of the $\langle \texttt{data|theory}\rangle$ \texttt{Universe-Lab} which is supported by the TIFR and the Department of Atomic Energy, Government of India.

\appendix

\section{Redshift conversion}
\label{sec:z_conversion}

We make use of the following formula to convert from the object's redshift to its Hubble velocity
\begin{align}
    v(z) = c \int_0^z \frac{dz'}{E(z')},
\end{align}
where $E(z)$ is defined as
\begin{align}
    E(z) \equiv \sqrt{ \Omega_{m0}(1+z)^3 + \Omega_{r0}(1+z)^4 + \Omega_\Lambda}. \label{eq:E_z}
\end{align}
The quantities $\Omega_{x0}$ represent the standard density parameters for species $x$ today, which are determined by our choice of cosmology. 

To convert velocities between frames (i.e. moving from $V$ in the heliocentric frame to $V_\mathrm{con}$ in the new (converted) frame), we use 
\begin{align}
    V_\mathrm{con} &= V + V_\mathrm{apex} [ \sin(b) \sin(b_\mathrm{apex})+ \cos(b) \cos(b_\mathrm{apex})\cos(l - l_\mathrm{apex})  ],
\end{align}
where $b$ and $l$ are the latitude and longitude coordinates of the source. In this expression, to convert the velocities to their values in the CMB frame we take values of $V_\mathrm{apex} = 371.0$ km/s, $l_\mathrm{apex} = 264.14$ deg, and $b_\mathrm{apex} = +48.26$ deg\cite{Fixsen:1996nj}.

\section{Testing uniform host-galaxy probability}
\label{sec:random}
\begin{figure}
\centering
\includegraphics[width=0.6\hsize]{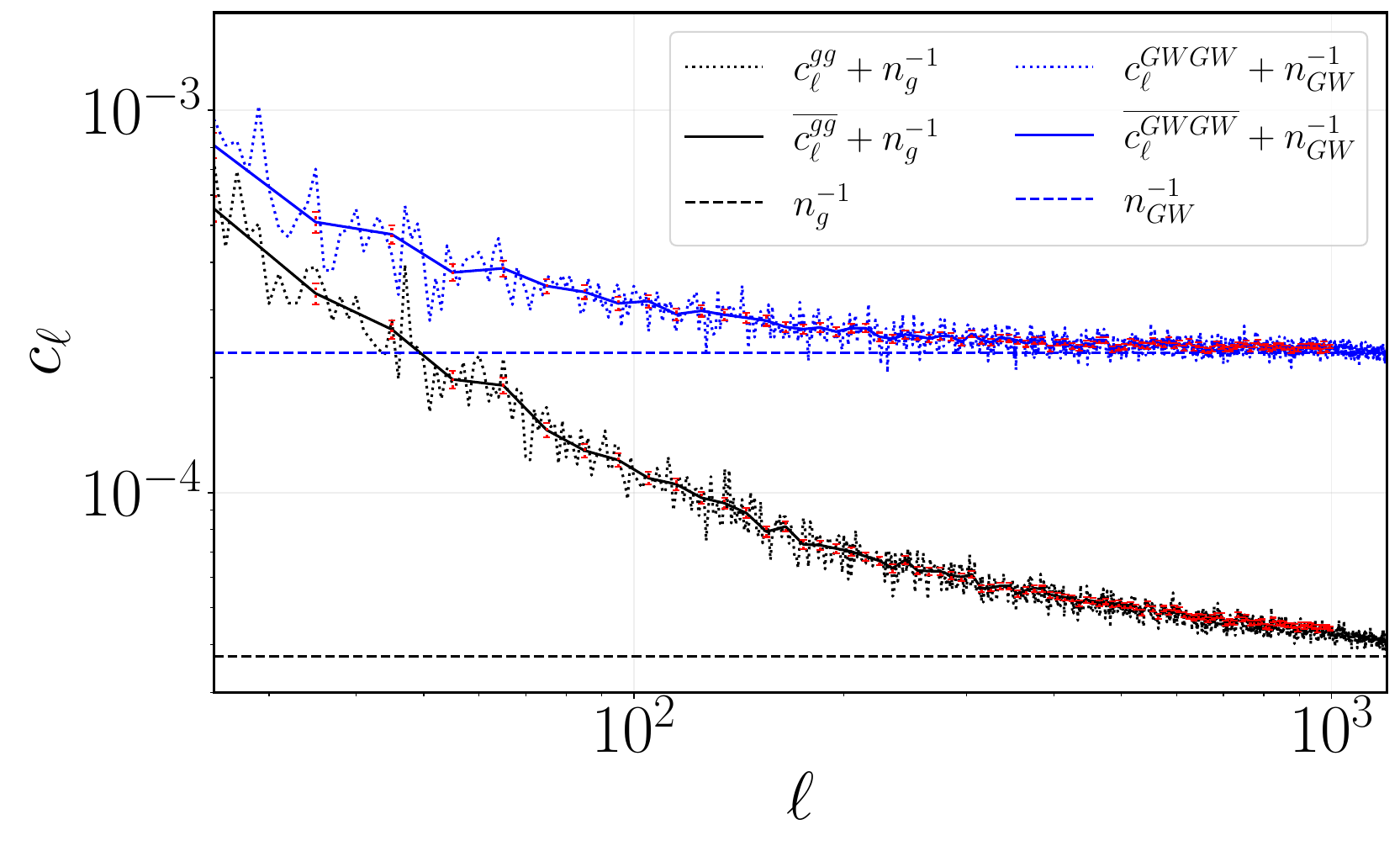}
\caption{Comparison of galaxy power spectrum (in black) and the mock siren catalog (in blue) based on uniform host-galaxy probability for 2MPZ within $0.1 < z < 0.2$. The dashed horizontal lines are the shot noise associated with each power spectrum. The red error bars are $1\sigma$ error for the power spectrum binned by $\Delta \ell = 10$.}
\label{fig:random_sel_pow}
\end{figure}

\begin{figure*}
\centering
\includegraphics[width=0.48\hsize]{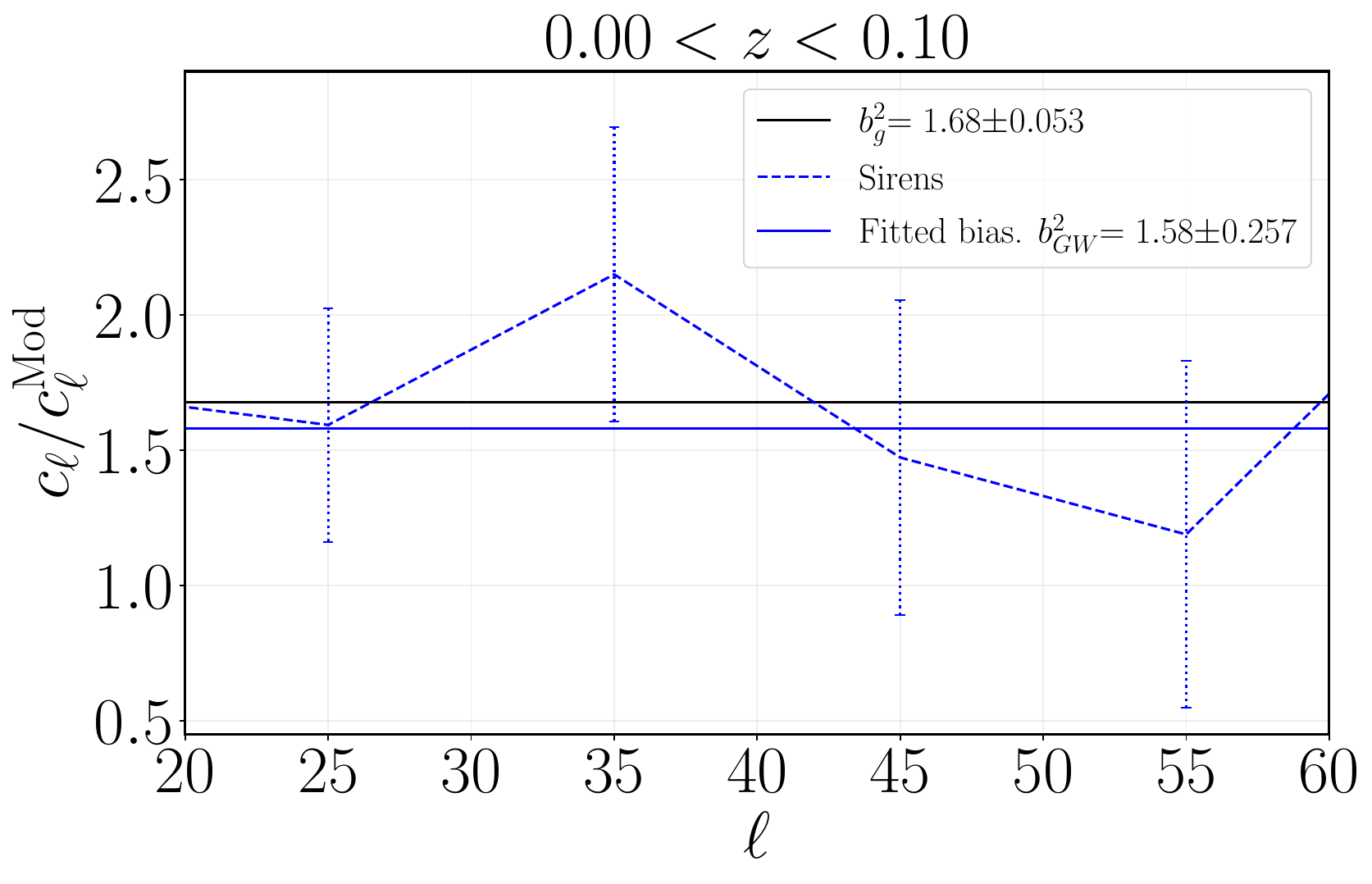}
\hspace{8pt}
\includegraphics[width=0.48\hsize]{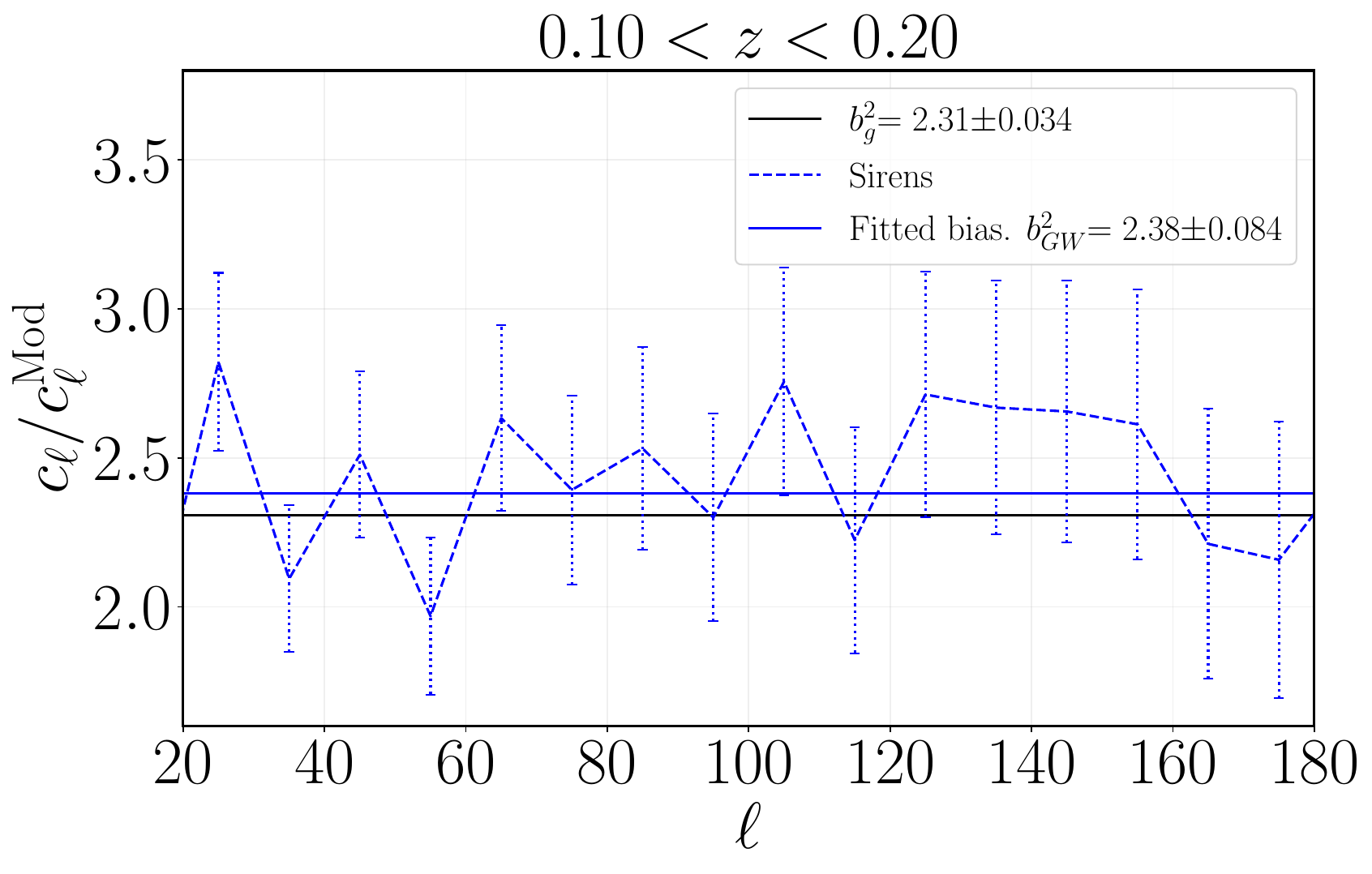}
\caption{Bias comparison for galaxy catalog and siren catalog based on uniform host-galaxy probability. The bars are $1\sigma$ error.}
\label{fig:random_sel}
\end{figure*}

In order to test our pipeline, we first carried out our GW bias calculations for a uniform host-galaxy probability (implemented by sending $\delta_l$, and $\delta_h$ in Eq.~\eqref{eq:selection} to infinity), so that the BBH hosts are chosen with equal probability irrespective of galaxy properties.  The uniform selection sampling was only implemented for the 2MPZ catalog. Figure \ref{fig:random_sel_pow} displays the calculated power spectra along with their shot noise for both the siren sample and galaxies within the second redshift bin. Note that if we subtract the shot-noise, the two power spectra statistically agree with each other, which is expected since the sirens catalog in this case is a uniform selection of the galaxy catalog and shares the same angular correlation. The resulting GW bias based on the uniform host-galaxy probability and the galaxy bias is shown in Figure \ref{fig:random_sel}. These plots show that the galaxy bias and the GW bias are statistically the same, which is expected for the uniform host-galaxy probability.

\section{Determining the best choice for $\ell_\mathrm{min}$}
\label{sec:l_min}
\begin{figure}
\centering
\includegraphics[width=0.55\hsize]{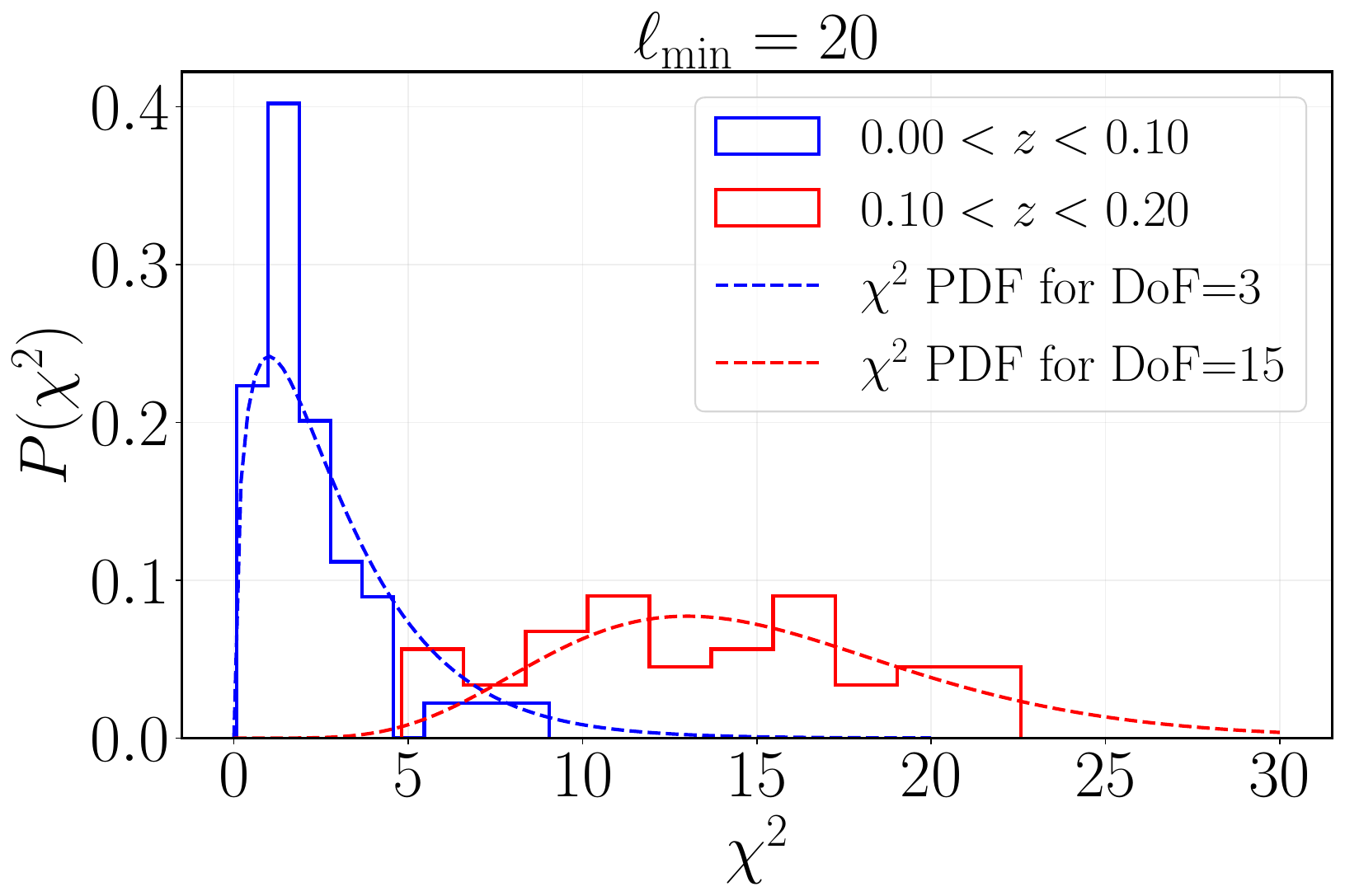}
\caption{The distribution of $\chi^2$ for $50$ host-galaxy probabilities with different values of $M_\mathcal{K}$ for 2MPZ (solid lines) vs. the theoretical distribution of $\chi^2$ with the given number of degrees of freedom (dashed lines), using $\ell_{\textrm{min}} = 20$ as indicated in the title.}
\label{fig:ell_2mpz}
\end{figure}

\begin{figure}
\centering
\includegraphics[width=0.49\hsize]{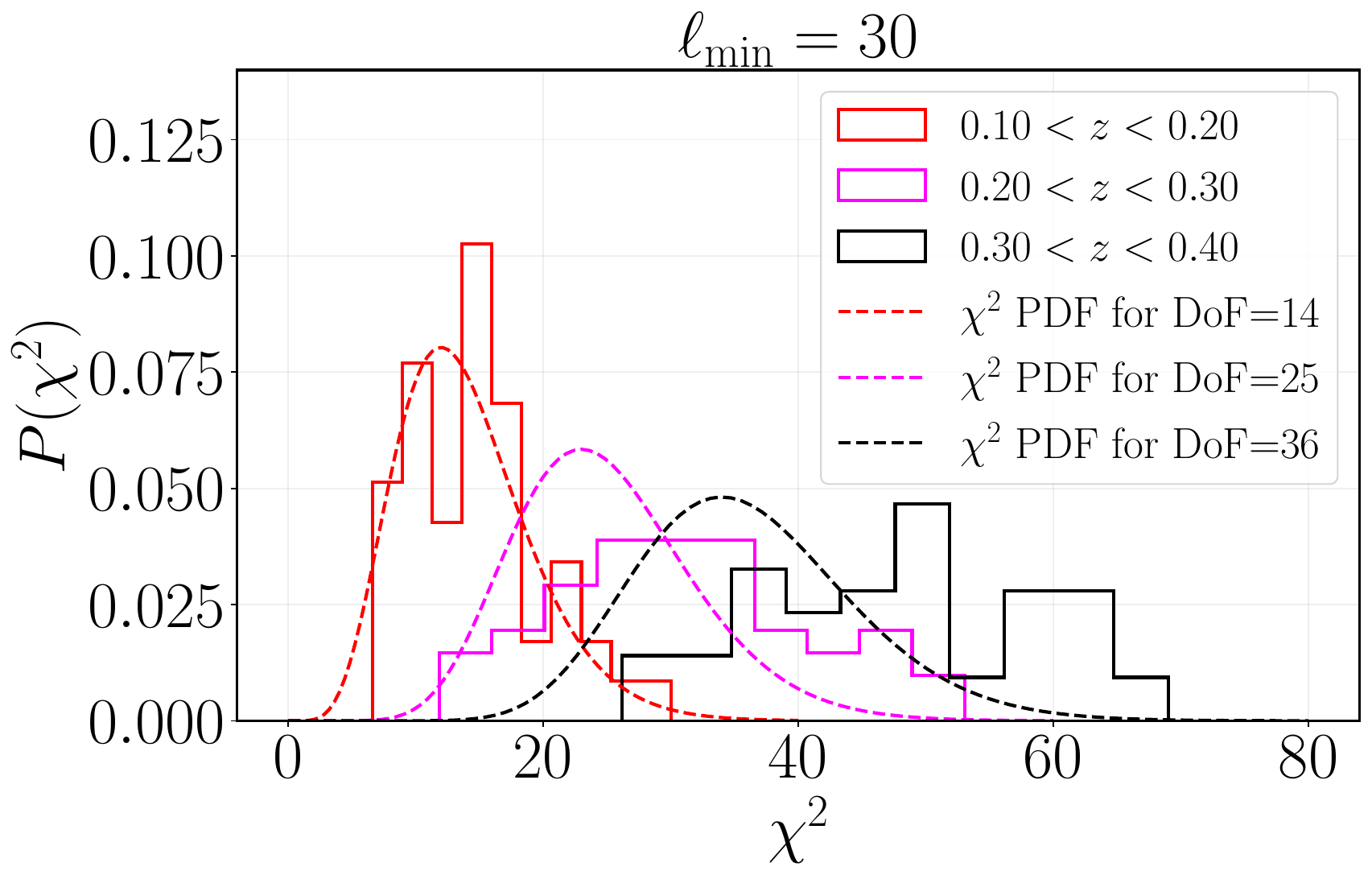}
\includegraphics[width=0.49\hsize]{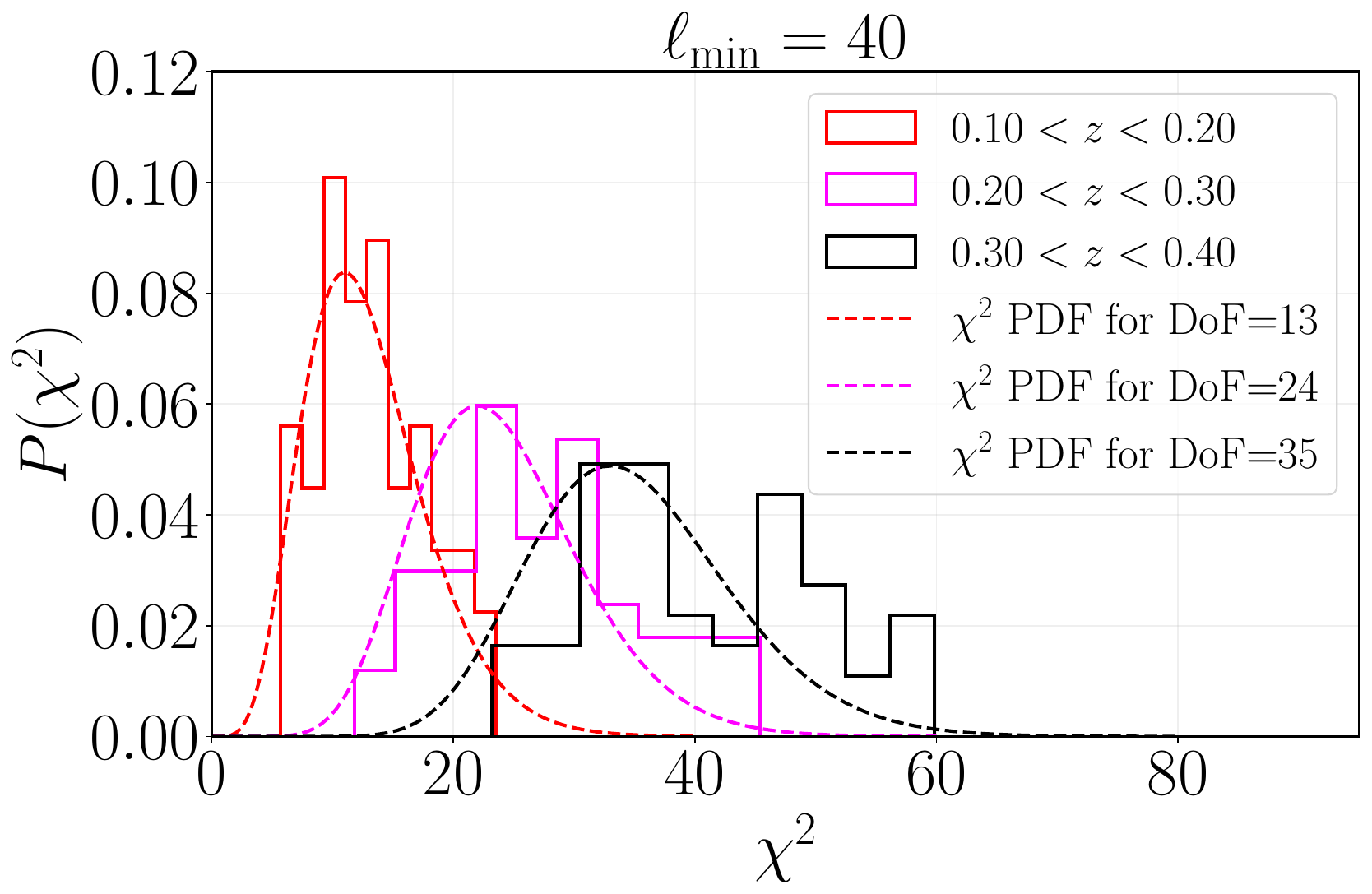}
\includegraphics[width=0.49\hsize]{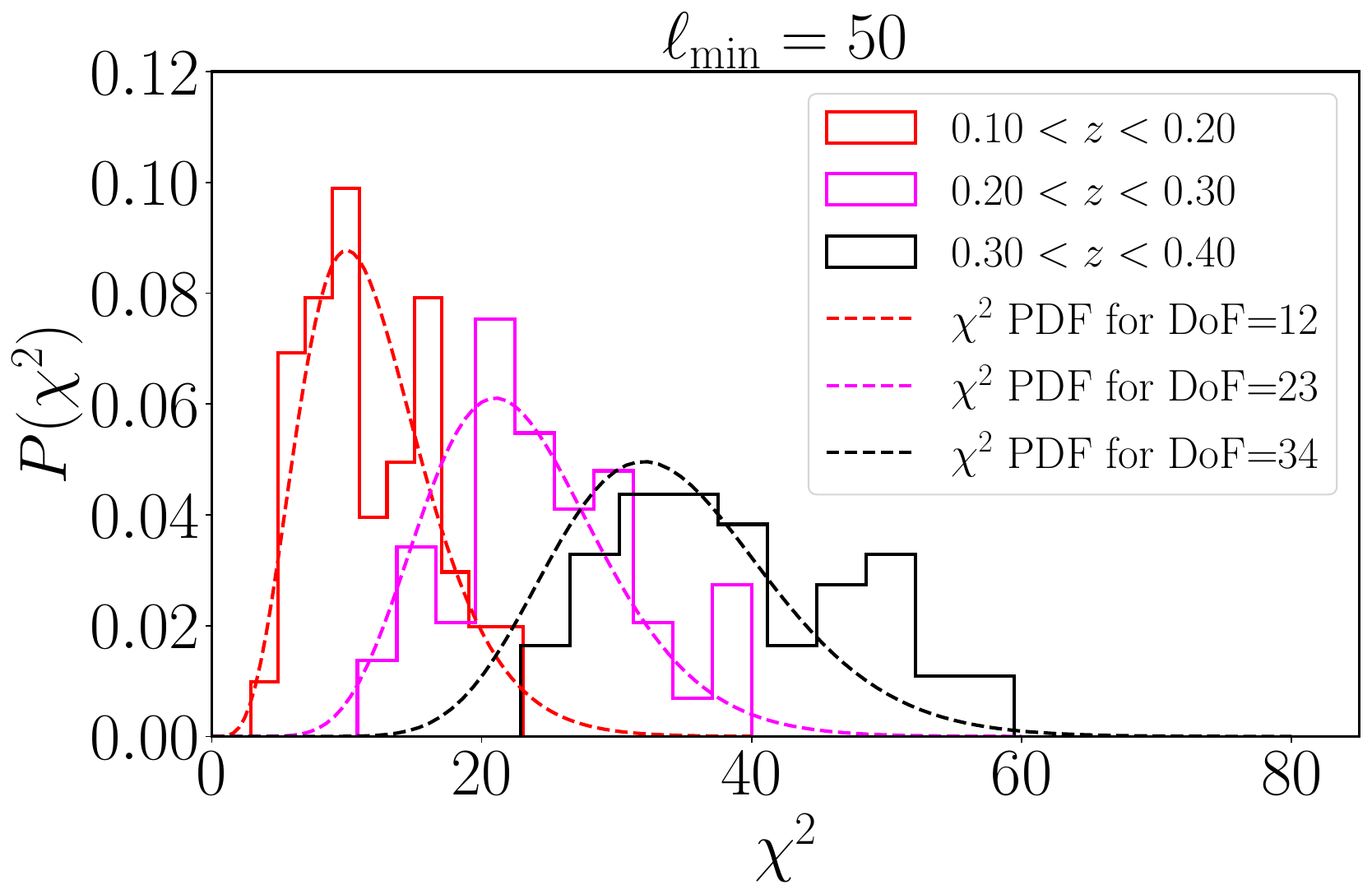}
\includegraphics[width=0.49\hsize]{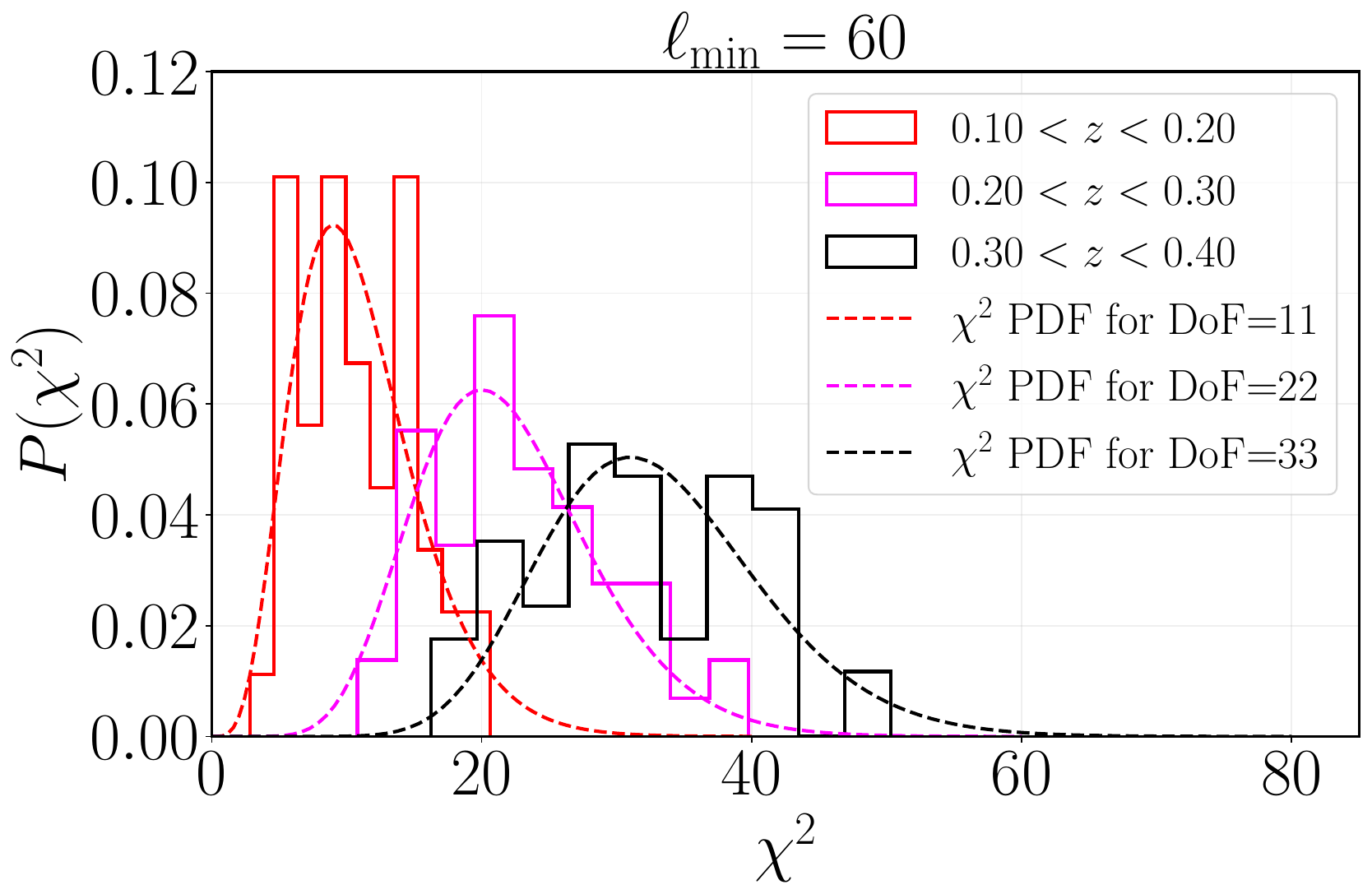}
\caption{Like Fig.~\ref{fig:ell_2mpz}, but for WISC.}
\label{fig:ell_wisc}
\end{figure}

As pointed out in Section \ref{sec:cl}, we find that the constant bias model is a valid approximation in some range for $\ell \in [\ell_{\mathrm{min}},\ell_{\mathrm{max}}]$. To determine $\ell_\mathrm{min}$, we evaluate the bias for $50$ different host-galaxy probabilities with varying $M_\mathcal{K}$ by fitting a constant bias. We repeat this process by changing $\ell_\mathrm{min}$ in $\Delta \ell = 10$ steps, as we had earlier computed the mean value of the power spectrum within the same $\Delta \ell = 10$ bins. We start with $\ell_\mathrm{min}=20$ because it simplifies our treatment of mask convolution (simply dividing by $f_{\textrm{sky}}$ rather than deconvolving the mask). To assess the goodness of fit, we compare the distribution of $\chi^2$ for the 50 host-galaxy probabilities with different $M_\mathcal{K}$ to the expected distribution of $\chi^2$ where the degree of freedom comes from the number of points between $\ell_\mathrm{min}$ and $\ell_\mathrm{max}$, minus one for the fitted bias parameter. The results are shown in Figure \ref{fig:ell_2mpz} and Figure \ref{fig:ell_wisc} for 2MPZ and WISC, respectively. In addition, we performed a Kolmogorov–Smirnov test between the two distributions, to verify that the measured $\chi^2$ followed the correct distribution. We required that the $p$ value $>0.05$ (i.e., that the measured $\chi^2$ distribution for our fits and theoretical values match), and found $\ell_\mathrm{min}=20$ to be a good choice for both redshift bins of 2MPZ, and $\ell_\mathrm{min}=[30,50,60]$ for the three redshift bins in WISC respectively in ascending order.  In the case of WISC we found that $\ell_{\mathrm{min}} = 20$ resulted in a poor fit for all bins, so we raised $\ell_{\mathrm{min}}$ until the $p > 0.05$ criterion was satisfied. Although we did a careful examination of scale cuts here, the results for bias do not change significantly with an alternative choice.
 
\section{Bias with a different photometric redshift convolution}
\label{sec:z-conv}
\begin{figure}
\centering
\includegraphics[width=0.48\hsize]{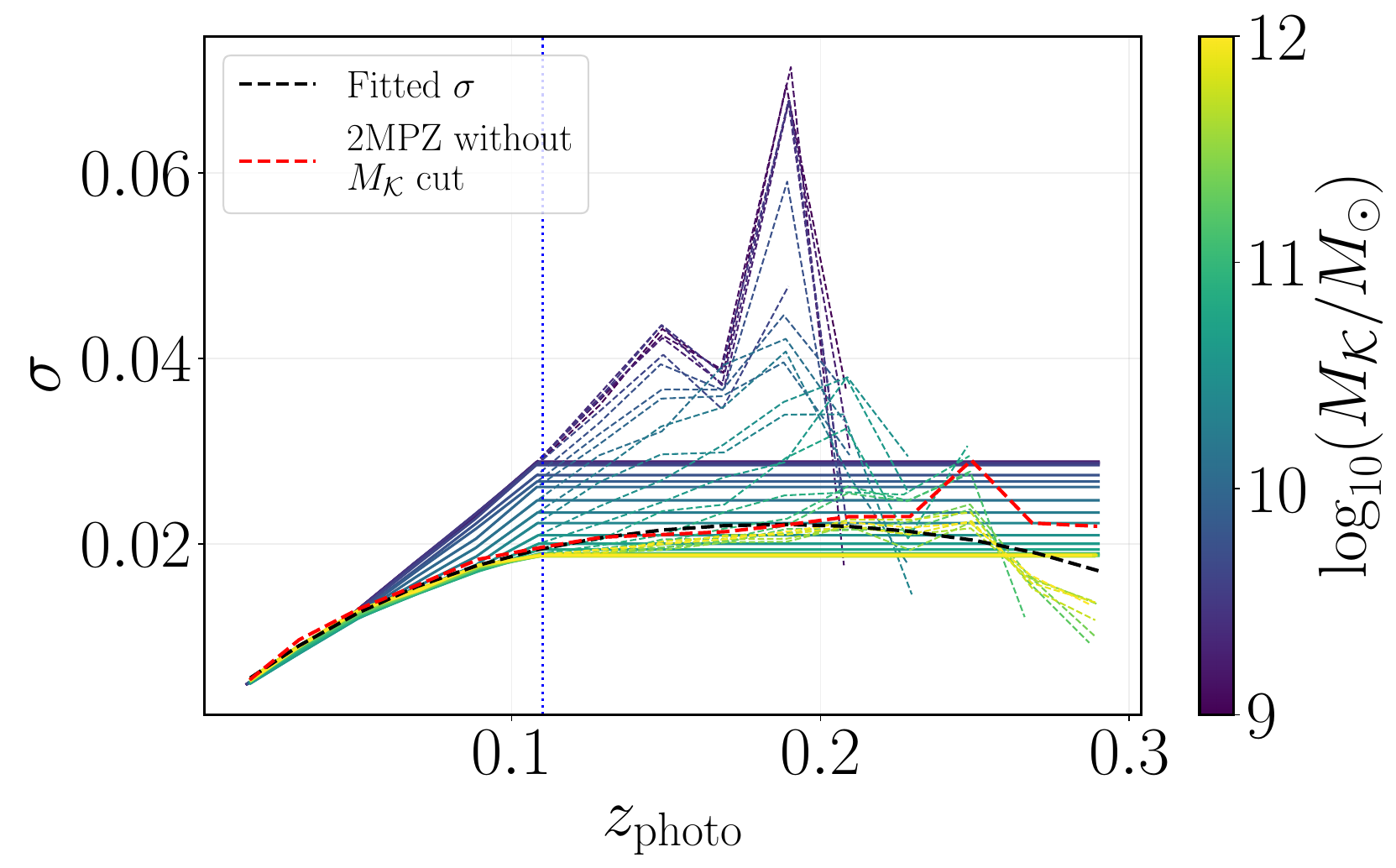}
\includegraphics[width=0.48\hsize]{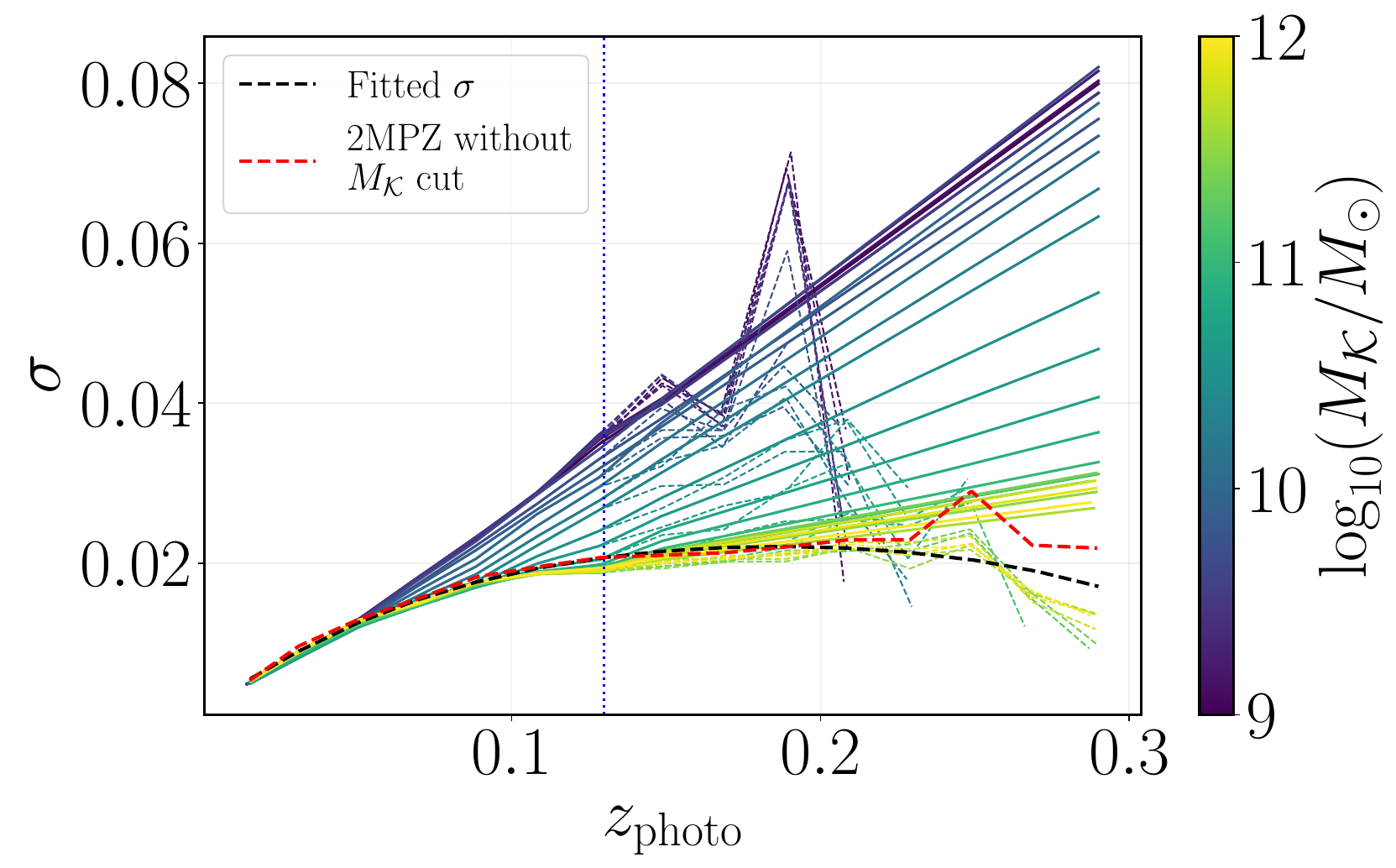}
\caption{The spread of spectroscopic redshift distribution around a particular photometric redshift, i.e. $\sigma$. This quantity enters Eq.~\eqref{eq:zconv}. The dashed lines represent the actual data. Since after $z \simeq 0.11$, there are not enough hosts (in the siren catalog), the data fluctuates, and we approximate them using (I) either a constant (left panel), (II) or a linear extrapolation (right panel). The red dashed line corresponds to the whole 2MPZ catalog. The black dashed line represents the fitted formula given in Eq.~\eqref{eqn:sigmaz}.}
\label{fig:2mpz_sigma}
\end{figure}

In Section \ref{sec:bias_measure}, we describe the convolution of the photometric redshift distribution with the redshift error. Here, we provide more details on the sensitivity of our results on the convolution scheme. We performed this analysis only for 2MPZ, which should also be more sensitive to changes in the redshift distribution, since the mean redshift is lower. First, as we discussed in Section \ref{sec:bias_measure}, we allow the mean of the Gaussian to deviate from the mean of the photometric distribution within the bin through a free parameter, $z_\mathrm{shift}$, which also depends on $M_\mathcal{K}$ , $\delta_l$ and $\delta_h$. We further check whether the variance of the scatter itself, that is, $\sigma$, also varies with $M_\mathcal{K}$ (setting $\delta_l=4$ and $\delta_h=0.5$) in addition to the dependence on the redshift. The results of our investigation are shown in Figure \ref{fig:2mpz_sigma}. The red dashed line corresponds to the whole 2MPZ catalog. The black dashed line represents the fitting formula given in Eq.~\eqref{eqn:sigmaz}.  Similarly to the analysis performed for $z_\mathrm{shift}$, we find the best fit to $\sigma$ from $z\simeq 0.11$ onward using both a constant and a linear extrapolation, since the actual data (dashed lines) fluctuate. Although we used two different fitting schemes here, the difference in their impact in the final bias calculation turned out to be insignificant. So in what follows, we discuss only the constant extrapolation version in comparison to the other models for $\sigma$.

\begin{figure}
\centering
\includegraphics[width=0.48\hsize]{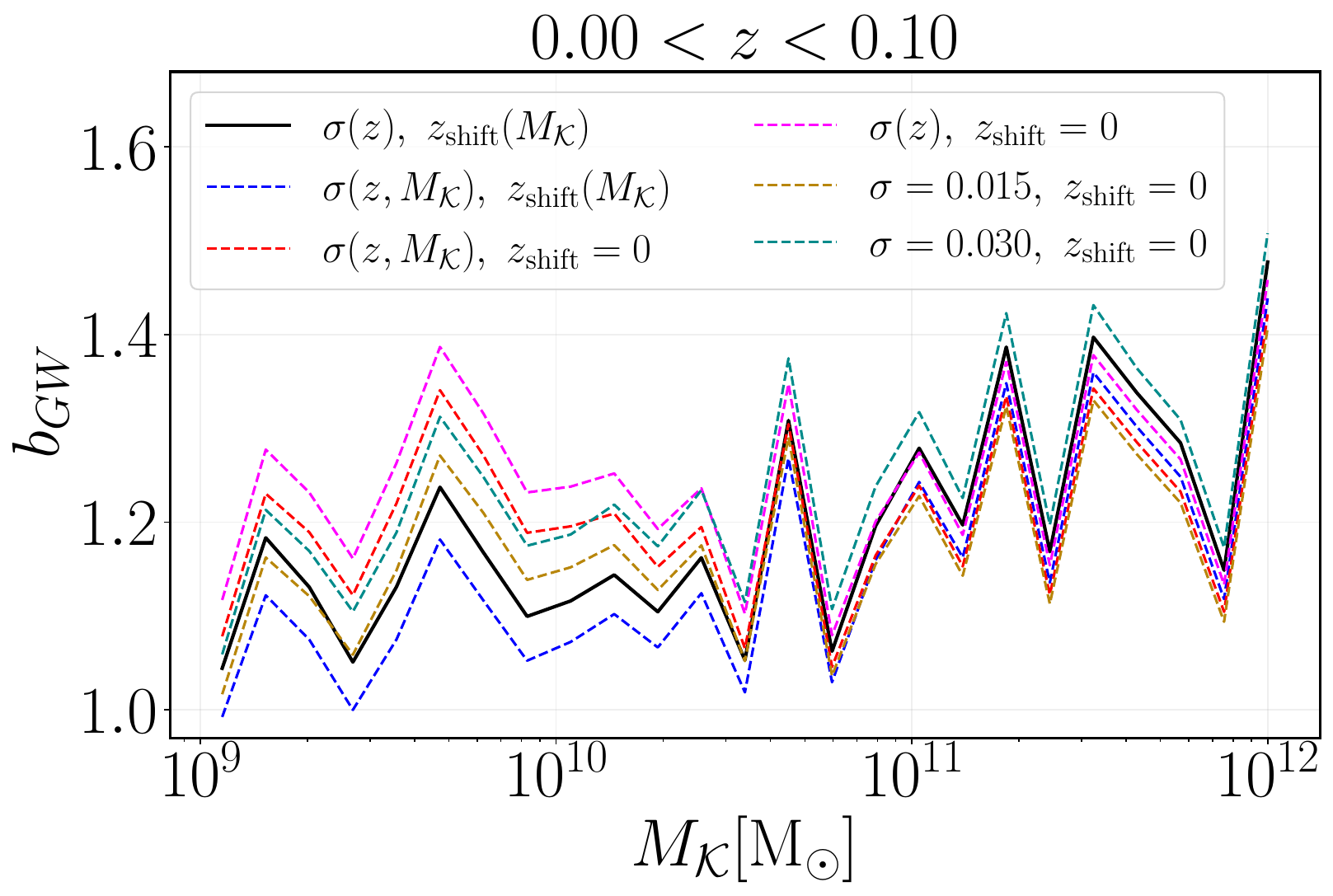}
\includegraphics[width=0.48\hsize]{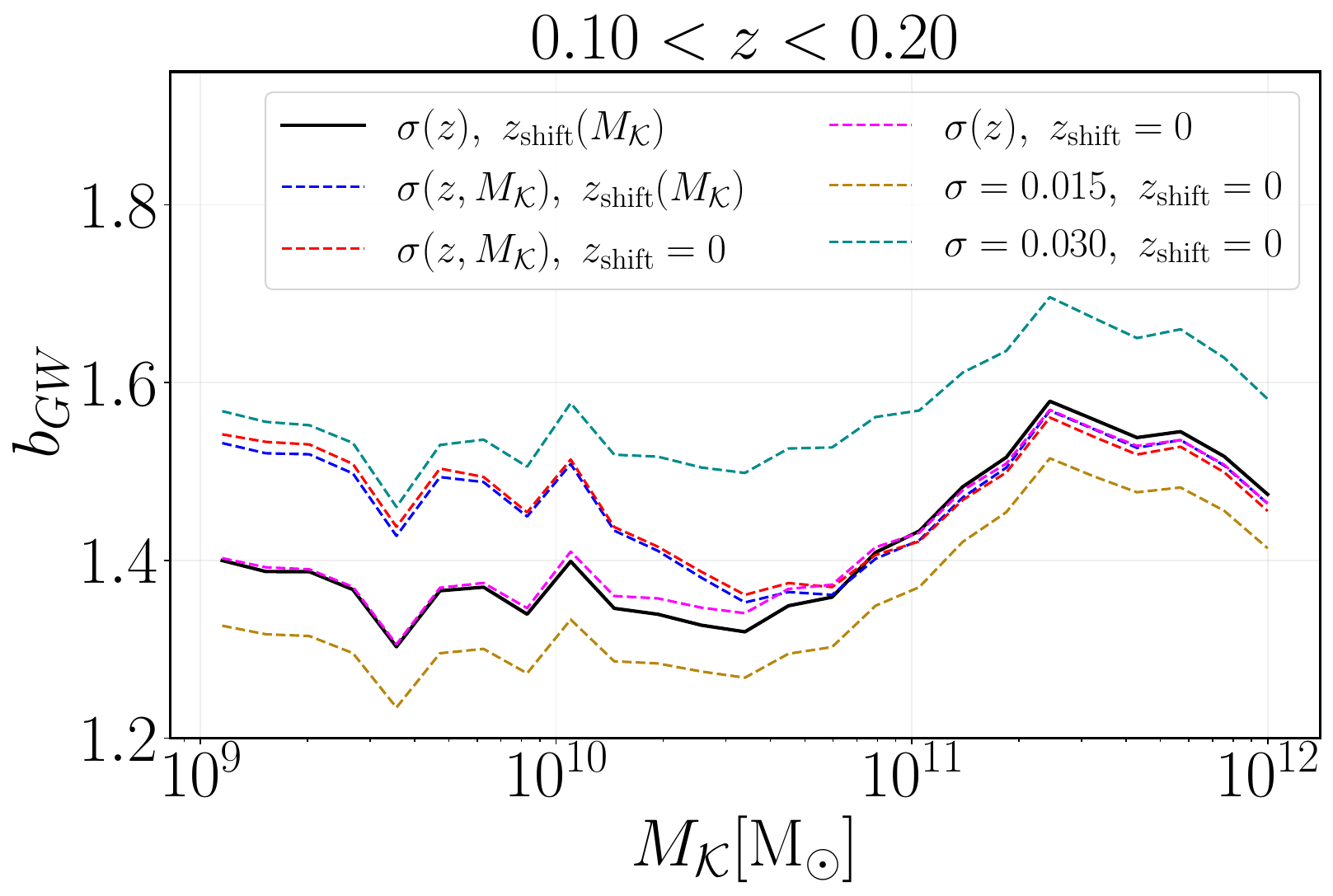}
\caption{Bias for 2MPZ catalog with six different photometric redshift convolution schemes. The solid black line represents the result for assuming the original $\sigma(z)$ but considering $z_\text{shift}(M_\mathcal{K})$ used in the analysis of Section \ref{sec:results}. Refer to the text for an explanation of other cases. The fluctuating behaviour is due to the statistical scatter in the results. We omit the corresponding error bars for the sake of clarity. The error bars can be seen by comparing this plot to the corresponding plot in the main text, i.e.\ Figure \ref{fig:m_k}.}
\label{fig:b_with_diff_conv}
\end{figure}

To see the effect of different convolution schemes, we evaluated the bias in six different cases (including the original error distribution properties derived based on entire 2MPZ catalog):
\begin{itemize}
    
    \item The original redshift dependent $\sigma(z)$ given in Eq.~\eqref{eqn:sigmaz} and $z_\mathrm{shift}=0$. The results for GW bias are shown by the pink dashed curves in Figure \ref{fig:b_with_diff_conv}.
    
    \item The original result for $\sigma(z)$ with redshift dependence given in Eq.~\eqref{eqn:sigmaz} but we take the fitted $M_\mathcal{K}$-dependent $z_\mathrm{shift}$ as displayed in Figure \ref{fig:2mpz-offset}. The resulting $b_{GW}$ is shown by the black solid curves in Figure \ref{fig:b_with_diff_conv}.
    
    \item A redshift and $M_\mathcal{K}$ dependent $\sigma(z, M_\mathcal{K})$ as shown in Figure \ref{fig:2mpz_sigma} in addition to $M_\mathcal{K}$ dependent $z_\mathrm{shift}$ as shown in Figure \ref{fig:2mpz-offset}.  The results for $b_{GW}$ are shown by the blue dashed curves in Figure \ref{fig:b_with_diff_conv}.
    
    \item A redshift and $M_\mathcal{K}$ dependent $\sigma(z,M_\mathcal{K})$ shown in Figure \ref{fig:2mpz_sigma} but setting  $z_\mathrm{shift}=0$. The results for $b_{GW}$ are shown by the red dashed curves in Figure \ref{fig:b_with_diff_conv}.

    \item Setting $\sigma=0.015$, and $z_\mathrm{shift}=0$. The results are shown by the gold dashed curves in Figure \ref{fig:b_with_diff_conv}
    
    \item Setting $\sigma=0.030$, and $z_\mathrm{shift}=0$. The results are shown by the green dashed curves in Figure \ref{fig:b_with_diff_conv}
\end{itemize}

\begin{figure}
\centering
\includegraphics[width=0.45\hsize]{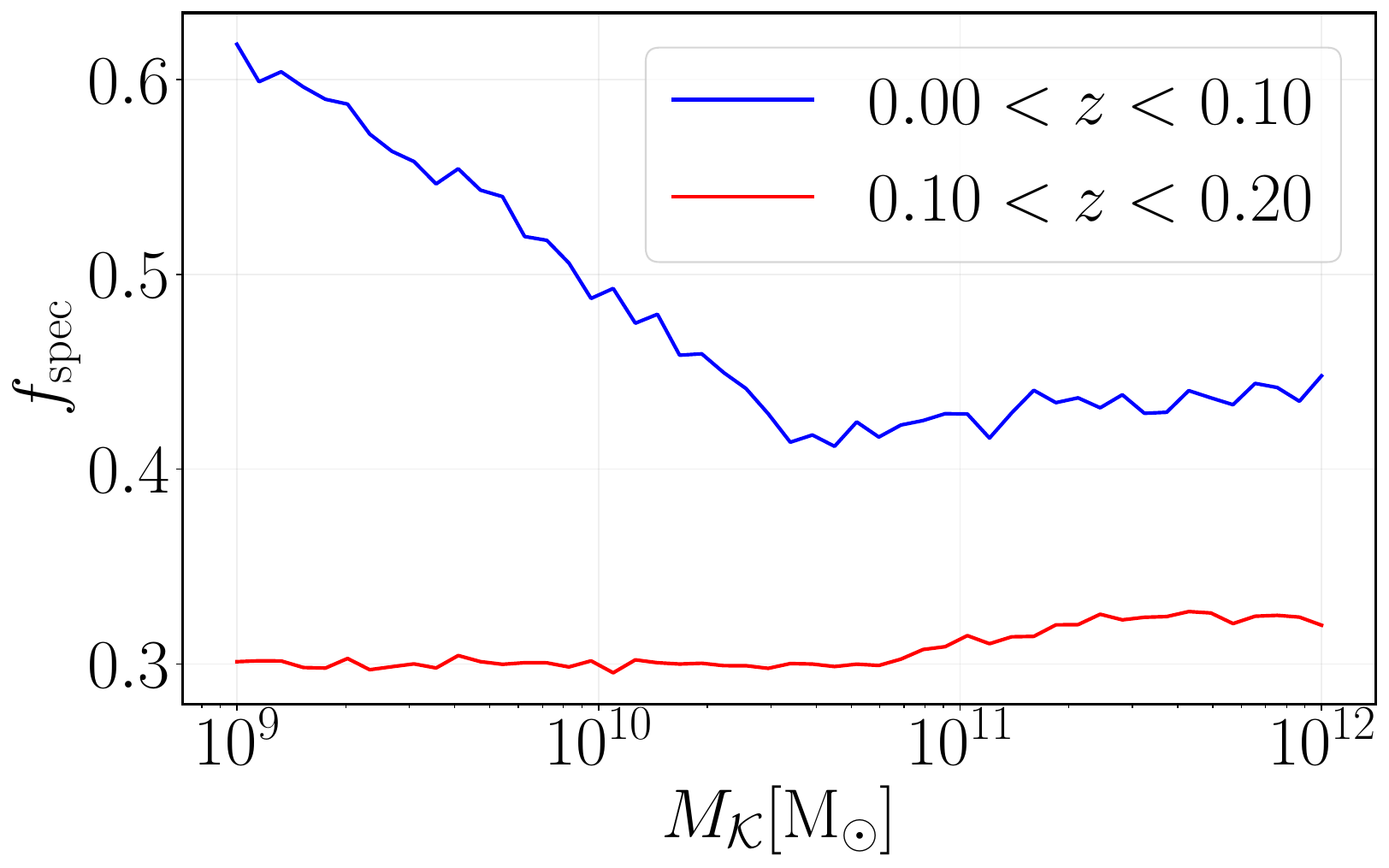}
\caption{Fraction of GW hosts with measured spectroscopic redshift, $f_\mathrm{spec}$, versus $M_\mathcal{K}$, for mock siren catalogs in 2MPZ. }
\label{fig:f_spec}
\end{figure}
Putting all together, these results show the bias calculation is somewhat affected by the details of the redshift convolution scheme, in particular the second redshift bin. In the first redshift bin, $z_\mathrm{shift}$ is the parameter that has a greater impact on $b_{GW}$ in the low $M_\mathcal{K}$ range. In the second redshift bin, $\sigma$ is more impactful and induces differences from the original error distribution assumption. This is to be expected, since for $z < 0.1$, $\sigma(z, M_{\mathcal{K}})$ agrees well with $\sigma(z)$ used for the galaxy catalogs, whereas for $z > 0.1$, we see much more variation with $M_{\mathcal{K}}$. Given that it is difficult to measure $\sigma(z, M_{\mathcal{K}})$ precisely for $z > 0.1$, we opt to use $\sigma(z)$ fitted to the galaxy catalog, without any variation with $M_{\mathcal{K}}$.

Note that the photometric redshift convolution affects only the part of the catalog that does not have spectroscopic redshift information. The fraction of GW hosts that have spectroscopic redshifts in each redshift bin of 2MPZ is shown in Figure \ref{fig:f_spec}.

\begin{figure*}
\centering
\includegraphics[width=0.49\hsize]{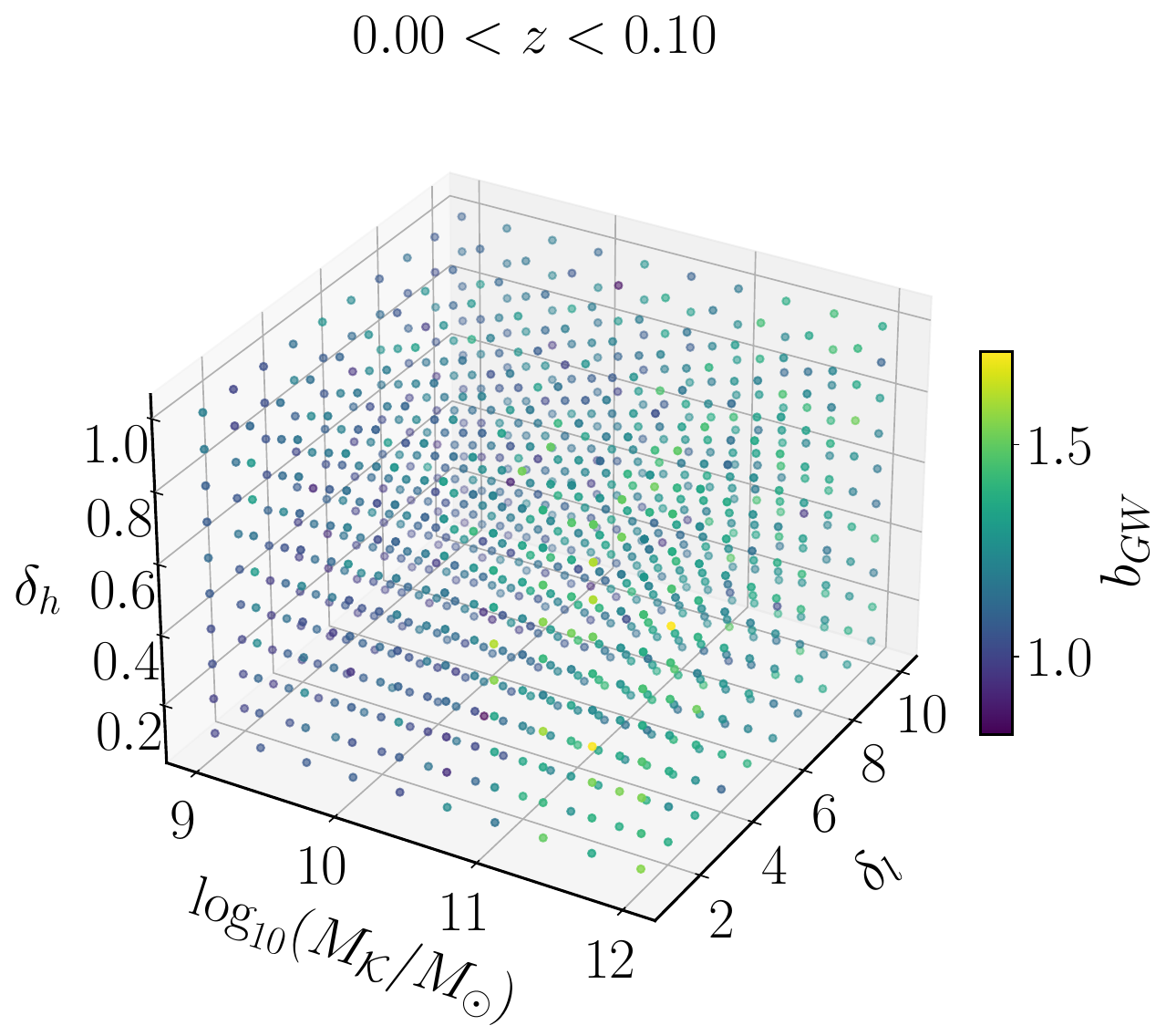}
\includegraphics[width=0.49\hsize]{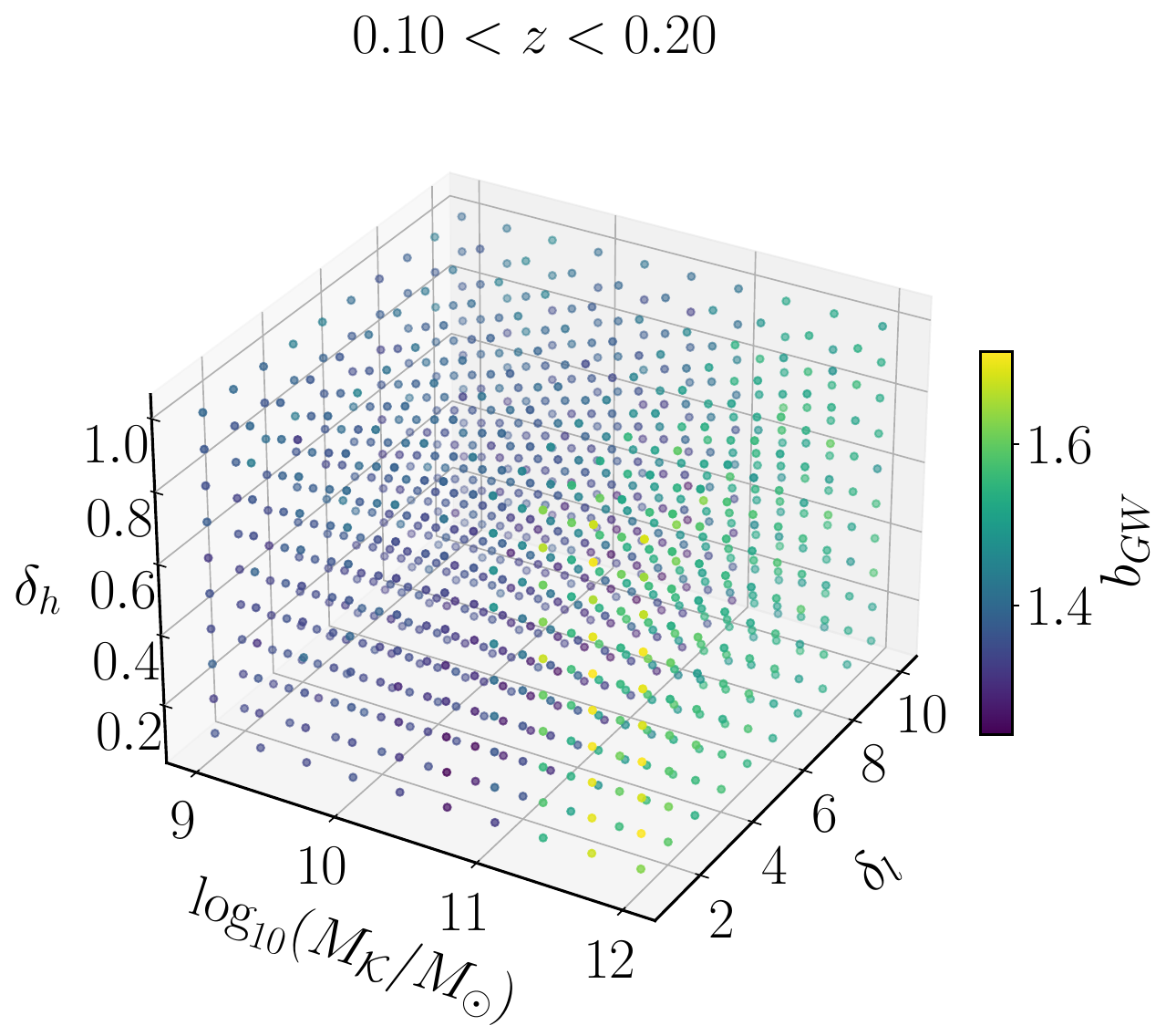}
\caption{Bias for varying $M_\mathcal{K}$, $\delta_l$, and $\delta_h$ in Eq.~\eqref{eq:selection} for 2MPZ catalog. The left and right panels are for the first and second redshift bins, respectively, i.e. $0<z<0.1$, and $0.1<z<0.2$. The rotating 3D GIF files are hyperlinked for \href{https://drive.google.com/file/d/16PYYnwr3V4i0VEpdhjrRCmBTrC0sZRk5/view?usp=drive_link}{$0<z<0.1$}, and  for \href{https://drive.google.com/file/d/1B4Jnffjagd1N92WhYkadkjSiDCFbVwzK/view?usp=drive_link}{$0.1<z<0.2$}.}
\label{fig:2mpz_3D}
\end{figure*}

\begin{figure*}
\centering
\includegraphics[width=0.49\hsize]{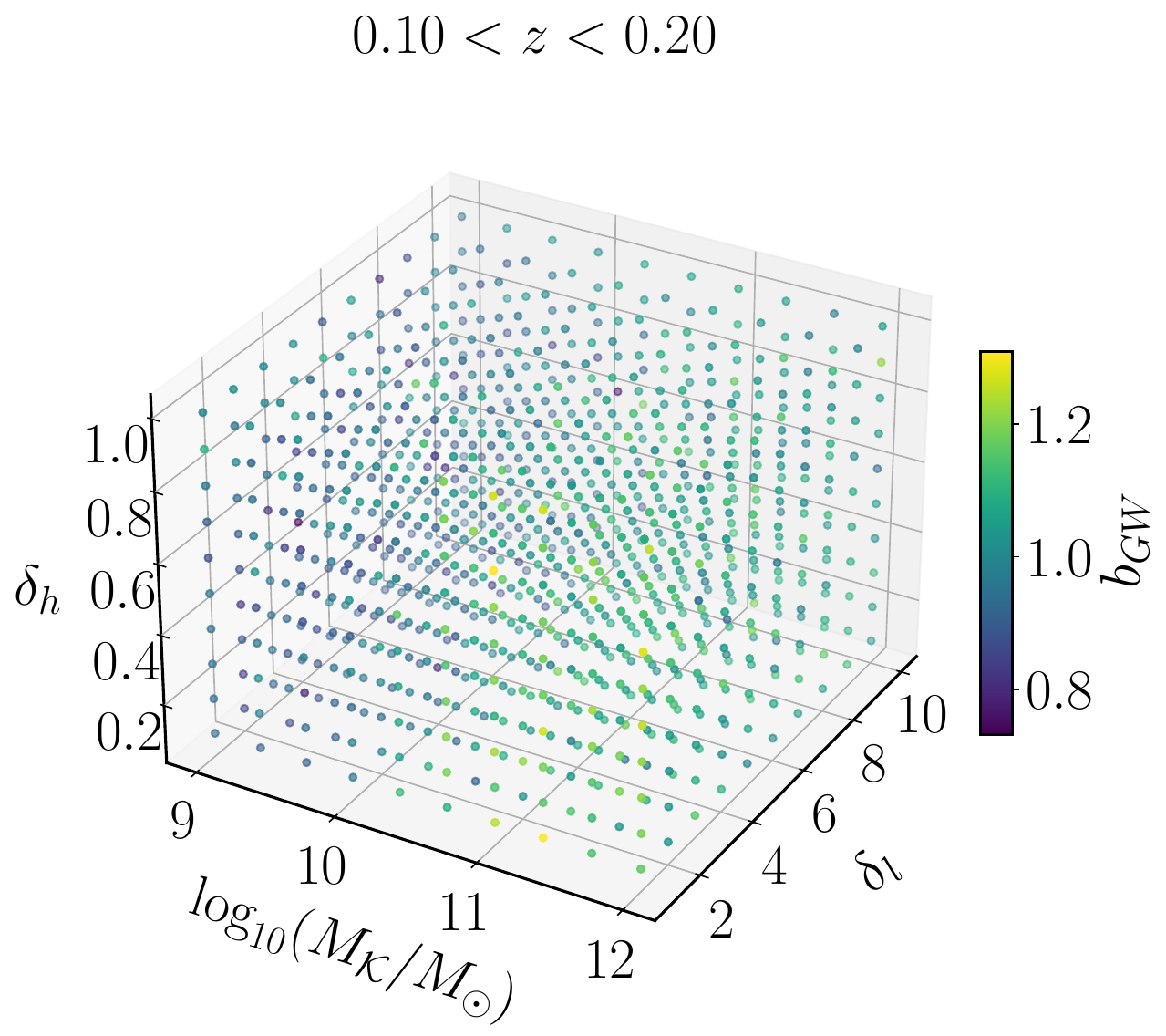}
\includegraphics[width=0.49\hsize]{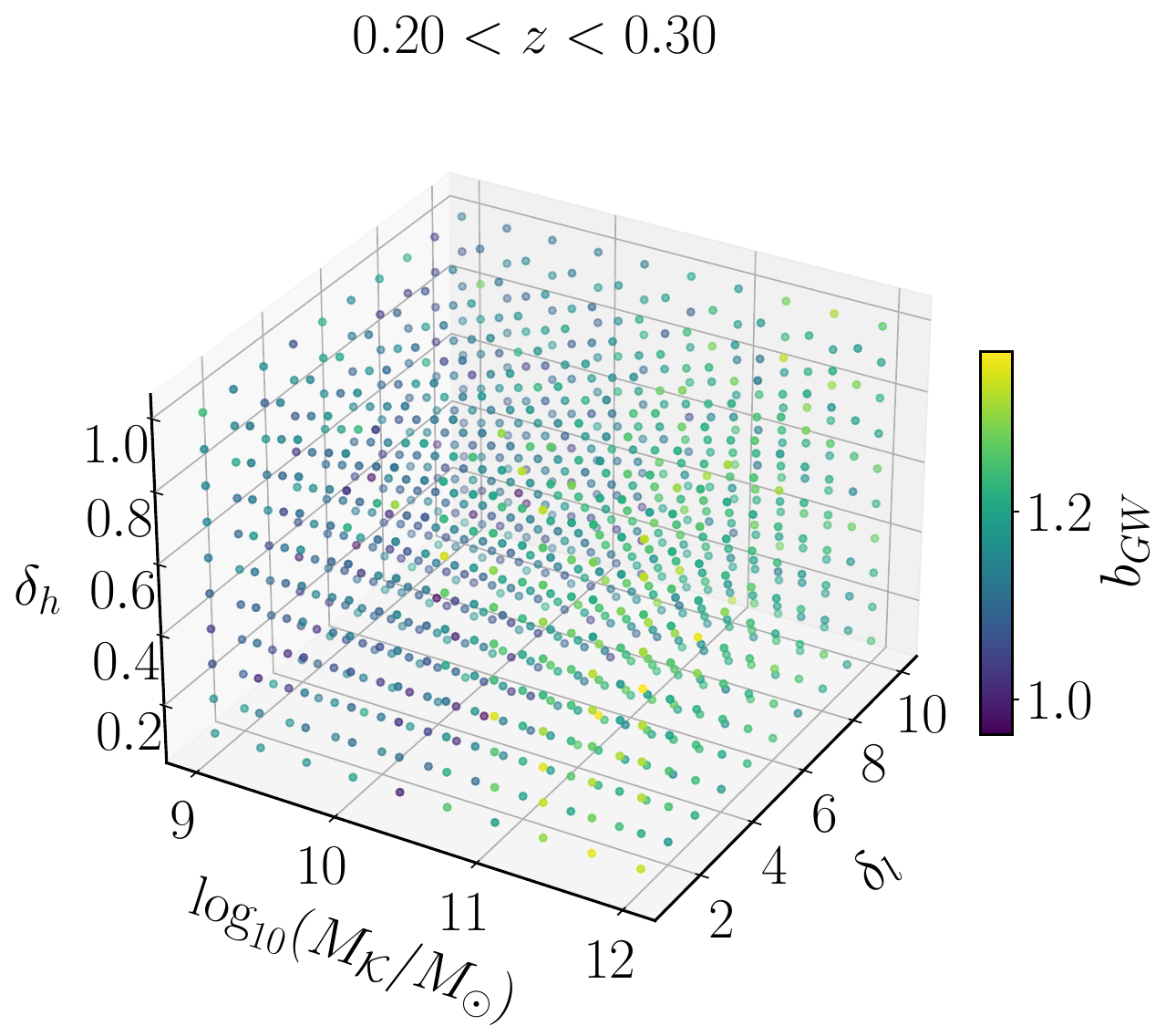}
\includegraphics[width=0.49\hsize]{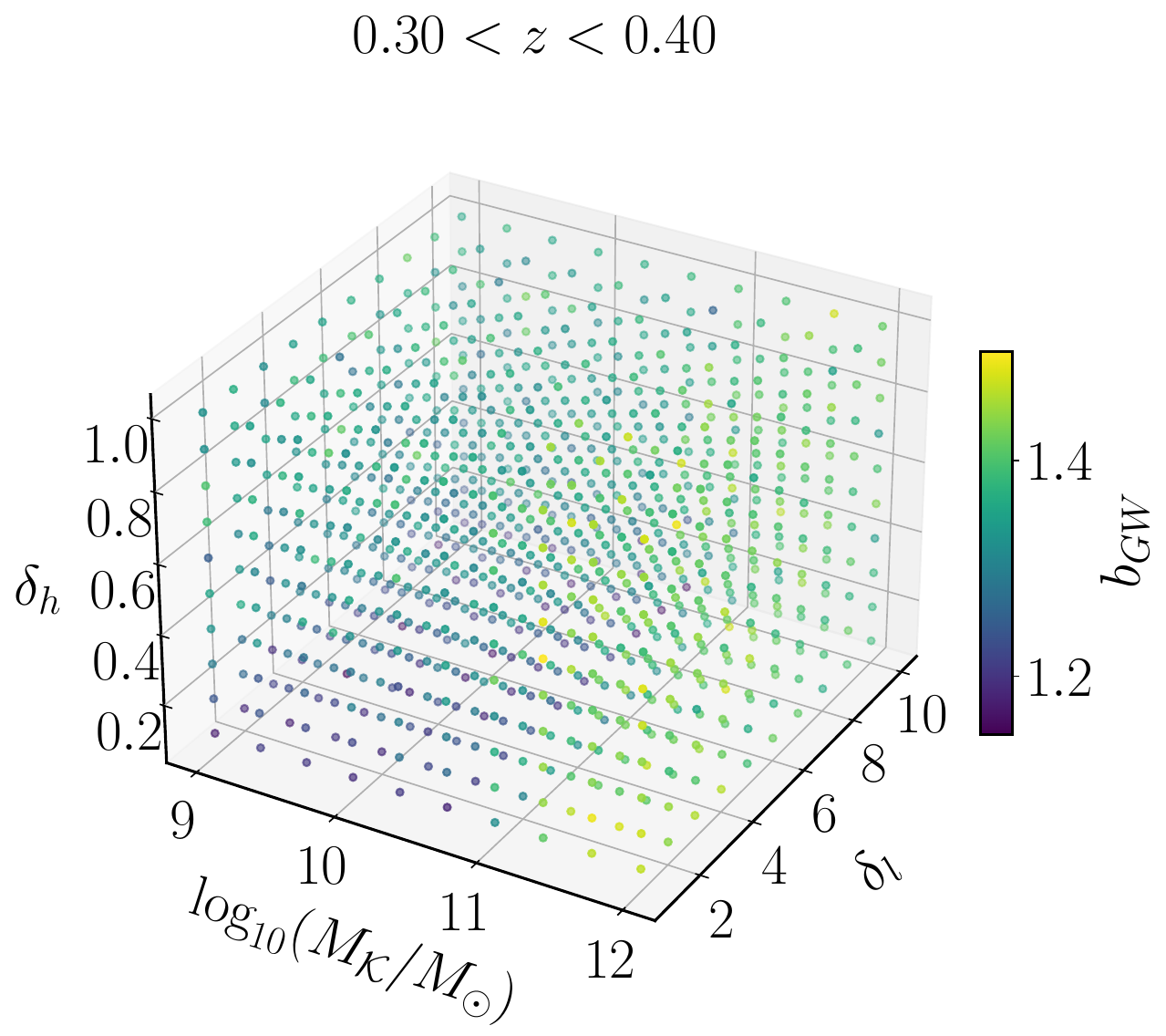}
\caption{Bias for varying $M_\mathcal{K}$, $\delta_l$, and $\delta_h$ in Eq.~\eqref{eq:selection} for WISC catalog. The rotating 3D GIF files are hyperlinked for \href{https://drive.google.com/file/d/1pZGS8YuILkuJ9tvoKOcuRSOj9JgL4b8o/view?usp=sharing}{$0.1<z<0.2$}, for
\href{https://drive.google.com/file/d/1FHaZUXGjZINz43Hkbn6MCP6ZL5E5Tt-m/view?usp=sharing}{$0.2<z<0.3$}, and
for \href{https://drive.google.com/file/d/1eFxY7Bj-E0uUSM_58kMZiHlpgYKzVnJu/view?usp=sharing}{$0.3<z<0.4$}.}
\label{fig:wisc_3D}
\end{figure*}

\section{3D plots of the combined effect of $\delta_l$, $\delta_h$ and $M_\mathcal{K}$ on GW bias}
\label{sec:3d}
Here, in addition to Figures \ref{fig:2mpz_deltas}, we offer a 3D visualization of the bias parameter for 2MPZ and WISC, respectively, across various selections of $M_\mathcal{K}$, $\delta_l$, and $\delta_h$ in Eq.~\eqref{eq:selection}, covering a large volume of the parameter space. A 3D rotating GIF file is also linked in the captions of Figure \ref{fig:2mpz_3D}, and Figure \ref{fig:wisc_3D} to allow the viewer to better see these plots.

\bibliographystyle{JHEP}
\bibliography{paper-I}

\providecommand{\href}[2]{#2}\begingroup\raggedright\begin{thebibliography}{100}

\bibitem{LIGOScientific:2016aoc}
{\scshape LIGO Scientific, Virgo} collaboration, \emph{{Observation of
  Gravitational Waves from a Binary Black Hole Merger}},
  \href{https://doi.org/10.1103/PhysRevLett.116.061102}{\emph{Phys. Rev. Lett.}
  {\bfseries 116} (2016) 061102}
  [\href{https://arxiv.org/abs/1602.03837}{{\ttfamily 1602.03837}}].

\bibitem{KAGRA:2013rdx}
{\scshape KAGRA, LIGO Scientific, Virgo} collaboration, \emph{{Prospects for
  observing and localizing gravitational-wave transients with Advanced LIGO,
  Advanced Virgo and KAGRA}},
  \href{https://doi.org/10.1007/s41114-020-00026-9}{\emph{Living Rev. Rel.}
  {\bfseries 19} (2016) 1} [\href{https://arxiv.org/abs/1304.0670}{{\ttfamily
  1304.0670}}].

\bibitem{LIGOScientific:2014pky}
{\scshape LIGO Scientific} collaboration, \emph{{Advanced LIGO}},
  \href{https://doi.org/10.1088/0264-9381/32/7/074001}{\emph{Class. Quant.
  Grav.} {\bfseries 32} (2015) 074001}
  [\href{https://arxiv.org/abs/1411.4547}{{\ttfamily 1411.4547}}].

\bibitem{VIRGO:2014yos}
{\scshape VIRGO} collaboration, \emph{{Advanced Virgo: a second-generation
  interferometric gravitational wave detector}},
  \href{https://doi.org/10.1088/0264-9381/32/2/024001}{\emph{Class. Quant.
  Grav.} {\bfseries 32} (2015) 024001}
  [\href{https://arxiv.org/abs/1408.3978}{{\ttfamily 1408.3978}}].

\bibitem{PhysRevD.88.043007}
{\scshape The KAGRA Collaboration} collaboration, \emph{Interferometer design
  of the kagra gravitational wave detector},
  \href{https://doi.org/10.1103/PhysRevD.88.043007}{\emph{Phys. Rev. D}
  {\bfseries 88} (2013) 043007}.

\bibitem{LIGOScientific:2018jsj}
{\scshape LIGO Scientific, Virgo} collaboration, \emph{{Binary Black Hole
  Population Properties Inferred from the First and Second Observing Runs of
  Advanced LIGO and Advanced Virgo}},
  \href{https://doi.org/10.3847/2041-8213/ab3800}{\emph{Astrophys. J. Lett.}
  {\bfseries 882} (2019) L24}
  [\href{https://arxiv.org/abs/1811.12940}{{\ttfamily 1811.12940}}].

\bibitem{LIGOScientific:2020kqk}
{\scshape LIGO Scientific, Virgo} collaboration, \emph{{Population Properties
  of Compact Objects from the Second LIGO-Virgo Gravitational-Wave Transient
  Catalog}}, \href{https://doi.org/10.3847/2041-8213/abe949}{\emph{Astrophys.
  J. Lett.} {\bfseries 913} (2021) L7}
  [\href{https://arxiv.org/abs/2010.14533}{{\ttfamily 2010.14533}}].

\bibitem{LIGOScientific:2021aug}
{\scshape LIGO Scientific, Virgo,, KAGRA, VIRGO} collaboration,
  \emph{{Constraints on the Cosmic Expansion History from GWTC\textendash{}3}},
  \href{https://doi.org/10.3847/1538-4357/ac74bb}{\emph{Astrophys. J.}
  {\bfseries 949} (2023) 76}
  [\href{https://arxiv.org/abs/2111.03604}{{\ttfamily 2111.03604}}].

\bibitem{KAGRA:2021duu}
{\scshape KAGRA, VIRGO, LIGO Scientific} collaboration, \emph{{Population of
  Merging Compact Binaries Inferred Using Gravitational Waves through GWTC-3}},
  \href{https://doi.org/10.1103/PhysRevX.13.011048}{\emph{Phys. Rev. X}
  {\bfseries 13} (2023) 011048}
  [\href{https://arxiv.org/abs/2111.03634}{{\ttfamily 2111.03634}}].

\bibitem{Meszaros:2019xej}
P.~M\'esz\'aros, D.B.~Fox, C.~Hanna and K.~Murase, \emph{{Multi-Messenger
  Astrophysics}}, \href{https://doi.org/10.1038/s42254-019-0101-z}{\emph{Nature
  Rev. Phys.} {\bfseries 1} (2019) 585}
  [\href{https://arxiv.org/abs/1906.10212}{{\ttfamily 1906.10212}}].

\bibitem{Berti:2022wzk}
E.~Berti, V.~Cardoso, Z.~Haiman, D.E.~Holz, E.~Mottola, S.~Mukherjee et~al.,
  \emph{{Snowmass2021 Cosmic Frontier White Paper: Fundamental Physics and
  Beyond the Standard Model}},  in \emph{{Snowmass 2021}}, 3, 2022
  [\href{https://arxiv.org/abs/2203.06240}{{\ttfamily 2203.06240}}].

\bibitem{Adhikari:2022sve}
R.X.~Adhikari et~al., \emph{{Report of the Topical Group on Cosmic Probes of
  Fundamental Physics for for Snowmass 2021}},
  \href{https://arxiv.org/abs/2209.11726}{{\ttfamily 2209.11726}}.

\bibitem{LIGOScientific:2018mvr}
{\scshape LIGO Scientific, Virgo} collaboration, \emph{{GWTC-1: A
  Gravitational-Wave Transient Catalog of Compact Binary Mergers Observed by
  LIGO and Virgo during the First and Second Observing Runs}},
  \href{https://doi.org/10.1103/PhysRevX.9.031040}{\emph{Phys. Rev. X}
  {\bfseries 9} (2019) 031040}
  [\href{https://arxiv.org/abs/1811.12907}{{\ttfamily 1811.12907}}].

\bibitem{LIGOScientific:2020ibl}
{\scshape LIGO Scientific, Virgo} collaboration, \emph{{GWTC-2: Compact Binary
  Coalescences Observed by LIGO and Virgo During the First Half of the Third
  Observing Run}},
  \href{https://doi.org/10.1103/PhysRevX.11.021053}{\emph{Phys. Rev. X}
  {\bfseries 11} (2021) 021053}
  [\href{https://arxiv.org/abs/2010.14527}{{\ttfamily 2010.14527}}].

\bibitem{LIGOScientific:2021djp}
{\scshape LIGO Scientific, VIRGO, KAGRA} collaboration, \emph{{GWTC-3: Compact
  Binary Coalescences Observed by LIGO and Virgo During the Second Part of the
  Third Observing Run}},  \href{https://arxiv.org/abs/2111.03606}{{\ttfamily
  2111.03606}}.

\bibitem{Schutz:1986gp}
B.F.~Schutz, \emph{{Determining the Hubble Constant from Gravitational Wave
  Observations}}, \href{https://doi.org/10.1038/323310a0}{\emph{Nature}
  {\bfseries 323} (1986) 310}.

\bibitem{Oguri:2016dgk}
M.~Oguri, \emph{{Measuring the distance-redshift relation with the
  cross-correlation of gravitational wave standard sirens and galaxies}},
  \href{https://doi.org/10.1103/PhysRevD.93.083511}{\emph{Phys. Rev. D}
  {\bfseries 93} (2016) 083511}
  [\href{https://arxiv.org/abs/1603.02356}{{\ttfamily 1603.02356}}].

\bibitem{LIGOScientific:2017adf}
{\scshape LIGO Scientific, Virgo, 1M2H, Dark Energy Camera GW-E, DES, DLT40,
  Las Cumbres Observatory, VINROUGE, MASTER} collaboration, \emph{{A
  gravitational-wave standard siren measurement of the Hubble constant}},
  \href{https://doi.org/10.1038/nature24471}{\emph{Nature} {\bfseries 551}
  (2017) 85} [\href{https://arxiv.org/abs/1710.05835}{{\ttfamily 1710.05835}}].

\bibitem{Chen:2017rfc}
H.-Y.~Chen, M.~Fishbach and D.E.~Holz, \emph{{A two per cent Hubble constant
  measurement from standard sirens within five years}},
  \href{https://doi.org/10.1038/s41586-018-0606-0}{\emph{Nature} {\bfseries
  562} (2018) 545} [\href{https://arxiv.org/abs/1712.06531}{{\ttfamily
  1712.06531}}].

\bibitem{LIGOScientific:2018gmd}
{\scshape LIGO Scientific, Virgo} collaboration, \emph{{A Standard Siren
  Measurement of the Hubble Constant from GW170817 without the Electromagnetic
  Counterpart}},
  \href{https://doi.org/10.3847/2041-8213/aaf96e}{\emph{Astrophys. J. Lett.}
  {\bfseries 871} (2019) L13}
  [\href{https://arxiv.org/abs/1807.05667}{{\ttfamily 1807.05667}}].

\bibitem{Mukherjee:2018ebj}
S.~Mukherjee and B.D.~Wandelt, \emph{{Beyond the classical distance-redshift
  test: cross-correlating redshift-free standard candles and sirens with
  redshift surveys}},  \href{https://arxiv.org/abs/1808.06615}{{\ttfamily
  1808.06615}}.

\bibitem{DES:2019ccw}
{\scshape DES, LIGO Scientific, Virgo} collaboration, \emph{{First Measurement
  of the Hubble Constant from a Dark Standard Siren using the Dark Energy
  Survey Galaxies and the LIGO/Virgo Binary\textendash{}Black-hole Merger
  GW170814}}, \href{https://doi.org/10.3847/2041-8213/ab14f1}{\emph{Astrophys.
  J. Lett.} {\bfseries 876} (2019) L7}
  [\href{https://arxiv.org/abs/1901.01540}{{\ttfamily 1901.01540}}].

\bibitem{Mukherjee:2020hyn}
S.~Mukherjee, B.D.~Wandelt, S.M.~Nissanke and A.~Silvestri, \emph{{Accurate
  precision Cosmology with redshift unknown gravitational wave sources}},
  \href{https://doi.org/10.1103/PhysRevD.103.043520}{\emph{Phys. Rev. D}
  {\bfseries 103} (2021) 043520}
  [\href{https://arxiv.org/abs/2007.02943}{{\ttfamily 2007.02943}}].

\bibitem{Bera:2020jhx}
S.~Bera, D.~Rana, S.~More and S.~Bose, \emph{{Incompleteness Matters Not:
  Inference of $H_0$ from Binary Black Hole\textendash{}Galaxy
  Cross-correlations}},
  \href{https://doi.org/10.3847/1538-4357/abb4e0}{\emph{Astrophys. J.}
  {\bfseries 902} (2020) 79}
  [\href{https://arxiv.org/abs/2007.04271}{{\ttfamily 2007.04271}}].

\bibitem{Mukherjee:2022afz}
S.~Mukherjee, A.~Krolewski, B.D.~Wandelt and J.~Silk, \emph{{Cross-correlating
  dark sirens and galaxies: measurement of $H_0$ from GWTC-3 of
  LIGO-Virgo-KAGRA}},  \href{https://arxiv.org/abs/2203.03643}{{\ttfamily
  2203.03643}}.

\bibitem{Afroz:2024joi}
S.~Afroz and S.~Mukherjee, \emph{{Prospect of Precision Cosmology and Testing
  General Relativity using Binary Black Holes- Galaxies Cross-correlation}},
  \href{https://arxiv.org/abs/2407.09262}{{\ttfamily 2407.09262}}.

\bibitem{NANOGrav:2023gor}
{\scshape NANOGrav} collaboration, \emph{{The NANOGrav 15 yr Data Set: Evidence
  for a Gravitational-wave Background}},
  \href{https://doi.org/10.3847/2041-8213/acdac6}{\emph{Astrophys. J. Lett.}
  {\bfseries 951} (2023) L8}
  [\href{https://arxiv.org/abs/2306.16213}{{\ttfamily 2306.16213}}].

\bibitem{EPTA:2023sfo}
{\scshape EPTA} collaboration, \emph{{The second data release from the European
  Pulsar Timing Array I. The dataset and timing analysis}},
  \href{https://doi.org/10.1051/0004-6361/202346841}{\emph{Astron. Astrophys.}
  (2023) } [\href{https://arxiv.org/abs/2306.16224}{{\ttfamily 2306.16224}}].

\bibitem{Xu:2023wog}
H.~Xu et~al., \emph{{Searching for the Nano-Hertz Stochastic Gravitational Wave
  Background with the Chinese Pulsar Timing Array Data Release I}},
  \href{https://doi.org/10.1088/1674-4527/acdfa5}{\emph{Res. Astron.
  Astrophys.} {\bfseries 23} (2023) 075024}
  [\href{https://arxiv.org/abs/2306.16216}{{\ttfamily 2306.16216}}].

\bibitem{Sah:2024etc}
M.R.~Sah and S.~Mukherjee, \emph{{Discovering the Cosmic Evolution of
  Supermassive Black Holes using Nano-Hertz Gravitational Waves and Galaxy
  Surveys}},  \href{https://arxiv.org/abs/2407.11669}{{\ttfamily 2407.11669}}.

\bibitem{Sah:2024oyg}
M.R.~Sah, S.~Mukherjee, V.~Saeedzadeh, A.~Babul, M.~Tremmel and T.R.~Quinn,
  \emph{{Imprints of Supermassive Black Hole Evolution on the Spectral and
  Spatial Anisotropy of Nano-Hertz Stochastic Gravitational-Wave Background}},
  \href{https://arxiv.org/abs/2404.14508}{{\ttfamily 2404.14508}}.

\bibitem{Belczynski:2001uc}
K.~Belczynski, V.~Kalogera and T.~Bulik, \emph{{A Comprehensive study of binary
  compact objects as gravitational wave sources: Evolutionary channels, rates,
  and physical properties}},
  \href{https://doi.org/10.1086/340304}{\emph{Astrophys. J.} {\bfseries 572}
  (2001) 407} [\href{https://arxiv.org/abs/astro-ph/0111452}{{\ttfamily
  astro-ph/0111452}}].

\bibitem{Neijssel2019}
C.J.~{Neijssel}, A.~{Vigna-G{\'o}mez}, S.~{Stevenson}, J.W.~{Barrett},
  S.M.~{Gaebel}, F.S.~{Broekgaarden} et~al., \emph{{The effect of the
  metallicity-specific star formation history on double compact object
  mergers}}, \href{https://doi.org/10.1093/mnras/stz2840}{\emph{\mnras}
  {\bfseries 490} (2019) 3740}
  [\href{https://arxiv.org/abs/1906.08136}{{\ttfamily 1906.08136}}].

\bibitem{OShaughnessy:2009szr}
R.~O'Shaughnessy, V.~Kalogera and K.~Belczynski, \emph{{Binary Compact Object
  Coalescence Rates: The Role of Elliptical Galaxies}},
  \href{https://doi.org/10.1088/0004-637X/716/1/615}{\emph{Astrophys. J.}
  {\bfseries 716} (2010) 615}
  [\href{https://arxiv.org/abs/0908.3635}{{\ttfamily 0908.3635}}].

\bibitem{Dominik:2014yma}
M.~Dominik, E.~Berti, R.~O'Shaughnessy, I.~Mandel, K.~Belczynski, C.~Fryer
  et~al., \emph{{Double Compact Objects III: Gravitational Wave Detection
  Rates}}, \href{https://doi.org/10.1088/0004-637X/806/2/263}{\emph{Astrophys.
  J.} {\bfseries 806} (2015) 263}
  [\href{https://arxiv.org/abs/1405.7016}{{\ttfamily 1405.7016}}].

\bibitem{Mapelli:2017hqk}
M.~Mapelli, N.~Giacobbo, E.~Ripamonti and M.~Spera, \emph{{The cosmic merger
  rate of stellar black hole binaries from the Illustris simulation}},
  \href{https://doi.org/10.1093/mnras/stx2123}{\emph{Mon. Not. Roy. Astron.
  Soc.} {\bfseries 472} (2017) 2422}
  [\href{https://arxiv.org/abs/1708.05722}{{\ttfamily 1708.05722}}].

\bibitem{Giacobbo:2017qhh}
N.~Giacobbo, M.~Mapelli and M.~Spera, \emph{{Merging black hole binaries: the
  effects of progenitor's metallicity, mass-loss rate and Eddington factor}},
  \href{https://doi.org/10.1093/mnras/stx2933}{\emph{Mon. Not. Roy. Astron.
  Soc.} {\bfseries 474} (2018) 2959}
  [\href{https://arxiv.org/abs/1711.03556}{{\ttfamily 1711.03556}}].

\bibitem{Cao2018}
L.~{Cao}, Y.~{Lu} and Y.~{Zhao}, \emph{{Host galaxy properties of mergers of
  stellar binary black holes and their implications for advanced LIGO
  gravitational wave sources}},
  \href{https://doi.org/10.1093/mnras/stx3087}{\emph{\mnras} {\bfseries 474}
  (2018) 4997} [\href{https://arxiv.org/abs/1711.09190}{{\ttfamily
  1711.09190}}].

\bibitem{Fishbach:2018edt}
M.~Fishbach, D.E.~Holz and W.M.~Farr, \emph{{Does the Black Hole Merger Rate
  Evolve with Redshift?}},
  \href{https://doi.org/10.3847/2041-8213/aad800}{\emph{Astrophys. J. Lett.}
  {\bfseries 863} (2018) L41}
  [\href{https://arxiv.org/abs/1805.10270}{{\ttfamily 1805.10270}}].

\bibitem{Santoliquido:2022kyu}
F.~Santoliquido, M.~Mapelli, M.C.~Artale and L.~Boco, \emph{{Modelling the host
  galaxies of binary compact object mergers with observational scaling
  relations}}, \href{https://doi.org/10.1093/mnras/stac2384}{\emph{Mon. Not.
  Roy. Astron. Soc.} {\bfseries 516} (2022) 3297}
  [\href{https://arxiv.org/abs/2205.05099}{{\ttfamily 2205.05099}}].

\bibitem{Dominik:2012kk}
M.~Dominik, K.~Belczynski, C.~Fryer, D.~Holz, E.~Berti, T.~Bulik et~al.,
  \emph{{Double Compact Objects I: The Significance of the Common Envelope on
  Merger Rates}},
  \href{https://doi.org/10.1088/0004-637X/759/1/52}{\emph{Astrophys. J.}
  {\bfseries 759} (2012) 52} [\href{https://arxiv.org/abs/1202.4901}{{\ttfamily
  1202.4901}}].

\bibitem{Toffano:2019ekp}
M.~Toffano, M.~Mapelli, N.~Giacobbo, M.C.~Artale and G.~Ghirlanda, \emph{{The
  host galaxies of double compact objects across cosmic time}},
  \href{https://doi.org/10.1093/mnras/stz2415}{\emph{Mon. Not. Roy. Astron.
  Soc.} {\bfseries 489} (2019) 4622}
  [\href{https://arxiv.org/abs/1906.01072}{{\ttfamily 1906.01072}}].

\bibitem{Artale:2019tfl}
M.C.~Artale, M.~Mapelli, Y.~Bouffanais, N.~Giacobbo, M.~Spera and M.~Pasquato,
  \emph{{Mass and star formation rate of the host galaxies of compact binary
  mergers across cosmic time}},
  \href{https://doi.org/10.1093/mnras/stz3190}{\emph{Mon. Not. Roy. Astron.
  Soc.} {\bfseries 491} (2020) 3419}
  [\href{https://arxiv.org/abs/1910.04890}{{\ttfamily 1910.04890}}].

\bibitem{McCarthy2020}
K.S.~{McCarthy}, Z.~{Zheng} and E.~{Ramirez-Ruiz}, \emph{{Constraining delay
  time distribution of binary neutron star mergers from host galaxy
  properties}}, \href{https://doi.org/10.1093/mnras/staa3206}{\emph{\mnras}
  {\bfseries 499} (2020) 5220}
  [\href{https://arxiv.org/abs/2007.15024}{{\ttfamily 2007.15024}}].

\bibitem{Afroz:2024fzp}
S.~Afroz and S.~Mukherjee, \emph{{Phase Space of Binary Black Holes from
  Gravitational Wave Observations to Unveil its Formation History}},
  \href{https://arxiv.org/abs/2411.07304}{{\ttfamily 2411.07304}}.

\bibitem{2017PhRvD..95l4046G}
D.~{Gerosa} and E.~{Berti}, \emph{{Are merging black holes born from stellar
  collapse or previous mergers?}},
  \href{https://doi.org/10.1103/PhysRevD.95.124046}{\emph{prd} {\bfseries 95}
  (2017) 124046} [\href{https://arxiv.org/abs/1703.06223}{{\ttfamily
  1703.06223}}].

\bibitem{2019PhRvL.123r1101Y}
Y.~{Yang}, I.~{Bartos}, V.~{Gayathri}, K.E.S.~{Ford}, Z.~{Haiman},
  S.~{Klimenko} et~al., \emph{{Hierarchical Black Hole Mergers in Active
  Galactic Nuclei}},
  \href{https://doi.org/10.1103/PhysRevLett.123.181101}{\emph{prl} {\bfseries
  123} (2019) 181101} [\href{https://arxiv.org/abs/1906.09281}{{\ttfamily
  1906.09281}}].

\bibitem{2020ApJ...893...35D}
Z.~{Doctor}, D.~{Wysocki}, R.~{O'Shaughnessy}, D.E.~{Holz} and B.~{Farr},
  \emph{{Black Hole Coagulation: Modeling Hierarchical Mergers in Black Hole
  Populations}}, \href{https://doi.org/10.3847/1538-4357/ab7fac}{\emph{apj}
  {\bfseries 893} (2020) 35}
  [\href{https://arxiv.org/abs/1911.04424}{{\ttfamily 1911.04424}}].

\bibitem{2011ApJS..192....3P}
B.~{Paxton}, L.~{Bildsten}, A.~{Dotter}, F.~{Herwig}, P.~{Lesaffre} and
  F.~{Timmes}, \emph{{Modules for Experiments in Stellar Astrophysics (MESA)}},
  \href{https://doi.org/10.1088/0067-0049/192/1/3}{\emph{apjs} {\bfseries 192}
  (2011) 3} [\href{https://arxiv.org/abs/1009.1622}{{\ttfamily 1009.1622}}].

\bibitem{2014MNRAS.444.1518V}
M.~{Vogelsberger}, S.~{Genel}, V.~{Springel}, P.~{Torrey}, D.~{Sijacki},
  D.~{Xu} et~al., \emph{{Introducing the Illustris Project: simulating the
  coevolution of dark and visible matter in the Universe}},
  \href{https://doi.org/10.1093/mnras/stu1536}{\emph{mnras} {\bfseries 444}
  (2014) 1518} [\href{https://arxiv.org/abs/1405.2921}{{\ttfamily 1405.2921}}].

\bibitem{2023MNRAS.519.3154H}
P.F.~{Hopkins}, A.~{Wetzel}, C.~{Wheeler}, R.~{Sanderson}, M.Y.~{Grudi{\'c}},
  O.~{Sameie} et~al., \emph{{FIRE-3: updated stellar evolution models, yields,
  and microphysics and fitting functions for applications in galaxy
  simulations}}, \href{https://doi.org/10.1093/mnras/stac3489}{\emph{mnras}
  {\bfseries 519} (2023) 3154}
  [\href{https://arxiv.org/abs/2203.00040}{{\ttfamily 2203.00040}}].

\bibitem{2019ApJ...887...53F}
R.~{Farmer}, M.~{Renzo}, S.E.~{de Mink}, P.~{Marchant} and S.~{Justham},
  \emph{{Mind the Gap: The Location of the Lower Edge of the Pair-instability
  Supernova Black Hole Mass Gap}},
  \href{https://doi.org/10.3847/1538-4357/ab518b}{\emph{\apj} {\bfseries 887}
  (2019) 53} [\href{https://arxiv.org/abs/1910.12874}{{\ttfamily 1910.12874}}].

\bibitem{2022MNRAS.515.5495M}
S.~{Mukherjee}, \emph{{The redshift dependence of black hole mass distribution:
  is it reliable for standard sirens cosmology?}},
  \href{https://doi.org/10.1093/mnras/stac2152}{\emph{\mnras} {\bfseries 515}
  (2022) 5495} [\href{https://arxiv.org/abs/2112.10256}{{\ttfamily
  2112.10256}}].

\bibitem{2023MNRAS.523.4539K}
C.~{Karathanasis}, S.~{Mukherjee} and S.~{Mastrogiovanni}, \emph{{Binary black
  holes population and cosmology in new lights: signature of PISN mass and
  formation channel in GWTC-3}},
  \href{https://doi.org/10.1093/mnras/stad1373}{\emph{\mnras} {\bfseries 523}
  (2023) 4539} [\href{https://arxiv.org/abs/2204.13495}{{\ttfamily
  2204.13495}}].

\bibitem{Wechsler:2018pic}
R.H.~Wechsler and J.L.~Tinker, \emph{{The Connection between Galaxies and their
  Dark Matter Halos}},
  \href{https://doi.org/10.1146/annurev-astro-081817-051756}{\emph{Ann. Rev.
  Astron. Astrophys.} {\bfseries 56} (2018) 435}
  [\href{https://arxiv.org/abs/1804.03097}{{\ttfamily 1804.03097}}].

\bibitem{Benson:1999mva}
A.J.~Benson, S.~Cole, C.S.~Frenk, C.M.~Baugh and C.G.~Lacey, \emph{{The Nature
  of galaxy bias and clustering}},
  \href{https://doi.org/10.1046/j.1365-8711.2000.03101.x}{\emph{Mon. Not. Roy.
  Astron. Soc.} {\bfseries 311} (2000) 793}
  [\href{https://arxiv.org/abs/astro-ph/9903343}{{\ttfamily
  astro-ph/9903343}}].

\bibitem{Sheth:1999mn}
R.K.~Sheth and G.~Tormen, \emph{{Large scale bias and the peak background
  split}}, \href{https://doi.org/10.1046/j.1365-8711.1999.02692.x}{\emph{Mon.
  Not. Roy. Astron. Soc.} {\bfseries 308} (1999) 119}
  [\href{https://arxiv.org/abs/astro-ph/9901122}{{\ttfamily
  astro-ph/9901122}}].

\bibitem{Cole2000}
S.~{Cole}, C.G.~{Lacey}, C.M.~{Baugh} and C.S.~{Frenk}, \emph{{Hierarchical
  galaxy formation}},
  \href{https://doi.org/10.1046/j.1365-8711.2000.03879.x}{\emph{\mnras}
  {\bfseries 319} (2000) 168}
  [\href{https://arxiv.org/abs/astro-ph/0007281}{{\ttfamily
  astro-ph/0007281}}].

\bibitem{Peacock:2000qk}
J.A.~Peacock and R.E.~Smith, \emph{{Halo occupation numbers and galaxy bias}},
  \href{https://doi.org/10.1046/j.1365-8711.2000.03779.x}{\emph{Mon. Not. Roy.
  Astron. Soc.} {\bfseries 318} (2000) 1144}
  [\href{https://arxiv.org/abs/astro-ph/0005010}{{\ttfamily
  astro-ph/0005010}}].

\bibitem{ValeOstriker04}
A.~{Vale} and J.P.~{Ostriker}, \emph{{Linking halo mass to galaxy luminosity}},
  \href{https://doi.org/10.1111/j.1365-2966.2004.08059.x}{\emph{\mnras}
  {\bfseries 353} (2004) 189}
  [\href{https://arxiv.org/abs/astro-ph/0402500}{{\ttfamily
  astro-ph/0402500}}].

\bibitem{Gao:2006qz}
L.~Gao and S.D.M.~White, \emph{{Assembly bias in the clustering of dark matter
  haloes}}, \href{https://doi.org/10.1111/j.1745-3933.2007.00292.x}{\emph{Mon.
  Not. Roy. Astron. Soc.} {\bfseries 377} (2007) L5}
  [\href{https://arxiv.org/abs/astro-ph/0611921}{{\ttfamily
  astro-ph/0611921}}].

\bibitem{Croton:2006ys}
D.J.~Croton, L.~Gao and S.D.M.~White, \emph{{Halo assembly bias and its effects
  on galaxy clustering}},
  \href{https://doi.org/10.1111/j.1365-2966.2006.11230.x}{\emph{Mon. Not. Roy.
  Astron. Soc.} {\bfseries 374} (2007) 1303}
  [\href{https://arxiv.org/abs/astro-ph/0605636}{{\ttfamily
  astro-ph/0605636}}].

\bibitem{Dalal08}
N.~{Dalal}, M.~{White}, J.R.~{Bond} and A.~{Shirokov}, \emph{{Halo Assembly
  Bias in Hierarchical Structure Formation}},
  \href{https://doi.org/10.1086/591512}{\emph{\apj} {\bfseries 687} (2008) 12}
  [\href{https://arxiv.org/abs/0803.3453}{{\ttfamily 0803.3453}}].

\bibitem{Moster_2010}
B.P.~{Moster}, R.S.~{Somerville}, C.~{Maulbetsch}, F.C.~{van den Bosch},
  A.V.~{Macci{\`o}}, T.~{Naab} et~al., \emph{{Constraints on the Relationship
  between Stellar Mass and Halo Mass at Low and High Redshift}},
  \href{https://doi.org/10.1088/0004-637X/710/2/903}{\emph{\apj} {\bfseries
  710} (2010) 903} [\href{https://arxiv.org/abs/0903.4682}{{\ttfamily
  0903.4682}}].

\bibitem{Tinker_2010}
J.L.~{Tinker}, B.E.~{Robertson}, A.V.~{Kravtsov}, A.~{Klypin}, M.S.~{Warren},
  G.~{Yepes} et~al., \emph{{The Large-scale Bias of Dark Matter Halos:
  Numerical Calibration and Model Tests}},
  \href{https://doi.org/10.1088/0004-637X/724/2/878}{\emph{\apj} {\bfseries
  724} (2010) 878} [\href{https://arxiv.org/abs/1001.3162}{{\ttfamily
  1001.3162}}].

\bibitem{Behroozi_2013}
P.S.~{Behroozi}, R.H.~{Wechsler} and C.~{Conroy}, \emph{{The Average Star
  Formation Histories of Galaxies in Dark Matter Halos from z = 0-8}},
  \href{https://doi.org/10.1088/0004-637X/770/1/57}{\emph{\apj} {\bfseries 770}
  (2013) 57} [\href{https://arxiv.org/abs/1207.6105}{{\ttfamily 1207.6105}}].

\bibitem{Behroozi2019}
P.~{Behroozi}, R.H.~{Wechsler}, A.P.~{Hearin} and C.~{Conroy},
  \emph{{UNIVERSEMACHINE: The correlation between galaxy growth and dark matter
  halo assembly from z = 0-10}},
  \href{https://doi.org/10.1093/mnras/stz1182}{\emph{\mnras} {\bfseries 488}
  (2019) 3143} [\href{https://arxiv.org/abs/1806.07893}{{\ttfamily
  1806.07893}}].

\bibitem{Scelfo:2018sny}
G.~Scelfo, N.~Bellomo, A.~Raccanelli, S.~Matarrese and L.~Verde,
  \emph{{GW$\times$LSS: chasing the progenitors of merging binary black
  holes}}, \href{https://doi.org/10.1088/1475-7516/2018/09/039}{\emph{JCAP}
  {\bfseries 09} (2018) 039}
  [\href{https://arxiv.org/abs/1809.03528}{{\ttfamily 1809.03528}}].

\bibitem{Mukherjee:2019oma}
S.~Mukherjee and J.~Silk, \emph{{Time-dependence of the astrophysical
  stochastic gravitational wave background}},
  \href{https://doi.org/10.1093/mnras/stz3226}{\emph{Mon. Not. Roy. Astron.
  Soc.} {\bfseries 491} (2020) 4690}
  [\href{https://arxiv.org/abs/1912.07657}{{\ttfamily 1912.07657}}].

\bibitem{Calore:2020bpd}
F.~Calore, A.~Cuoco, T.~Regimbau, S.~Sachdev and P.D.~Serpico,
  \emph{{Cross-correlating galaxy catalogs and gravitational waves: a
  tomographic approach}},
  \href{https://doi.org/10.1103/PhysRevResearch.2.023314}{\emph{Phys. Rev.
  Res.} {\bfseries 2} (2020) 023314}
  [\href{https://arxiv.org/abs/2002.02466}{{\ttfamily 2002.02466}}].

\bibitem{Diaz:2021pem}
C.C.~Diaz and S.~Mukherjee, \emph{{Mapping the cosmic expansion history from
  LIGO-Virgo-KAGRA in synergy with DESI and SPHEREx}},
  \href{https://doi.org/10.1093/mnras/stac208}{\emph{Mon. Not. Roy. Astron.
  Soc.} {\bfseries 511} (2022) 2782}
  [\href{https://arxiv.org/abs/2107.12787}{{\ttfamily 2107.12787}}].

\bibitem{Libanore:2021jqv}
S.~Libanore, M.C.~Artale, D.~Karagiannis, M.~Liguori, N.~Bartolo, Y.~Bouffanais
  et~al., \emph{{Clustering of Gravitational Wave and Supernovae events: a
  multitracer analysis in Luminosity Distance Space}},
  \href{https://doi.org/10.1088/1475-7516/2022/02/003}{\emph{JCAP} {\bfseries
  02} (2022) 003} [\href{https://arxiv.org/abs/2109.10857}{{\ttfamily
  2109.10857}}].

\bibitem{Naab17}
T.~Naab and J.P.~Ostriker, \emph{Theoretical challenges in galaxy formation},
  \href{https://doi.org/10.1146/annurev-astro-081913-040019}{\emph{Annual
  Review of Astronomy and Astrophysics} {\bfseries 55} (2017) 59–109}.

\bibitem{2023IAUS..373..283N}
K.~{Nagamine}, \emph{{Feedback models in galaxy simulations and probing their
  impact by cosmological hydrodynamic simulations}},  in \emph{Resolving the
  Rise and Fall of Star Formation in Galaxies}, T.~{Wong} and W.-T.~{Kim},
  eds., vol.~373 of \emph{IAU Symposium}, pp.~283--292, Jan., 2023,
  \href{https://doi.org/10.1017/S1743921323000133}{DOI}
  [\href{https://arxiv.org/abs/2307.09576}{{\ttfamily 2307.09576}}].

\bibitem{Sales:2022ich}
L.V.~Sales, A.~Wetzel and A.~Fattahi, \emph{{Baryonic solutions and challenges
  for cosmological models of dwarf galaxies}},
  \href{https://doi.org/10.1038/s41550-022-01689-w}{\emph{Nature Astron.}
  {\bfseries 6} (2022) 897} [\href{https://arxiv.org/abs/2206.05295}{{\ttfamily
  2206.05295}}].

\bibitem{Libanore21}
S.~Libanore, M.C.~Artale, D.~Karagiannis, M.~Liguori, N.~Bartolo, Y.~Bouffanais
  et~al., \emph{Gravitational wave mergers as tracers of large scale
  structures},
  \href{https://doi.org/10.1088/1475-7516/2021/02/035}{\emph{Journal of
  Cosmology and Astroparticle Physics} {\bfseries 2021} (2021) 035–035}.

\bibitem{2024MNRAS.530.1129P}
M.~{Peron}, A.~{Ravenni}, S.~{Libanore}, M.~{Liguori} and M.C.~{Artale},
  \emph{{Clustering of binary black hole mergers: a detailed analysis of the
  EAGLE + MOBSE simulation}},
  \href{https://doi.org/10.1093/mnras/stae893}{\emph{\mnras} {\bfseries 530}
  (2024) 1129} [\href{https://arxiv.org/abs/2305.18003}{{\ttfamily
  2305.18003}}].

\bibitem{Scelfo20}
G.~Scelfo, L.~Boco, A.~Lapi and M.~Viel, \emph{{Exploring
  galaxies-gravitational waves cross-correlations as an astrophysical probe}},
  \href{https://doi.org/10.1088/1475-7516/2020/10/045}{\emph{JCAP} {\bfseries
  10} (2020) 045} [\href{https://arxiv.org/abs/2007.08534}{{\ttfamily
  2007.08534}}].

\bibitem{paper-II}
D.S.~Hosseini, A.~Dehghani, J.L.~Kim, A.~Krolewski, S.~Mukherjee and
  G.~Geshnizjani, ``{Bridging galaxies and binary black holes: estimating the
  gravitational wave bias parameter from 3D power spectra using spectroscopic
  survey}.'' In Preparation.

\bibitem{Mukherjee:2021bmw}
S.~Mukherjee and A.~Moradinezhad~Dizgah, \emph{{Toward a Precision Measurement
  of Binary Black Holes Formation Channels Using Gravitational Waves and
  Emission Lines}},
  \href{https://doi.org/10.3847/2041-8213/ac903b}{\emph{Astrophys. J. Lett.}
  {\bfseries 937} (2022) L27}
  [\href{https://arxiv.org/abs/2111.13166}{{\ttfamily 2111.13166}}].

\bibitem{Namikawa16}
T.~{Namikawa}, A.~{Nishizawa} and A.~{Taruya}, \emph{{Detecting black-hole
  binary clustering via the second-generation gravitational-wave detectors}},
  \href{https://doi.org/10.1103/PhysRevD.94.024013}{\emph{\prd} {\bfseries 94}
  (2016) 024013} [\href{https://arxiv.org/abs/1603.08072}{{\ttfamily
  1603.08072}}].

\bibitem{Mukherjee:2019wcg}
S.~Mukherjee, B.D.~Wandelt and J.~Silk, \emph{{Probing the theory of gravity
  with gravitational lensing of gravitational waves and galaxy surveys}},
  \href{https://doi.org/10.1093/mnras/staa827}{\emph{Mon. Not. Roy. Astron.
  Soc.} {\bfseries 494} (2020) 1956}
  [\href{https://arxiv.org/abs/1908.08951}{{\ttfamily 1908.08951}}].

\bibitem{Mukherjee:2020mha}
S.~Mukherjee, B.D.~Wandelt and J.~Silk, \emph{{Testing the general theory of
  relativity using gravitational wave propagation from dark standard sirens}},
  \href{https://doi.org/10.1093/mnras/stab001}{\emph{Mon. Not. Roy. Astron.
  Soc.} {\bfseries 502} (2021) 1136}
  [\href{https://arxiv.org/abs/2012.15316}{{\ttfamily 2012.15316}}].

\bibitem{Libanore22}
S.~Libanore, M.~Artale, D.~Karagiannis, M.~Liguori, N.~Bartolo, Y.~Bouffanais
  et~al., \emph{Clustering of gravitational wave and supernovae events: a
  multitracer analysis in luminosity distance space},
  \href{https://doi.org/10.1088/1475-7516/2022/02/003}{\emph{Journal of
  Cosmology and Astroparticle Physics} {\bfseries 2022} (2022) 003}.

\bibitem{Gagnon23}
E.L.~{Gagnon}, D.~{Anbajagane}, J.~{Prat}, C.~{Chang} and J.~{Frieman},
  \emph{{Cosmological Constraints from Combining Galaxy Surveys and
  Gravitational Wave Observatories}},
  \href{https://doi.org/10.48550/arXiv.2312.16289}{\emph{arXiv e-prints} (2023)
  arXiv:2312.16289} [\href{https://arxiv.org/abs/2312.16289}{{\ttfamily
  2312.16289}}].

\bibitem{Raccanelli16}
A.~{Raccanelli}, E.D.~{Kovetz}, S.~{Bird}, I.~{Cholis} and J.B.~{Mu{\~n}oz},
  \emph{{Determining the progenitors of merging black-hole binaries}},
  \href{https://doi.org/10.1103/PhysRevD.94.023516}{\emph{\prd} {\bfseries 94}
  (2016) 023516} [\href{https://arxiv.org/abs/1605.01405}{{\ttfamily
  1605.01405}}].

\bibitem{Scelfo18}
G.~{Scelfo}, N.~{Bellomo}, A.~{Raccanelli}, S.~{Matarrese} and L.~{Verde},
  \emph{{GW{\texttimes}LSS: chasing the progenitors of merging binary black
  holes}}, \href{https://doi.org/10.1088/1475-7516/2018/09/039}{\emph{"jcap"}
  {\bfseries 2018} (2018) 039}
  [\href{https://arxiv.org/abs/1809.03528}{{\ttfamily 1809.03528}}].

\bibitem{Libanore23}
S.~Libanore, M.~Liguori and A.~Raccanelli, \emph{Signatures of primordial black
  holes in gravitational wave clustering},
  \href{https://doi.org/10.1088/1475-7516/2023/08/055}{\emph{Journal of
  Cosmology and Astroparticle Physics} {\bfseries 2023} (2023) 055}.

\bibitem{2011PhRvD..84d3516C}
A.~{Challinor} and A.~{Lewis}, \emph{{Linear power spectrum of observed source
  number counts}},
  \href{https://doi.org/10.1103/PhysRevD.84.043516}{\emph{\prd} {\bfseries 84}
  (2011) 043516} [\href{https://arxiv.org/abs/1105.5292}{{\ttfamily
  1105.5292}}].

\bibitem{Lewis:1999bs}
A.~Lewis, A.~Challinor and A.~Lasenby, \emph{{Efficient computation of CMB
  anisotropies in closed FRW models}},
  \href{https://doi.org/10.1086/309179}{\emph{\apj} {\bfseries 538} (2000) 473}
  [\href{https://arxiv.org/abs/astro-ph/9911177}{{\ttfamily
  astro-ph/9911177}}].

\bibitem{Howlett:2012mh}
C.~Howlett, A.~Lewis, A.~Hall and A.~Challinor, \emph{Cmb power spectrum
  parameter degeneracies in the era of precision cosmology},
  \href{https://doi.org/10.1088/1475-7516/2012/04/027}{\emph{Journal of
  Cosmology and Astroparticle Physics} {\bfseries 2012} (2012) 027–027}.

\bibitem{Challinor05}
A.~{Challinor} and A.~{Lewis}, \emph{{Lensed CMB power spectra from all-sky
  correlation functions}},
  \href{https://doi.org/10.1103/PhysRevD.71.103010}{\emph{\prd} {\bfseries 71}
  (2005) 103010} [\href{https://arxiv.org/abs/astro-ph/0502425}{{\ttfamily
  astro-ph/0502425}}].

\bibitem{Planck:2018vyg}
{\scshape Planck} collaboration, \emph{{Planck 2018 results. VI. Cosmological
  parameters}},
  \href{https://doi.org/10.1051/0004-6361/201833910}{\emph{Astron. Astrophys.}
  {\bfseries 641} (2020) A6}
  [\href{https://arxiv.org/abs/1807.06209}{{\ttfamily 1807.06209}}].

\bibitem{madau2014cosmic}
P.~Madau and M.~Dickinson, \emph{Cosmic star-formation history}, {\emph{Annual
  Review of Astronomy and Astrophysics} {\bfseries 52} (2014) 415}.

\bibitem{Ross_2015}
A.J.~Ross, L.~Samushia, C.~Howlett, W.J.~Percival, A.~Burden and M.~Manera,
  \emph{The clustering of the {SDSS} {DR}7 main galaxy sample {\textendash} i.
  a 4~per cent distance measure at z~=~0.15},
  \href{https://doi.org/10.1093/mnras/stv154}{\emph{Monthly Notices of the
  Royal Astronomical Society} {\bfseries 449} (2015) 835}.

\bibitem{Dalya:2021ewn}
G.~D\'alya et~al., \emph{{GLADE+~: an extended galaxy catalogue for
  multimessenger searches with advanced gravitational-wave detectors}},
  \href{https://doi.org/10.1093/mnras/stac1443}{\emph{Mon. Not. Roy. Astron.
  Soc.} {\bfseries 514} (2022) 1403}
  [\href{https://arxiv.org/abs/2110.06184}{{\ttfamily 2110.06184}}].

\bibitem{Bilicki:2013sza}
M.~Bilicki, T.H.~Jarrett, J.A.~Peacock, M.E.~Cluver and L.~Steward,
  \emph{{2MASS Photometric Redshift catalog: a comprehensive three-dimensional
  census of the whole sky}},
  \href{https://doi.org/10.1088/0067-0049/210/1/9}{\emph{Astrophys. J. Suppl.}
  {\bfseries 210} (2014) 9} [\href{https://arxiv.org/abs/1311.5246}{{\ttfamily
  1311.5246}}].

\bibitem{2MASSXSC}
T.H.~Jarrett, T.~Chester, R.~Cutri, S.~Schneider, M.~Skrutskie and J.P.~Huchra,
  \emph{{2mass extended source catalog: overview and algorithms}},
  \href{https://doi.org/10.1086/301330}{\emph{Astron. J.} {\bfseries 119}
  (2000) 2498} [\href{https://arxiv.org/abs/astro-ph/0004318}{{\ttfamily
  astro-ph/0004318}}].

\bibitem{2MASS}
M.F.~{Skrutskie}, R.M.~{Cutri}, R.~{Stiening}, M.D.~{Weinberg}, S.~{Schneider},
  J.M.~{Carpenter} et~al., \emph{{The Two Micron All Sky Survey (2MASS)}},
  \href{https://doi.org/10.1086/498708}{\emph{\aj} {\bfseries 131} (2006)
  1163}.

\bibitem{WISE2010}
E.L.~{Wright}, P.R.M.~{Eisenhardt}, A.K.~{Mainzer}, M.E.~{Ressler},
  R.M.~{Cutri}, T.~{Jarrett} et~al., \emph{{The Wide-field Infrared Survey
  Explorer (WISE): Mission Description and Initial On-orbit Performance}},
  \href{https://doi.org/10.1088/0004-6256/140/6/1868}{\emph{\aj} {\bfseries
  140} (2010) 1868} [\href{https://arxiv.org/abs/1008.0031}{{\ttfamily
  1008.0031}}].

\bibitem{WISEALLSKY}
R.M.~{Cutri} and {et al.}, \emph{{VizieR Online Data Catalog: WISE All-Sky Data
  Release (Cutri+ 2012)}}, {\emph{VizieR Online Data Catalog} (2012) II/311}.

\bibitem{Hambly:2001yj}
N.~Hambly et~al., \emph{{The supercosmos sky survey. paper I: introduction and
  description}},
  \href{https://doi.org/10.1111/j.1365-2966.2001.04660.x}{\emph{Mon. Not. Roy.
  Astron. Soc.} {\bfseries 326} (2001) 1279}
  [\href{https://arxiv.org/abs/astro-ph/0108286}{{\ttfamily
  astro-ph/0108286}}].

\bibitem{Hambly:2001yi}
N.~Hambly, M.~Irwin and H.~MacGillivray, \emph{{The supercosmos sky survey.
  paper. 2. Image detection, parameterisation, classification and photometry}},
  \href{https://doi.org/10.1111/j.1365-2966.2001.04661.x}{\emph{Mon. Not. Roy.
  Astron. Soc.} {\bfseries 326} (2001) 1295}
  [\href{https://arxiv.org/abs/astro-ph/0108290}{{\ttfamily
  astro-ph/0108290}}].

\bibitem{Hambly:2001yh}
N.~Hambly, C.~Davenhall, M.~Irwin and H.~MacGillivray, \emph{{The supercosmos
  sky survey. paper. 3. Astrometry}},
  \href{https://doi.org/10.1111/j.1365-2966.2001.04662.x}{\emph{Mon. Not. Roy.
  Astron. Soc.} {\bfseries 326} (2001) 1315}
  [\href{https://arxiv.org/abs/astro-ph/0108291}{{\ttfamily
  astro-ph/0108291}}].

\bibitem{ANNz}
A.A.~{Collister} and O.~{Lahav}, \emph{{ANNz: Estimating Photometric Redshifts
  Using Artificial Neural Networks}},
  \href{https://doi.org/10.1086/383254}{\emph{\pasp} {\bfseries 116} (2004)
  345} [\href{https://arxiv.org/abs/astro-ph/0311058}{{\ttfamily
  astro-ph/0311058}}].

\bibitem{sdss9}
C.P.~{Ahn}, R.~{Alexandroff}, C.~{Allende Prieto}, S.F.~{Anderson},
  T.~{Anderton}, B.H.~{Andrews} et~al., \emph{{The Ninth Data Release of the
  Sloan Digital Sky Survey: First Spectroscopic Data from the SDSS-III Baryon
  Oscillation Spectroscopic Survey}},
  \href{https://doi.org/10.1088/0067-0049/203/2/21}{\emph{\apjs} {\bfseries
  203} (2012) 21} [\href{https://arxiv.org/abs/1207.7137}{{\ttfamily
  1207.7137}}].

\bibitem{6df}
D.H.~{Jones}, M.A.~{Read}, W.~{Saunders}, M.~{Colless}, T.~{Jarrett},
  Q.A.~{Parker} et~al., \emph{{The 6dF Galaxy Survey: final redshift release
  (DR3) and southern large-scale structures}},
  \href{https://doi.org/10.1111/j.1365-2966.2009.15338.x}{\emph{\mnras}
  {\bfseries 399} (2009) 683}
  [\href{https://arxiv.org/abs/0903.5451}{{\ttfamily 0903.5451}}].

\bibitem{2DFGRS:2001zay}
{\scshape 2DFGRS} collaboration, \emph{{The 2dF Galaxy Redshift Survey: Spectra
  and redshifts}},
  \href{https://doi.org/10.1046/j.1365-8711.2001.04902.x}{\emph{Mon. Not. Roy.
  Astron. Soc.} {\bfseries 328} (2001) 1039}
  [\href{https://arxiv.org/abs/astro-ph/0106498}{{\ttfamily
  astro-ph/0106498}}].

\bibitem{Jarrett04}
T.~{Jarrett}, \emph{{Large Scale Structure in the Local Universe - The 2MASS
  Galaxy Catalog}}, \href{https://doi.org/10.1071/AS04050}{\emph{Publ. Astron.
  Soc. Austral.} {\bfseries 21} (2004) 396}
  [\href{https://arxiv.org/abs/astro-ph/0405069}{{\ttfamily
  astro-ph/0405069}}].

\bibitem{Bell:2003cj}
E.F.~Bell, D.H.~McIntosh, N.~Katz and M.D.~Weinberg, \emph{{The optical and
  near-infrared properties of galaxies. 1. Luminosity and stellar mass
  functions}}, \href{https://doi.org/10.1086/378847}{\emph{Astrophys. J.
  Suppl.} {\bfseries 149} (2003) 289}
  [\href{https://arxiv.org/abs/astro-ph/0302543}{{\ttfamily
  astro-ph/0302543}}].

\bibitem{Bilicki:2016irk}
M.~Bilicki et~al., \emph{{WISE x SuperCOSMOS photometric redshift catalog: 20
  million galaxies over 3pi steradians}},
  \href{https://doi.org/10.3847/0067-0049/225/1/5}{\emph{Astrophys. J. Suppl.}
  {\bfseries 225} (2016) 5} [\href{https://arxiv.org/abs/1607.01182}{{\ttfamily
  1607.01182}}].

\bibitem{ALLWISE}
R.M.~{Cutri}, E.L.~{Wright}, T.~{Conrow}, J.W.~{Fowler}, P.R.M.~{Eisenhardt},
  C.~{Grillmair} et~al., \emph{{VizieR Online Data Catalog: AllWISE Data
  Release (Cutri+ 2013)}}, {\emph{VizieR Online Data Catalog} (2021) II/328}.

\bibitem{GAMA}
S.P.~{Driver}, P.~{Norberg}, I.K.~{Baldry}, S.P.~{Bamford}, A.M.~{Hopkins},
  J.~{Liske} et~al., \emph{{GAMA: towards a physical understanding of galaxy
  formation}},
  \href{https://doi.org/10.1111/j.1468-4004.2009.50512.x}{\emph{Astronomy and
  Geophysics} {\bfseries 50} (2009) 5.12}
  [\href{https://arxiv.org/abs/0910.5123}{{\ttfamily 0910.5123}}].

\bibitem{GAMAII}
J.~{Liske}, I.K.~{Baldry}, S.P.~{Driver}, R.J.~{Tuffs}, M.~{Alpaslan},
  E.~{Andrae} et~al., \emph{{Galaxy And Mass Assembly (GAMA): end of survey
  report and data release 2}},
  \href{https://doi.org/10.1093/mnras/stv1436}{\emph{\mnras} {\bfseries 452}
  (2015) 2087} [\href{https://arxiv.org/abs/1506.08222}{{\ttfamily
  1506.08222}}].

\bibitem{HEALPix}
K.M.~{G{\'o}rski}, E.~{Hivon}, A.J.~{Banday}, B.D.~{Wandelt}, F.K.~{Hansen},
  M.~{Reinecke} et~al., \emph{{HEALPix: A Framework for High-Resolution
  Discretization and Fast Analysis of Data Distributed on the Sphere}},
  \href{https://doi.org/10.1086/427976}{\emph{\apj} {\bfseries 622} (2005) 759}
  [\href{https://arxiv.org/abs/arXiv:astro-ph/0409513}{{\ttfamily
  arXiv:astro-ph/0409513}}].

\bibitem{Zonca2019}
A.~Zonca, L.~Singer, D.~Lenz, M.~Reinecke, C.~Rosset, E.~Hivon et~al.,
  \emph{healpy: equal area pixelization and spherical harmonics transforms for
  data on the sphere in python},
  \href{https://doi.org/10.21105/joss.01298}{\emph{Journal of Open Source
  Software} {\bfseries 4} (2019) 1298}.

\bibitem{Balaguera-Antolinez:2017dpm}
A.~Balaguera-Antol\'\i{}nez, M.~Bilicki, E.~Branchini and A.~Postiglione,
  \emph{{Extracting cosmological information from the angular power spectrum of
  the 2MASS Photometric Redshift catalogue}},
  \href{https://doi.org/10.1093/mnras/sty262}{\emph{Mon. Not. Roy. Astron.
  Soc.} {\bfseries 476} (2018) 1050}
  [\href{https://arxiv.org/abs/1711.04583}{{\ttfamily 1711.04583}}].

\bibitem{Xavier:2018owe}
H.S.~Xavier, M.V.~Costa-Duarte, A.~Balaguera-Antol\'\i{}nez and M.~Bilicki,
  \emph{{All-sky angular power spectra from cleaned WISE\texttimes{}SuperCOSMOS
  galaxy number counts}},
  \href{https://doi.org/10.1088/1475-7516/2019/08/037}{\emph{JCAP} {\bfseries
  08} (2019) 037} [\href{https://arxiv.org/abs/1812.08182}{{\ttfamily
  1812.08182}}].

\bibitem{Kettlety2018}
T.~{Kettlety}, J.~{Hesling}, S.~{Phillipps}, M.N.~{Bremer}, M.E.~{Cluver},
  E.N.~{Taylor} et~al., \emph{{Galaxy and mass assembly (GAMA): the consistency
  of GAMA and WISE derived mass-to-light ratios}},
  \href{https://doi.org/10.1093/mnras/stx2379}{\emph{\mnras} {\bfseries 473}
  (2018) 776} [\href{https://arxiv.org/abs/1709.08316}{{\ttfamily
  1709.08316}}].

\bibitem{Fitzpatrick1999}
E.L.~{Fitzpatrick}, \emph{{Correcting for the Effects of Interstellar
  Extinction}}, \href{https://doi.org/10.1086/316293}{\emph{\pasp} {\bfseries
  111} (1999) 63} [\href{https://arxiv.org/abs/astro-ph/9809387}{{\ttfamily
  astro-ph/9809387}}].

\bibitem{Banerjee2010}
S.~{Banerjee}, H.~{Baumgardt} and P.~{Kroupa}, \emph{{Stellar-mass black holes
  in star clusters: implications for gravitational wave radiation}},
  \href{https://doi.org/10.1111/j.1365-2966.2009.15880.x}{\emph{\mnras}
  {\bfseries 402} (2010) 371}
  [\href{https://arxiv.org/abs/0910.3954}{{\ttfamily 0910.3954}}].

\bibitem{mukherjee2021impact}
S.~Mukherjee, T.~Broadhurst, J.M.~Diego, J.~Silk and G.F.~Smoot, \emph{Impact
  of astrophysical binary coalescence time-scales on the rate of lensed
  gravitational wave events}, {\emph{Mon. Not. Roy. Astron. Soc.} {\bfseries
  506} (2021) 3751}.

\bibitem{Fishbach:2021mhp}
M.~Fishbach and V.~Kalogera, \emph{{The Time Delay Distribution and Formation
  Metallicity of LIGO-Virgo\textquoteright{}s Binary Black Holes}},
  \href{https://doi.org/10.3847/2041-8213/ac05c4}{\emph{Astrophys. J. Lett.}
  {\bfseries 914} (2021) L30}
  [\href{https://arxiv.org/abs/2105.06491}{{\ttfamily 2105.06491}}].

\bibitem{abbott2020gw190425}
{\scshape LIGO Scientific, Virgo} collaboration, \emph{{GW190425: Observation
  of a Compact Binary Coalescence with Total Mass $\sim 3.4 M_{\odot}$}},
  \href{https://doi.org/10.3847/2041-8213/ab75f5}{\emph{Astrophys. J. Lett.}
  {\bfseries 892} (2020) L3}
  [\href{https://arxiv.org/abs/2001.01761}{{\ttfamily 2001.01761}}].

\bibitem{Chruslinska:2018hrb}
M.~Chruslinska, G.~Nelemans and K.~Belczynski, \emph{{The influence of the
  distribution of cosmic star formation at different metallicities on the
  properties of merging double compact objects}},
  \href{https://doi.org/10.1093/mnras/sty3087}{\emph{Mon. Not. Roy. Astron.
  Soc.} {\bfseries 482} (2019) 5012}
  [\href{https://arxiv.org/abs/1811.03565}{{\ttfamily 1811.03565}}].

\bibitem{Chruslinska:2022ovf}
M.~Chru\'sli\'nska, \emph{{Chemical Evolution of the Universe and its
  Consequences for Gravitational-Wave Astrophysics}},
  \href{https://doi.org/10.1002/andp.202200170}{\emph{Annalen Phys.} {\bfseries
  536} (2024) 2200170} [\href{https://arxiv.org/abs/2206.10622}{{\ttfamily
  2206.10622}}].

\bibitem{Artale_2020}
M.C.~Artale, Y.~Bouffanais, M.~Mapelli, N.~Giacobbo, N.B.~Sabha,
  F.~Santoliquido et~al., \emph{An astrophysically motivated ranking criterion
  for low-latency electromagnetic follow-up of gravitational wave events},
  \href{https://doi.org/10.1093/mnras/staa1252}{\emph{Monthly Notices of the
  Royal Astronomical Society} {\bfseries 495} (2020) 1841–1852}.

\bibitem{2023MNRAS.523.5719R}
L.~{Rauf}, C.~{Howlett}, T.M.~{Davis} and C.D.P.~{Lagos}, \emph{{Exploring
  binary black hole mergers and host galaxies with SHARK and COMPAS}},
  \href{https://doi.org/10.1093/mnras/stad1757}{\emph{\mnras} {\bfseries 523}
  (2023) 5719} [\href{https://arxiv.org/abs/2302.08172}{{\ttfamily
  2302.08172}}].

\bibitem{Salim07}
S.~{Salim}, R.M.~{Rich}, S.~{Charlot}, J.~{Brinchmann}, B.D.~{Johnson},
  D.~{Schiminovich} et~al., \emph{{UV Star Formation Rates in the Local
  Universe}}, \href{https://doi.org/10.1086/519218}{\emph{\apjs} {\bfseries
  173} (2007) 267} [\href{https://arxiv.org/abs/0704.3611}{{\ttfamily
  0704.3611}}].

\bibitem{Mannucci2010}
F.~{Mannucci}, G.~{Cresci}, R.~{Maiolino}, A.~{Marconi} and A.~{Gnerucci},
  \emph{{A fundamental relation between mass, star formation rate and
  metallicity in local and high-redshift galaxies}},
  \href{https://doi.org/10.1111/j.1365-2966.2010.17291.x}{\emph{\mnras}
  {\bfseries 408} (2010) 2115}
  [\href{https://arxiv.org/abs/1005.0006}{{\ttfamily 1005.0006}}].

\bibitem{Peng10}
Y.-j.~{Peng}, S.J.~{Lilly}, K.~{Kova{\v{c}}}, M.~{Bolzonella}, L.~{Pozzetti},
  A.~{Renzini} et~al., \emph{{Mass and Environment as Drivers of Galaxy
  Evolution in SDSS and zCOSMOS and the Origin of the Schechter Function}},
  \href{https://doi.org/10.1088/0004-637X/721/1/193}{\emph{\apj} {\bfseries
  721} (2010) 193} [\href{https://arxiv.org/abs/1003.4747}{{\ttfamily
  1003.4747}}].

\bibitem{Mapelli_2018}
M.~{Mapelli}, N.~{Giacobbo}, M.~{Toffano}, E.~{Ripamonti}, A.~{Bressan},
  M.~{Spera} et~al., \emph{{The host galaxies of double compact objects merging
  in the local Universe}},
  \href{https://doi.org/10.1093/mnras/sty2663}{\emph{\mnras} {\bfseries 481}
  (2018) 5324} [\href{https://arxiv.org/abs/1809.03521}{{\ttfamily
  1809.03521}}].

\bibitem{Guo2011}
Q.~{Guo}, S.~{White}, M.~{Boylan-Kolchin}, G.~{De Lucia}, G.~{Kauffmann},
  G.~{Lemson} et~al., \emph{{From dwarf spheroidals to cD galaxies: simulating
  the galaxy population in a {\ensuremath{\Lambda}}CDM cosmology}},
  \href{https://doi.org/10.1111/j.1365-2966.2010.18114.x}{\emph{\mnras}
  {\bfseries 413} (2011) 101}
  [\href{https://arxiv.org/abs/1006.0106}{{\ttfamily 1006.0106}}].

\bibitem{2001MNRAS.322..231K}
P.~{Kroupa}, \emph{{On the variation of the initial mass function}},
  \href{https://doi.org/10.1046/j.1365-8711.2001.04022.x}{\emph{\mnras}
  {\bfseries 322} (2001) 231}
  [\href{https://arxiv.org/abs/astro-ph/0009005}{{\ttfamily
  astro-ph/0009005}}].

\bibitem{2006csxs.book..157M}
J.E.~{McClintock} and R.A.~{Remillard}, \emph{{Black hole binaries}},  in
  \emph{Compact stellar X-ray sources}, vol.~39, pp.~157--213 (2006),
  \href{https://doi.org/10.48550/arXiv.astro-ph/0306213}{DOI}.

\bibitem{Gorski_2005}
K.M.~{G{\'o}rski}, E.~{Hivon}, A.J.~{Banday}, B.D.~{Wandelt}, F.K.~{Hansen},
  M.~{Reinecke} et~al., \emph{{HEALPix: A Framework for High-Resolution
  Discretization and Fast Analysis of Data Distributed on the Sphere}},
  \href{https://doi.org/10.1086/427976}{\emph{\apj} {\bfseries 622} (2005) 759}
  [\href{https://arxiv.org/abs/astro-ph/0409513}{{\ttfamily
  astro-ph/0409513}}].

\bibitem{2004MNRAS.348..885E}
G.~{Efstathiou}, \emph{{A maximum likelihood analysis of the low cosmic
  microwave background multipoles from the Wilkinson Microwave Anisotropy
  Probe}},
  \href{https://doi.org/10.1111/j.1365-2966.2004.07409.x}{\emph{\mnras}
  {\bfseries 348} (2004) 885}
  [\href{https://arxiv.org/abs/astro-ph/0310207}{{\ttfamily
  astro-ph/0310207}}].

\bibitem{Knox95}
L.~{Knox}, \emph{{Determination of inflationary observables by cosmic microwave
  background anisotropy experiments}},
  \href{https://doi.org/10.1103/PhysRevD.52.4307}{\emph{\prd} {\bfseries 52}
  (1995) 4307} [\href{https://arxiv.org/abs/astro-ph/9504054}{{\ttfamily
  astro-ph/9504054}}].

\bibitem{Leandro_2022}
H.~Leandro, V.~Marra and R.~Sturani, \emph{Measuring the hubble constant with
  black sirens},
  \href{https://doi.org/10.1103/physrevd.105.023523}{\emph{Physical Review D}
  {\bfseries 105} (2022) }.

\bibitem{2004ApJ...601....1W}
D.H.~{Weinberg}, R.~{Dav{\'e}}, N.~{Katz} and L.~{Hernquist}, \emph{{Galaxy
  Clustering and Galaxy Bias in a {\ensuremath{\Lambda}}CDM Universe}},
  \href{https://doi.org/10.1086/380481}{\emph{\apj} {\bfseries 601} (2004) 1}
  [\href{https://arxiv.org/abs/astro-ph/0212356}{{\ttfamily
  astro-ph/0212356}}].

\bibitem{2018MNRAS.481.1133P}
J.A.~{Peacock} and M.~{Bilicki}, \emph{{Wide-area tomography of CMB lensing and
  the growth of cosmological density fluctuations}},
  \href{https://doi.org/10.1093/mnras/sty2314}{\emph{\mnras} {\bfseries 481}
  (2018) 1133} [\href{https://arxiv.org/abs/1805.11525}{{\ttfamily
  1805.11525}}].

\bibitem{2018MNRAS.476.1050B}
A.~{Balaguera-Antol{\'\i}nez}, M.~{Bilicki}, E.~{Branchini} and
  A.~{Postiglione}, \emph{{Extracting cosmological information from the angular
  power spectrum of the 2MASS Photometric Redshift catalogue}},
  \href{https://doi.org/10.1093/mnras/sty262}{\emph{\mnras} {\bfseries 476}
  (2018) 1050} [\href{https://arxiv.org/abs/1711.04583}{{\ttfamily
  1711.04583}}].

\bibitem{2010ApJ...724..878T}
J.L.~{Tinker}, B.E.~{Robertson}, A.V.~{Kravtsov}, A.~{Klypin}, M.S.~{Warren},
  G.~{Yepes} et~al., \emph{{The Large-scale Bias of Dark Matter Halos:
  Numerical Calibration and Model Tests}},
  \href{https://doi.org/10.1088/0004-637X/724/2/878}{\emph{\apj} {\bfseries
  724} (2010) 878} [\href{https://arxiv.org/abs/1001.3162}{{\ttfamily
  1001.3162}}].

\bibitem{2010ApJ...717..379B}
P.S.~{Behroozi}, C.~{Conroy} and R.H.~{Wechsler}, \emph{{A Comprehensive
  Analysis of Uncertainties Affecting the Stellar Mass-Halo Mass Relation for 0
  < z < 4}}, \href{https://doi.org/10.1088/0004-637X/717/1/379}{\emph{\apj}
  {\bfseries 717} (2010) 379}
  [\href{https://arxiv.org/abs/1001.0015}{{\ttfamily 1001.0015}}].

\bibitem{Schechter76}
P.~{Schechter}, \emph{{An analytic expression for the luminosity function for
  galaxies.}}, \href{https://doi.org/10.1086/154079}{\emph{\apj} {\bfseries
  203} (1976) 297}.

\bibitem{Cole01}
S.~{Cole}, P.~{Norberg}, C.M.~{Baugh}, C.S.~{Frenk}, J.~{Bland-Hawthorn},
  T.~{Bridges} et~al., \emph{{The 2dF galaxy redshift survey: near-infrared
  galaxy luminosity functions}},
  \href{https://doi.org/10.1046/j.1365-8711.2001.04591.x}{\emph{\mnras}
  {\bfseries 326} (2001) 255}
  [\href{https://arxiv.org/abs/astro-ph/0012429}{{\ttfamily
  astro-ph/0012429}}].

\bibitem{dey2019overview}
A.~Dey, D.J.~Schlegel, D.~Lang, R.~Blum, K.~Burleigh, X.~Fan et~al.,
  \emph{Overview of the desi legacy imaging surveys}, {\emph{The Astronomical
  Journal} {\bfseries 157} (2019) 168}.

\bibitem{collaboration2022euclid}
E.~Collaboration, R.~Scaramella et~al., \emph{Euclid preparation-i. the euclid
  wide survey}, {\emph{A\&A} {\bfseries 662} (2022) A112}.

\bibitem{ivezic2019lsst}
{\v{Z}}.~Ivezi{\'c}, S.M.~Kahn, J.A.~Tyson, B.~Abel, E.~Acosta, R.~Allsman
  et~al., \emph{Lsst: from science drivers to reference design and anticipated
  data products}, {\emph{The Astrophysical Journal} {\bfseries 873} (2019)
  111}.

\bibitem{Dore:2019pld}
O.~Dor\'e et~al., \emph{{WFIRST: The Essential Cosmology Space Observatory for
  the Coming Decade}},  \href{https://arxiv.org/abs/1904.01174}{{\ttfamily
  1904.01174}}.

\bibitem{astropy:2013}
{Astropy Collaboration}, T.P.~{Robitaille}, E.J.~{Tollerud}, P.~{Greenfield},
  M.~{Droettboom}, E.~{Bray} et~al., \emph{{Astropy: A community Python package
  for astronomy}},
  \href{https://doi.org/10.1051/0004-6361/201322068}{\emph{\aap} {\bfseries
  558} (2013) A33} [\href{https://arxiv.org/abs/1307.6212}{{\ttfamily
  1307.6212}}].

\bibitem{astropy:2018}
{Astropy Collaboration}, A.M.~{Price-Whelan}, B.M.~{Sip{\H{o}}cz},
  H.M.~{G{\"u}nther}, P.L.~{Lim}, S.M.~{Crawford} et~al., \emph{{The Astropy
  Project: Building an Open-science Project and Status of the v2.0 Core
  Package}}, \href{https://doi.org/10.3847/1538-3881/aabc4f}{\emph{\aj}
  {\bfseries 156} (2018) 123}
  [\href{https://arxiv.org/abs/1801.02634}{{\ttfamily 1801.02634}}].

\bibitem{astropy:2022}
{Astropy Collaboration}, A.M.~{Price-Whelan}, P.L.~{Lim}, N.~{Earl},
  N.~{Starkman}, L.~{Bradley} et~al., \emph{{The Astropy Project: Sustaining
  and Growing a Community-oriented Open-source Project and the Latest Major
  Release (v5.0) of the Core Package}},
  \href{https://doi.org/10.3847/1538-4357/ac7c74}{\emph{apj} {\bfseries 935}
  (2022) 167} [\href{https://arxiv.org/abs/2206.14220}{{\ttfamily
  2206.14220}}].

\bibitem{Fixsen:1996nj}
D.J.~Fixsen, E.S.~Cheng, J.M.~Gales, J.C.~Mather, R.A.~Shafer and E.L.~Wright,
  \emph{{The Cosmic Microwave Background spectrum from the full COBE FIRAS data
  set}}, \href{https://doi.org/10.1086/178173}{\emph{Astrophys. J.} {\bfseries
  473} (1996) 576} [\href{https://arxiv.org/abs/astro-ph/9605054}{{\ttfamily
  astro-ph/9605054}}].

\end{thebibliography}\endgroup

\end{document}